\documentclass[aps,pre,twocolumn,nofootinbib,showkeys]{revtex4-1}
\pdfoutput=1
\usepackage[colorlinks,bookmarksopen,bookmarksnumbered,citecolor=red,urlcolor=red]{hyperref}

\newcommand\BibTeX{{\rmfamily B\kern-.05em \textsc{i\kern-.025em b}\kern-.08em
T\kern-.1667em\lower.7ex\hbox{E}\kern-.125emX}}

\usepackage{moreverb}

\usepackage[utf8]{inputenc}
\usepackage[english]{babel}
\usepackage{amsfonts,amssymb,amsbsy,amsmath}
\usepackage{graphicx}
\usepackage{placeins}
\usepackage{caption}
\usepackage{subcaption}
\usepackage{natbib}
\setcitestyle{authoryear}

\newcommand{\bsrm}[1]{\boldsymbol{\rm #1}}

\newcommand{\dbsrm}[1]{\dot{\boldsymbol{\rm #1}}}
\newcommand{\dd}{\mathrm{d}}
\newcommand{\GWN}{{\textsc{\tiny GWN}}}

\let\citein=\cite
\renewcommand{\cite}[1]{(\citein{#1})}

\begin{document}

\title{Stochastic parameterization of subgrid-scale processes in coupled ocean-atmosphere systems: Benefits and limitations of response theory}

\author{Jonathan Demaeyer}
\email{E-mail: Jonathan.Demaeyer@meteo.be}
\author{St\'{e}phane Vannitsem}
\affiliation{Royal Meteorological Institute of Belgium, Avenue Circulaire, 3, 1180 Brussels, Belgium}

\begin{abstract}
A stochastic subgrid-scale parameterization based on the Ruelle's response theory and proposed in~\citein{WL2012} is tested in the context of a low-order coupled ocean-atmosphere model for which a part of the atmospheric modes are considered as unresolved. A natural separation of the phase-space into an invariant set and its complement allows for an analytical derivation of the different terms involved in the parameterization, namely the average, the fluctuation and the long memory terms. In this case, the fluctuation term is an additive stochastic noise. Its application to the low-order system reveals that a considerable correction of the low-frequency variability along the invariant subset can be obtained, provided that the coupling is sufficiently weak. This new approach of scale separation opens new avenues of subgrid-scale parameterizations in multiscale systems used for climate forecasts.
\end{abstract}


\maketitle

\section{Introduction}
\label{sec:intro}

Atmospheric weather prediction and climate models display a limited predictability due to the intrinsic property of sensitivity to initial conditions. Additional factors also limit this predictability, such as the resolution limitation induced by the discretization of the numerical model implying that \emph{subgrid} scale processes are not resolved. The latter is usually considered as one of the most important source of \emph{model error}~\cite{L1978pred,DK1987,L82}. Other factors affecting the forecasts are the lack of representation of some processes at play or of the boundary conditions of the systems (see e.g.~\citein{N2005,N2007,NPV2009}). This limitation is also a crucial problem for climate prediction at seasonal, interannual and decadal timescales for which the model is considerably drifting from the reality, as discussed for instance in~\citein{Da2009}. \\

Besides increasing the resolution to include subgrid scales processes, another approach consists in representing their effects through an appropriate parameterization. It was soon recognized that for systems with multiple timescales, the rapidly varying nature of the faster component implies that a deterministic parameterization (described by a deterministic function) is not sufficient to take into account their variability. This is particularly true for climate systems, for which stochastic models were proposed to describe the impact of the fast components of the system on the slow variables~\cite{H1976,NN1981}. Since then, multiple methods have been explored to derive stochastic parameterizations: basic empirical schemes whose purpose is to increase the variability of the models and of ensemble forecasts~\cite{BMP1999,AMP2013,BD2015}, stochastic backscatter schemes allowing for a reduction of the impact of small scale dissipation~\cite{FD1997,F1999,S2005}, various averaging methods~\cite{AIW2003,CKM2011,V2014} in the spirit of~\citein{H1976}, allowing for a systematic derivation of the noise for a large timescale separation, coarse graining and modeling based on conditional Markov chains~\cite{CV2008}, singular perturbation theory for Markov processes~\cite{MTV2001}, and techniques based on the fluctuation-dissipation theorem~\cite{A2012,A2013}.\\

Recently, a new framework based on the Ruelle response theory~\cite{R1997,R2009} of statistical mechanics has been proposed~\cite{WL2012} (See also the discussion in \citein{FOBWL2015}.). Initially derived for systems possessing a Sinai-Ruelle-Bowen (SRB) measure~\cite{Y2002}, this framework can be applied to a broader class of systems provided that the chaotic hypothesis~\cite{GC1995} is fulfilled. This approach has the interesting property that it does not require a coarse-graining of the phase space nor a clear timescale separation between the resolved and unresolved components of the system under consideration. The latter point is of particular interest in the case of atmospheric modeling, for which no clear spectral gaps between the different scales is present~\cite{D2004,LS2013}. Moreover, it can be shown to give the same result as the classical projection theory of statistical mechanics~\cite{WL2013}, which states that an optimal parameterization should include deterministic, stochastic and non-Markovian effects, as also discussed in~\citein{CLW2015}.

 The study of coupled ocean-atmosphere climatic models is an important topic, aiming at the understanding of phenomena based on the local ocean-atmosphere coupling, like the El-Ni\~{n}o Southern Oscillation (ENSO) or possibly the North-Atlantic Oscillation (NAO). Several efforts have been made to build a stochastic parameterization to model the impact of the fast atmosphere on the slow ocean, e.g.~\citein{AIW2003}. For instance, \citein{V2014} considered a low-order ocean model mechanically coupled to the atmosphere, and featuring a double-gyre. The parameterization was done through an averaging method and yielded an additive stochastic process to model the atmosphere. In the present paper, we assess the performance of the response theory framework in the context of a low-order ocean-atmosphere coupled system, by parameterizing a subset of the atmospheric variables and studying the effect of the parameterization on the rest of the atmopshere and on the ocean. This question is much more involved than the one considered in~\citein{V2014} due to the fact that there is not a clear separation of timescales anymore, and therefore one needs to consider alternative parameterization schemes than the averaging approach of~\citein{AIW2003}. The approach of~\citein{WL2012} offers a very interesting alternative that does not require timescale separation but rather a weak coupling.\\

The model is a 36-dimensional quasi-geostrophic ocean-atmosphere model proposed in~\citein{VDDG2015}. The coupling between the ocean and the atmosphere is mechanical (through the wind stress) as well as thermal (with radiative and heat fluxes). This model displays a low-frequency variability (LFV) on the bidecadal timescale, typical of the dynamics effectively observed in the atmosphere~\cite{DM2000,KDWBMG2007}. This LFV is related to the presence of a long-period periodic orbit that forms the backbone of the attractor. Indeed, this orbit emerges for a sufficiently large value of the radiative input via a Hopf bifurcation and then develops in a subspace of the full phase space. This subspace is invariant under the action of the Jacobian defining the advection term typically appearing in the quasi-geostrophic equations. Therefore, the orbits living in this subspace remain stable while they may destabilize in the other directions. It offers thus a partition of the phase space where typically periodic orbits arise via Hopf bifurcations~\cite{CP1996,CP1997,CP1999}. In the system considered here, the invariant subspace is related to the long time-scale and the fast dynamics is thus organized around the solution generated in this ``slow'' subspace. This natural decomposition of the phase space into an invariant slow subspace and its complement is particularly suitable to test the response theory parameterization because it allows for an analytic derivation of the quantities involved in the theory. In addition, as stated above, this parameterization method requires a weak coupling between the resolved and unresolved components. In the model considered here, it is present because the unresolved atmospheric part is mechanically weakly coupled to the resolved ocean through the wind stress forcing and, as we will show, it induces noticeable changes to the low-frequency variability. The response theory method is thus convenient to address this problem and correct this low-frequency signal.\\

In Section~\ref{sec:WL}, we introduce briefly the method based on the linear response theory developed in~\citein{WL2012}. In Section~\ref{sec:oa-mod}, we describe the coupled ocean-atmosphere model on which the method is applied. The results of this procedure for a particular subgrid configuration, leading to an additive noise, are presented in Section~\ref{sec:res}. Finally, the conclusions and implications of these results are provided in Section~\ref{sec:conc}.

\section{The Response Theory approach to subgrid parameterization}
\label{sec:WL}

Given a dynamical system $\dbsrm{Z} = \bsrm{F}(\bsrm{Z})$ which represents a real system, one way to assess subgrid parameterization is to assumes that its variables can be decomposed into two sets $\bsrm{X}$ and $\bsrm{Y}$, where the former is the resolved part and the latter is the unresolved part. The parameterization method can thus be tested on a well-defined closed problem, for which $\bsrm{X}$ are the variables of interest and $\bsrm{Y}$ the source of model error.\\

In the response theory approach, the system is conveniently rewritten as :
\begin{equation}
  \label{eq:wlcmod}
  \left\{
    \begin{array}{lcl}
      \dbsrm{X} & = & \bsrm{F}_{X}(\bsrm{X}) + \bsrm\Psi_{X}(\bsrm{X},\bsrm{Y})  \\
      \dbsrm{Y} & = & \bsrm{F}_{Y}(\bsrm{Y}) + \bsrm\Psi_{Y}(\bsrm{X},\bsrm{Y}) 
    \end{array}
    \right.
\end{equation}
The couplings $\bsrm\Psi_{X}$, $\bsrm\Psi_{Y}$ are seen as perturbations to $\bsrm{F}_{X}$ and $\bsrm{F}_{Y}$, and they are assumed to be sufficiently weak so that the formal setting presented in this section gives good results. In this context, a subgrid parameterization is a method that aims at obtaining the reduced system:
\begin{equation}
  \label{eq:redmod}
  \dbsrm{X} = \bsrm{F}_{X}(\bsrm{X}) + \bsrm{\Xi}(\bsrm{X},t) 
\end{equation}
that displays statistical properties similar to the $\bsrm{X}$-component of the full system and where $\bsrm{\Xi}$ is a process to be determined by the method. We now  briefly sketch how the Ruelle response theory~\cite{R1997,R2009} can be used to derive such a parameterization.\\

The response theory characterizes the contribution of the ``perturbation'' $\bsrm\Psi_X$, $\bsrm\Psi_Y$ to the invariant measure\footnote{The theory assumes that for the system under consideration, a SRB measure exists (e.g. an Axiom-A system).} $\tilde\rho$ of the coupled system as:
\begin{equation}
  \label{eq:pertrho}
  \tilde\rho = \rho_0 + \delta_\Psi \rho^{(1)} + \delta_{\Psi,\Psi} \rho^{(2)} + O(\Psi^3)
\end{equation}
where $\rho_0$ is the invariant measure of the uncoupled system which is also supposed to be an existing, well defined SRB measure. The terms $\delta_\Psi \rho^{(1)}$ and $\delta_{\Psi,\Psi} \rho^{(2)}$ are respectively the first and second order responses to the perturbation.
As shown in~\citein{WL2012}, this theory gives the framework to parameterize the effect of the coupling on the component $\bsrm{X}$. Indeed, the authors derived a parameterization $\bsrm\Xi(\bsrm{X},t)$ composed of three different terms having a response similar, up to order two, to the couplings $\bsrm\Psi_X$ and $\bsrm\Psi_Y$:
\begin{equation}
  \label{eq:WLres}
  \bsrm{\Xi}(\bsrm{X},t) = \bsrm{M}_1(\bsrm{X}) + \bsrm{M}_2(\bsrm{X},t) + \bsrm{M}_3(\bsrm{X},t)
\end{equation}
with
\begin{align}
  \bsrm{M}_1(\bsrm{X}) & = \Big\langle \bsrm\Psi_X(\bsrm{X},\bsrm{Y}) \Big\rangle_{\rho_{0,Y}} \label{eq:M1def} \\
  \bsrm{M}_2(\bsrm{X},t) & = \bsrm\Psi_{X,1}(\bsrm{X}) \, \bsrm\sigma(t) \label{eq:M2def}\\
  \bsrm{M}_3(\bsrm{X},t) & = \int_0^\infty \dd s \, \bsrm{h}(\bsrm{X}(t-s),s). \label{eq:M3def}
\end{align}
with $\rho_{0,Y}$ the invariant measure of the unperturbed system $\dbsrm{Y} = \bsrm{F}_Y(\bsrm{Y})$, and where 
\begin{equation}
  \label{eq:psipdef}
  \bsrm\Psi_X^\prime(\bsrm{X},\bsrm{Y})=\bsrm\Psi_X(\bsrm{X},\bsrm{Y})-\bsrm{M}_1(\bsrm{X})
\end{equation}
is assumed to be separable\footnote{It is in principle always possible to find a basis of functions over which such a decomposition is possible.},
\begin{equation}
  \bsrm\Psi_X^\prime(\bsrm{X},\bsrm{Y})= \bsrm\Psi_{X,1}(\bsrm{X}) \,\bsrm\Psi_{X,2}(\bsrm{Y})
\end{equation}
The process $\bsrm\sigma(t)$ is a stochastic process such that:
\begin{equation}
  \label{eq:sigma}
  \Big\langle \bsrm\sigma(t_1) \otimes \bsrm\sigma(t_2) \Big\rangle = \quad \bsrm{g}(t_1-t_2)
\end{equation}
with the \emph{cross-correlation}
\begin{equation}
  \label{eq:ggendef}
  \bsrm{g}(s) = \Big \langle \bsrm\Psi_{X,2}^\prime(\bsrm{Y})  \otimes \bsrm\Psi_{X,2}^\prime\Big(\bsrm\phi^s_Y(\bsrm{Y})\Big) \Big \rangle_{\rho_{0,Y}}
\end{equation}
where $\otimes$ is the outer product and $\bsrm\phi_Y^s$ is the flow of the unperturbed system $\dbsrm{Y} = \bsrm{F}_Y(\bsrm{Y})$.
The function $\bsrm{h}$ inside the term $\bsrm{M}_3$ is the memory kernel and is written :
\begin{equation}
  \label{eq:M3gen}
  \bsrm{h} (\bsrm{X},s) = \Big\langle \bsrm\Psi_Y(\bsrm{X},\bsrm{Y}) \cdot \bsrm\nabla_Y \bsrm\Psi_X\Big(\bsrm\phi^s_X(\bsrm{X}),\bsrm\phi^s_Y(\bsrm{Y})\Big) \Big\rangle_{\rho_{0,Y}}
\end{equation}
where $\bsrm\phi_X^s$ is the flow of the unperturbed system $\dbsrm{X} = \bsrm{F}_X(\bsrm{X})$.
We note that the terms $\bsrm{M}_1$, $\bsrm{M}_2$ and $\bsrm{M}_3$ are respectively an averaging, a fluctuation and a memory term and that their responses up to order two match the response of the perturbation. Consequently, this ensures that for a \emph{weak coupling}, the response of the parameterization~(\ref{eq:WLres}) on the observables will be roughly the same as the coupling.

A key point about the applicability of the method is that only the measure of the uncoupled $\bsrm{Y}$-dynamics is needed to compute the quantities involved in the perturbative approach, in contrast to the averaging method~\cite{AIW2003,CKM2011,V2014} for which a measure conditional on the value of $\bsrm{X}$ is necessary. Finally, we note that no clear timescale separations between the $\bsrm{X}$ and $\bsrm{Y}$ components are assumed in this approach. The only requirement here is a weak coupling between the resolved and unresolved components. We now turn to a brief description of the coupled ocean-atmosphere model and its properties.

\section{The ocean-atmosphere coupled model}
\label{sec:oa-mod}

The coupled ocean-atmosphere model for midlatitudes is composed of a two-layer atmosphere over a shallow-water ocean layer on a $\beta$-plane~\cite{VDDG2015}. The ocean is considered as a closed basin with no-flux boundary conditions, while the atmosphere is defined in a channel, periodic in the zonal direction and with free-slip boundary conditions along the meridional boundaries. The model incorporates both a weak frictional coupling and an energy balance scheme which accounts for radiative and heat fluxes coupling between the ocean and the atmosphere. The latter coupling is typically several orders of magnitude greater than the former. The ocean temperature field in the model is a passively advected scalar, meaning that its impact on the oceanic transport features is only indirect, through the atmospheric feedbacks. Therefore, the oceanic and atmospheric components are subtly intertwined.\\

The dynamical fields of the model include the atmospheric barotropic streamfunction $\psi_{\rm a}$ and temperature anomaly $T_{\rm a}=2 \frac{f_0}{R} \theta_{\rm a}$ (with $f_0$ the Coriolis parameter at midlatitude and $R$ the Earth radius) as well as the oceanic streamfunction $\psi_{\rm o}$ and temperature anomaly $T_{\rm o}$. In order to compute the time evolution of these fields, they are expanded in Fourier series with a basis of functions satisfying orthogonality and the aforementioned boundary conditions. The result is a set of thirty-six ODEs for the coefficients of the expansion. The phase-space dimension of the atmosphere is 20, with 10 variables\footnote{We use here the same notations for the variables as in~\citein{VDDG2015}.} $\psi_{{\rm a},i}$ representing the barotropic streamfunction and 10 variables $\theta_{{\rm a},i}$ representing the temperature. The phase-space dimension of the ocean is 16, with 8 variables $\psi_{{\rm o},i}$ for the ocean streamfunction and 8 variables $\theta_{{\rm o},i}$ for the temperature. The main parameters of the model are listed in Tables~\ref{tab:descr}.\\

\begin{table*}
  \centering
  \begin{tabular}{| @{\hspace{.5cm}} c @{\hspace{0.5cm}} | | @{\hspace{.5cm}} c @{\hspace{.5cm}} |}
    \hline
    Parameter & Description (Unit)\\
    \hline 
    $\lambda$ & Heat exchange between the ocean and atmosphere. (Wm$^{-2}$K$^{-1}$) \\
    $r$ & Friction coefficient at the ocean bottom. (s$^{-1}$) \\
    $d$ & Friction coefficient between the ocean and the atmosphere (s$^{-1}$) \\
    $C_{\rm o}$ & Net short-wave radiation input for the ocean (Wm$^{-2}$) \\
    $k_d$ & Friction coefficient at the bottom of the atmosphere (s$^{-1}$) \\
    $k_d^\prime$ & Internal friction between the atmospheric layers  (s$^{-1}$) \\
    $H$ & Depth of the ocean layer (m) \\
    $G_{\rm o}$ & Specific heat capacity of the ocean (Jm$^{-2}$K$^{-1}$) \\
    $G_{\rm a}$ & Specific heat capacity of the atmosphere (Jm$^{-2}$K$^{-1}$) \\
    $q_{\rm a}$, $q_{\rm o}$, $q_Y$ & Variance of the Gaussian white noise for each component\\
    \hline
  \end{tabular}
  \caption{Description of the main parameters of the model. \label{tab:descr}}
\end{table*}

Because we retain only ten atmospheric and eight oceanic modes, we add a weak stochastic noise to the atmospheric equations in order to take into account the missing subgrid scales that do not destroy the deterministic dynamics found in~\citein{VDDG2015}. Moreover, on a more general perspective, the addition of noise can be seen as a way to regularize the measure of the system and to avoid the rigorous requirement of the theory of a well-defined (SRB) measure in deterministic systems~\cite{LC2014}.\\

In~\citein{VDDG2015}, it was shown that this model (without stochastic noise) displays a pronounced low-frequency variability (LFV), with a typical bidecadal period. This LFV is due to the presence of a long periodic orbit which remains stable in a 17-dimensional invariant manifold of the dynamics. This invariant manifold in the dynamical system is related to an invariant subspace of the Jacobian appearing in the partial differential equations of the model:
\begin{equation}
  \label{eq:Jac}
  J(\psi,\phi) = \frac{\partial \psi}{\partial x} \; \frac{\partial \phi}{\partial y} - \frac{\partial \psi}{\partial y} \; \frac{\partial \phi}{\partial x} \quad ,
\end{equation}
as shown in~\citein{CP1996}. The invariance can be stated as follow: for two fields $\psi$ and $\phi$ of the system\footnote{i.e. satifying the mass conservation and the boundary conditions.} belonging to this invariant subspace, the Jacobian maps these functions on another function of the subspace. Consequently, the variables of the ODEs corresponding to the basis functions of the invariant subspace form an invariant manifold. The Jacobian being related to the advection and to the nonlinear interactions, we shall see in the following section that the invariance thus generates a particular ODEs structure.\\

In the full phase space, the long periodic orbit is unstable and chaos develops in its vicinity. Therefore, the orbit forms the ``backbone'' of the chaotic attractor, superposing a fast dynamics on the slow evolution along the cycle. Moreover, the slow dynamics is not purely oceanic, but a ocean-atmosphere coupled mode of low-frequency variability. These properties make this model an interesting candidate to test parameterization methods, especially with the presence of this intricate coupled mode of variability. The idea is then to keep the modes pertaining to the slow subspace and model as noise the atmospheric variables falling outside this subset.

\section{Results}
\label{sec:res}

\subsection{Subgrid parameterization for the model}
\label{sec:red_add}

We consider in the following that the unresolved component contains only a subset of the atmospheric modes. The general structure of the equations can then be decomposed as
\begin{equation}
  \label{eq:oastruc}
  \left\{
    \begin{array}{lcl}
      \dbsrm{X}_{\rm a} & = & \bsrm{f}_{\rm a}(\bsrm{X}) + q_{\rm a} \, \bsrm\xi_{\rm a}(t) + \varepsilon \, \big(\bsrm{R}^{\rm a} \cdot \bsrm{Y} \\ 
      & & \qquad \qquad + \bsrm{Y}^{\rm T} \cdot \bsrm{C}^{\rm a} \cdot \bsrm{Y} + \bsrm{X}^{\rm T} \cdot \bsrm{V}^{\rm a} \cdot \bsrm{Y}\big) \\ \\
      \dbsrm{X}_{\rm o} & = & \bsrm{f}_{\rm o}(\bsrm{X}) + q_{\rm o} \, \bsrm\xi_{\rm o}(t) + \varepsilon \,\, \bsrm{R}^{\rm o} \cdot \bsrm{Y} \\ \\
      \dbsrm{Y} & = & \bsrm{A} \cdot \bsrm{Y} + q_Y \,  \bsrm\xi_Y (t) + \varepsilon \, \big( \bsrm{R}^Y \cdot \bsrm{X} \\
      & & \qquad \qquad + \bsrm{X}^{\rm T} \cdot \bsrm{V}^Y \cdot \bsrm{Y}  + \bsrm{X}^{\rm T} \cdot \bsrm{C}^Y \cdot \bsrm{X}\big)
    \end{array}
    \right.
\end{equation}
with $\bsrm{X}=(\bsrm{X}_{\rm a},\bsrm{X}_{\rm o})$ and where $\bsrm{X}_{\rm a}$, $\bsrm{X}_{\rm o}$ and $\bsrm{Y}$ are respectively the resolved atmospheric variables, the oceanic variables (which are thus fully resolved), and the unresolved atmospheric variables. The symbols $\bsrm{R}^{{\rm a},{\rm o},Y}$ and $\bsrm{C}^{{\rm a},Y}$ denote respectively matrices and tensors which encode the linear and quadratic dependence in the other component variables. The symbols $\bsrm{V}^{{\rm a},Y}$ denote tensors which represent the quadratic interaction terms with the other component. The product of two vectors $\bsrm{u}$ and $\bsrm{v}$ with a tensor $\bsrm{T}$ is here defined as:
\[\bsrm{u}^{\rm T}\cdot\bsrm{T}\cdot\bsrm{v}_{i} = \sum_{jk} T_{ijk} \, u_j \, v_k \quad .\]
The vectors of uncorrelated standard Gaussian white noise process added to each component are denoted $\bsrm\xi_{{\rm a},{\rm o},Y}$. As stated in the previous section, the presence of noise is justified to account for the subgrid scales that are not accounted for by the model and to regularize the measures. The $q$'s give the amplitude of the noise in each component. To study the dependence of the method on the coupling strength, we have also introduced the coupling parameter $\varepsilon$, with $\varepsilon=1$ being the original ocean-atmosphere system.\\

Now, we take the particular choice to select only atmospheric variables that are not in the 17-dimensional invariant manifold mentioned in the previous section. This allows to keep the fundamental structure of the long periodic orbit at the origin of the low-frequency variability. With this constraint, the subgrid configurations have to be defined by selecting the unresolved variables:
\[\psi_{{\rm a},2},\,\psi_{{\rm a},3},\,\psi_{{\rm a},4},\,\psi_{{\rm a},7},\,\psi_{{\rm a},8},\,\theta_{{\rm a},2},\,\theta_{{\rm a},3},\,\theta_{{\rm a},4},\,\theta_{{\rm a},7},\,\theta_{{\rm a},8}\]
The other atmospheric variable are considered as resolved:
\[\psi_{{\rm a},1},\,\psi_{{\rm a},5},\,\psi_{{\rm a},6},\,\psi_{{\rm a},9},\,\psi_{{\rm a},10},\,\theta_{{\rm a},1},\,\theta_{{\rm a},5},\,\theta_{{\rm a},6},\,\theta_{{\rm a},9},\,\theta_{{\rm a},10}\]
Among all possible, this specific choice is a severe truncation but it has the important advantage to simplify the structure of the coupling between the variables $\bsrm{X}$ and $\bsrm{Y}$. Indeed, with this decomposition
\begin{equation}
  \label{eq:oamod}
  \left\{
    \begin{array}{lcl}
      \dbsrm{X}_{\rm a} & = & \bsrm{f}_{\rm a}(\bsrm{X}) + q_{\rm a} \, \bsrm\xi_{\rm a}(t) + \varepsilon \,\, \bsrm{Y}^{\rm T} \cdot \bsrm{C}^{\rm a} \cdot \bsrm{Y} \\
      \dbsrm{X}_{\rm o} & = & \bsrm{f}_{\rm o}(\bsrm{X}) + q_{\rm o} \, \bsrm\xi_{\rm o}(t) + \varepsilon \,\, \bsrm{R}^{\rm o} \cdot \bsrm{Y} \\
      \dbsrm{Y} & = & \bsrm{A} \cdot \bsrm{Y} + q_Y \,  \bsrm\xi_Y (t) + \varepsilon \, \big( \bsrm{R}^Y \cdot \bsrm{X} + \bsrm{X}^{\rm T} \cdot \bsrm{V}^Y \cdot \bsrm{Y}\big)
    \end{array}
  \right.
\end{equation}
It is interesting to note the different orders of magnitude (in non-dimensional units) of the coupling between each components :
\begin{align}
  |\bsrm{R}^{\rm o}_\psi |& \sim d \sim 10^{-8} \label{eq:Ropsi}\\
  |\bsrm{R}^{\rm o}_T| & \sim \lambda^\prime_{\rm o} = \lambda/(G_{\rm o} f_0) \sim 10^{-4} \label{eq:RoT}\\
  |\bsrm{R}^Y| & \sim k_d, \lambda^\prime_{\rm a} \sim 10^{-2} \label{eq:RY}\\
  |\bsrm{C}^{\rm a}| & \sim |\bsrm{V}^Y| \sim 1 \label{eq:CVY}
\end{align}
where $\bsrm{R}^{\rm o}_\psi$ is associated to the wind stress coupling and $\bsrm{R}^{\rm o}_T$ is associated to the heat exchange coupling, $\bsrm{R}^{\rm o}_\psi + \bsrm{R}^{\rm o}_T = \bsrm{R}^{\rm o}$. While the resolved atmosphere is not weakly coupled to unresolved component, the transport in the ocean (associated with the wind stress) can be considered as weakly coupled to it. The heat and radiative exchange coupling (Eq.~(\ref{eq:RoT})) plays an intermediate role between the ocean transport coupling (Eq.~(\ref{eq:Ropsi})) and the strong atmospheric coupling (Eq.~(\ref{eq:CVY})). We thus expect that the best corrected feature of the model will be the transport, directly at the origin of the low-frequency variability which is a remarkable characteristic of the model.

In the framework of the response theory approach, the decomposition~(\ref{eq:oamod}) of the system reads as:
\begin{equation}
  \label{eq:roamod}
  \left\{
  \begin{array}{lcl}
    \dbsrm{X}_{\rm a} & = & \bsrm{F}_{\rm a}(\bsrm{X}) + \bsrm\Psi_{X_{\rm a}}(\bsrm{Y})  \\
    \dbsrm{X}_{\rm o} & = & \bsrm{F}_{\rm o}(\bsrm{X}) + \bsrm\Psi_{X_{\rm o}}(\bsrm{Y}) \\
    \dbsrm{Y} & = & \bsrm{F}_Y(\bsrm{Y}) + \bsrm\Psi_Y(\bsrm{X},\bsrm{Y})  
  \end{array}
  \right.
\end{equation}
where the noise terms are considered as included in the unperturbed tendencies $\bsrm{F}_{\rm a}$, $\bsrm{F}_{\rm o}$ and $\bsrm{F}_Y$, including the noise terms $q_{\rm a} \,  \bsrm\xi_{\rm a}$, $q_{\rm o} \,  \bsrm\xi_{\rm o}$ and $q_Y \,  \bsrm\xi_Y$. This choice is natural since these terms do not couple the three components. Defining
 \[\bsrm\Psi_X(\bsrm{Y})= \left[
 \begin{array}{c}
   \bsrm\Psi_{X_{\rm a}}(\bsrm{Y}) \\ \bsrm\Psi_{X_{\rm o}}(\bsrm{Y})
 \end{array}
 \right]
 \]
Eqs.~(\ref{eq:M1def}), (\ref{eq:M2def}) and (\ref{eq:M3def}) become:
\begin{align}
  \bsrm{M}_1 & = \Big\langle \bsrm\Psi_X(\bsrm{Y}) \Big\rangle_{\rho_{0,Y}} \label{eq:M1Y}\\
  \bsrm{M}_2(t) & = \bsrm{\sigma}(t) \\
  \bsrm{M}_3(\bsrm{X},t) & = \int_0^\infty \dd s \, \bsrm{h}(\bsrm{X}(t-s),s). \label{eq:M3adef}
\end{align}
with $\big\langle \bsrm{\sigma}(t) \otimes \bsrm{\sigma}(t^\prime)\big\rangle = \bsrm{g}(t-t^\prime)$ and 
\begin{equation}
  \label{eq:gdef}
  \bsrm{g}(s) = \Big \langle \bsrm\Psi_X^\prime(\bsrm{Y}) \otimes \bsrm\Psi_X^\prime\Big(\bsrm\phi^s_Y(\bsrm{Y})\Big) \Big \rangle_{\rho_{0,Y}}
\end{equation}
where $\bsrm\Psi_X^\prime(\bsrm{Y}) = \bsrm\Psi_X(\bsrm{Y})-\bsrm{M}_1$.
Therefore, as a direct consequence of the structure of the system (\ref{eq:oamod}), the fluctuation term $\bsrm{M}_2$ in the response theory parameterization is an \emph{additive stochastic process}.\\
Finally, the memory kernel is given by:
\begin{equation}
  \label{eq:M3}
  \bsrm{h} (\bsrm{X},s) = \Big\langle \bsrm\Psi_Y(\bsrm{X},\bsrm{Y}) \cdot \bsrm\nabla_Y \bsrm\Psi_X\Big(\bsrm\phi^s_Y(\bsrm{Y})\Big) \Big\rangle_{\rho_{0,Y}}
\end{equation}
and as another consequence of this particular decomposition, only the dynamics $\dbsrm{Y} = \bsrm{F}_Y(\bsrm{Y})$ of the unperturbed $\bsrm{Y}$-component have to be computed during the time evolution. This dynamics reduces to an Ornstein-Uhlenbeck process:
\begin{equation}
  \label{eq:OU}
  \dd \bsrm{Y} = \bsrm{A}\cdot \bsrm{Y}\, \dd t + q_Y \, \dd \bsrm{W}_Y(t)
\end{equation}
whose moments, correlations and invariant measure are well-known. Consequently, the computation of the averages (\ref{eq:M1Y}), (\ref{eq:gdef}) and (\ref{eq:M3}) over this measure is tractable analytically:
\begin{itemize}
\item The averaging term $\bsrm{M}_1$ is given by:
\[  \bsrm{M}_1 =\left[
  \begin{array}{c}
    \bsrm{M}_1^{\rm a} \\ \bsrm{M}_1^{\rm o}
  \end{array}
  \right]\]
with
\begin{align}
  \bsrm{M}_{1}^{\rm a} & = \varepsilon \; \mathrm{Tr}_{24} \mathrm{Tr}_{35} \, \left(\bsrm{C}^{\rm a}  \otimes \bsrm\sigma^Y \right) \label{eq:M1a} \\
  \bsrm{M}_{1}^{\rm o} & = 0 \label{eq:M1o}
\end{align}
where we have thus separated the atmospheric and oceanic parts. The symbol $\otimes$ is here the tensor outer product and the symbol $\mathrm{Tr}_{ij}$ indicates that the trace has been taken on the indices $i$ and $j$ of a tensor with more than two indices. The matrix $\bsrm\sigma^Y = \big\langle \bsrm{Y}^2 \big\rangle_{\rho_{0,Y}}$ is the covariance matrix of the unresolved dynamics.
\item The correlation function (\ref{eq:gdef}) defining the process $\bsrm{M}_2$ is given by:
\[  \bsrm{g}(s) = \left[
    \begin{array}[c]{cc}
      \bsrm{g}_{\rm a}(s) & 0 \\
      0 & \bsrm{g}_{\rm o}(s)
    \end{array}
    \right]
\]
with
\begin{multline}
  \label{eq:gsa}
  g_{{\rm a},ij}(s) = \\ \varepsilon^2 \; \mathrm{Tr} \left(\bsrm\sigma^Y \cdot \bsrm{E}^{\rm T}({s}) \cdot \bsrm{C}^{\rm a}_j \cdot \bsrm{E}({s})  \cdot \bsrm\sigma^Y \cdot \left(\bsrm{C}^{\rm a}_i + \bsrm{C}^{\rm aT}_i\right)\right)
\end{multline}
\begin{multline}
  \label{eq:gso}
  \bsrm{g}_{{\rm o}}(s) = \varepsilon^2 \; \bsrm{R}^{\rm o}  \cdot \bsrm\sigma^Y \cdot \bsrm{E}^{\rm T}(s) \cdot {\bsrm{R}^{\rm o}}^{\rm T}
\end{multline}
where $\bsrm{E}(t) = \exp(\bsrm{A} t)$ and $\bsrm{E}^{\rm T}(t)$ is the transpose of $\bsrm{E}(t)$. $\mathrm{Tr}$ is the usual matrix trace operation. The matrix $\bsrm{C}^{\rm a}_i$ is given by the $i$-th component of the tensor $\bsrm{C}^{\rm a}$. Thus, we see that $\bsrm{M}_2$ can as well be decomposed into two terms $\bsrm{M}_2^{\rm a}(t)=\bsrm\sigma_{\rm a}(t)$ and $\bsrm{M}_2^{\rm o}(t)=\bsrm\sigma_{\rm o}(t)$ with
\[  \big\langle \bsrm\sigma_\alpha(t) \, \bsrm\sigma_\beta(t^\prime)\big\rangle = \delta_{\alpha\beta} \, \bsrm{g}_\alpha(t-t^\prime)  \quad \! ; \quad \alpha,\beta \in \{{\rm a},{\rm o}\}\]
\item The memory kernel (\ref{eq:M3}) is given by:
\[  \bsrm{h}(\bsrm{X},s) =\left[
  \begin{array}{c}
    \bsrm{h}_{\rm a}(\bsrm{X}_{\rm a},s) \\ \bsrm{h}_{\rm o}(\bsrm{X}_{\rm o},s)
  \end{array}
  \right]\]
with
\begin{multline}
  \label{eq:hXsa}
  h_{{\rm a,i}}(\bsrm{X}_{\rm a},s) =   \varepsilon^2 \, \mathrm{Tr} \left(  \bsrm{X}_{{\rm a}}^{\rm T}\cdot \bsrm{V}^Y \cdot \bsrm\sigma^Y \right. 
  \\ \left. \cdot \bsrm{E}^{\rm T}(s) \cdot \left(\bsrm{C}^{\rm a}_i + \bsrm{C}^{\rm aT}_i\right) \cdot \bsrm{E}(s) \right)
\end{multline}
\begin{multline}
  \label{eq:hXso}
  \bsrm{h}_{{\rm o}}(\bsrm{X}_{\rm o},s) = \varepsilon^2 \; \bsrm{R}^{\rm o} \cdot \bsrm{E}(s) \cdot \bsrm{R}^Y \cdot \bsrm{X}_{{\rm o}} 
\end{multline}

and the integral in Eq.~(\ref{eq:M3adef}) gives two terms $\bsrm{M}_3^{\rm a}$ and $\bsrm{M}_3^{\rm o}$.
\end{itemize}
The resulting parameterization can be written as:
\begin{equation}
  \label{eq:paramsys}
    \left\{
  \begin{array}{lcl}
    \dbsrm{X}_{\rm a} & = & \bsrm{F}_{\rm a}(\bsrm{X}) + \bsrm{M}^{\rm a}_1 + \bsrm{M}^{\rm a}_2(t) + \bsrm{M}^{\rm a}_3(\bsrm{X}_{\rm a},t)  \\ \\
    \dbsrm{X}_{\rm o} & = & \bsrm{F}_{\rm o}(\bsrm{X}) + \bsrm{M}^{\rm o}_1 + \bsrm{M}^{\rm o}_2(t) + \bsrm{M}^{\rm o}_3(X_{\rm o},t)
  \end{array}
  \right.
\end{equation}
The calculations leading to these formula are explicited in Appendix~\ref{sec:Appoa} and an example is presented in Appendix~\ref{sec:simpex}. In this example, the number of variables is reduced to 3, one resolved and 2 unresolved. As revealed on Figure~\ref{fig:appsimp}, the subgrid-scale parameterization improves the description of the variable $x$. We will apply the same procedure to the low-order ocean-atmosphere coupled model in the following section.\\

\begin{table}
  \centering
  \begin{tabular}{| @{\hspace{.1cm}} c @{\hspace{0.1cm}} | |  @{\hspace{.2cm}} c @{\hspace{.2cm}} |  @{\hspace{.2cm}} c @{\hspace{.2cm}} |  @{\hspace{.2cm}} c @{\hspace{.2cm}} |}
    \hline
    Parameter & Case 1 & Case 2 & Case 3 \\
    \hline 
    $\lambda$ & $20$ & $100$ & $15.06$ \\
    $r$  & $10^{-8}$ & $10^{-8}$ & $10^{-7}$\\
    $d$  &$7.5 \times 10^{-8}$ & $6.0 \times 10^{-8}$ & $1.1 \times 10^{-7}$ \\
    $C_{\rm o}$  & $280$ & $350$ & $310$ \\
    $k_d$  & $4.128 \times 10^{-6}$ & $4.128 \times 10^{-6}$ & $2.972 \times 10^{-6}$ \\
    $k_d^\prime $  & $4.128 \times 10^{-6}$ & $4.128 \times 10^{-6}$ & $2.972 \times 10^{-6}$ \\
    $H$  & $500$ & $500$ & $136.5$ \\
    $G_{\rm o}$  & $2.00 \times 10^8$ & $2.00 \times 10^8$ & $5.46 \times 10^8$ \\
    $G_{\rm a}$  & $10^7$ & $10^7$ & $10^7$ \\
    $q_{\rm a},\, q_Y$ & $5 \times 10^{-4}$ & $5 \times 10^{-4}$ & $5 \times 10^{-4}$ \\
    $q_{\rm o}$ & $0$ & $0$ & $0$ \\    
    \hline
  \end{tabular}
  \caption{Values of the main parameters of the model for the three cases studied.   \label{tab:params}}
\end{table}

\subsection{Impact on the low-order model}
\label{sec:impact}

We tested the approach on three different parameter sets, each case being different typical situations of the LFV geometry explored so far. The last one (case 3) is a particular set of parameters derived in~\citein{V2015}. The parameters of the model are detailed in Table~\ref{tab:params}. On the practical side, we have considered two parameterizations for the term $\bsrm{M}_2$. For the first parameterization, we have modeled the term $\bsrm{M}_2$ with a Gaussian white noise (GWN) for which the covariance matrix $\bsrm\sigma_\GWN^2$ is given by:
\begin{equation}
  \label{eq:swn_amp}
  \bsrm\sigma_\GWN^2 = 2 \, \int_0^\infty  \bsrm{g}(s) \,\dd s
\end{equation}
 as in~\citein{AIW2003}. The second parameterization uses directly the Ornstein-Uhlenbeck (O-U) process (\ref{eq:OU}). Its advantage is that it generates automatically the correct correlations (\ref{eq:gsa}) and (\ref{eq:gso}).  The function $\bsrm{h}(\bsrm{X},s)$ in Eq.~(\ref{eq:M3}) has been sampled twenty times per day and the term $\bsrm{M}_3$ has been computed at the same frequency. Additionally, the time series used in the computations are also sampled at this frequency.

The small amplitudes of the noise coming from the model error are set to $q_{\rm a}=q_Y=5\times 10^{-4}$ and $q_{\rm o}=0$. Hence, we assume that the model error is negligible in the ocean. We integrated the model with a Heun stochastic scheme~\cite{GSH1988} over 1536 years with a timestep\footnote{This corresponds to  $\Delta t=0.01$ in adimensional time-unit.} $\Delta t=96.9$ s. This long timespan guarantees that we sample the LFV several times. Finally, to assess the impact of the coupling strength, we have considered two versions of the model (\ref{eq:oamod}) for each case : one with a weak coupling ($\varepsilon=0.5$), and one with a strong, normal coupling ($\varepsilon=1$). \\

\begin{figure*}
  \begin{subfigure}{\textwidth}
    \centering
    \includegraphics[width=.45\linewidth]{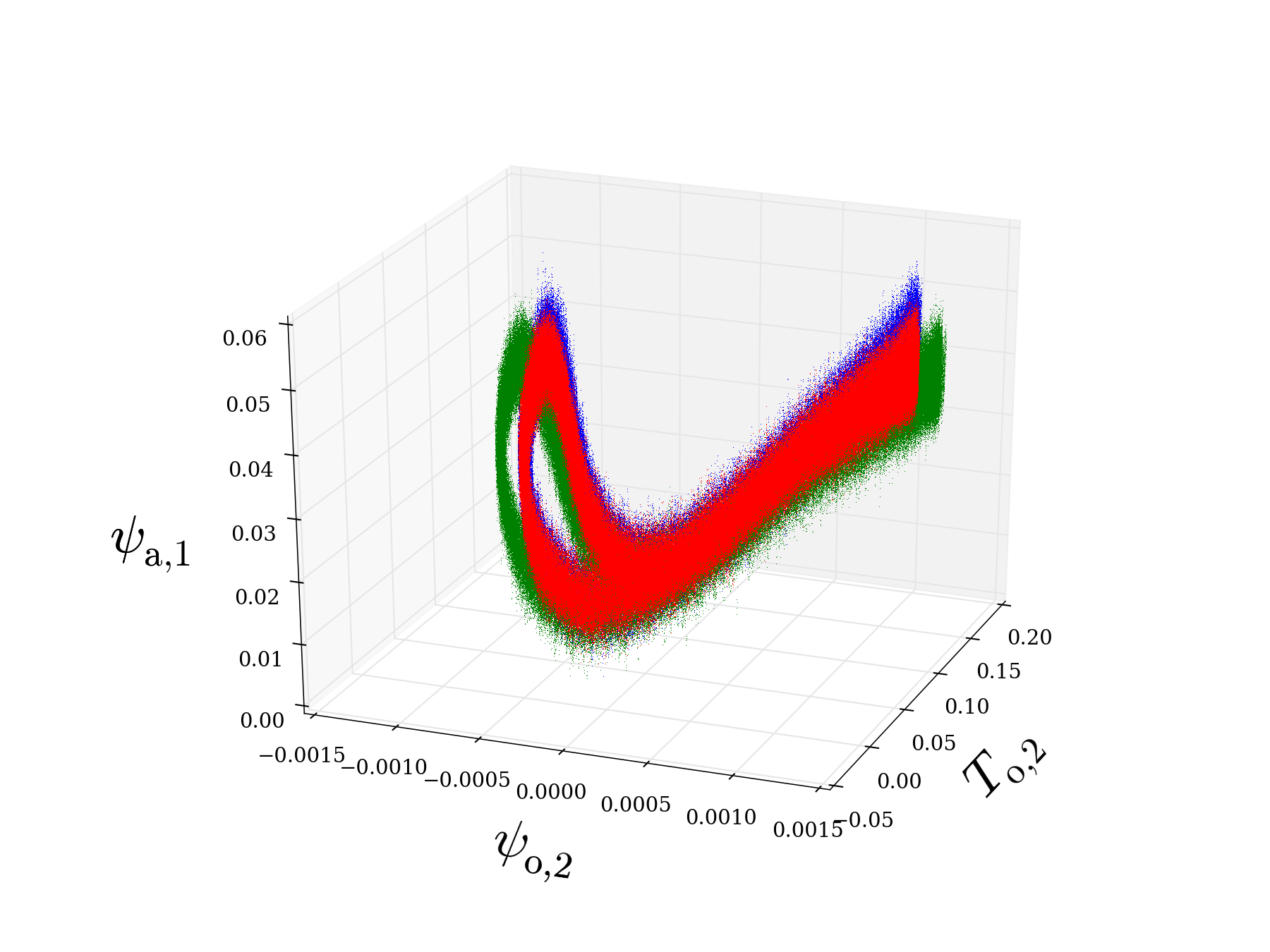}
    \includegraphics[width=.45\linewidth]{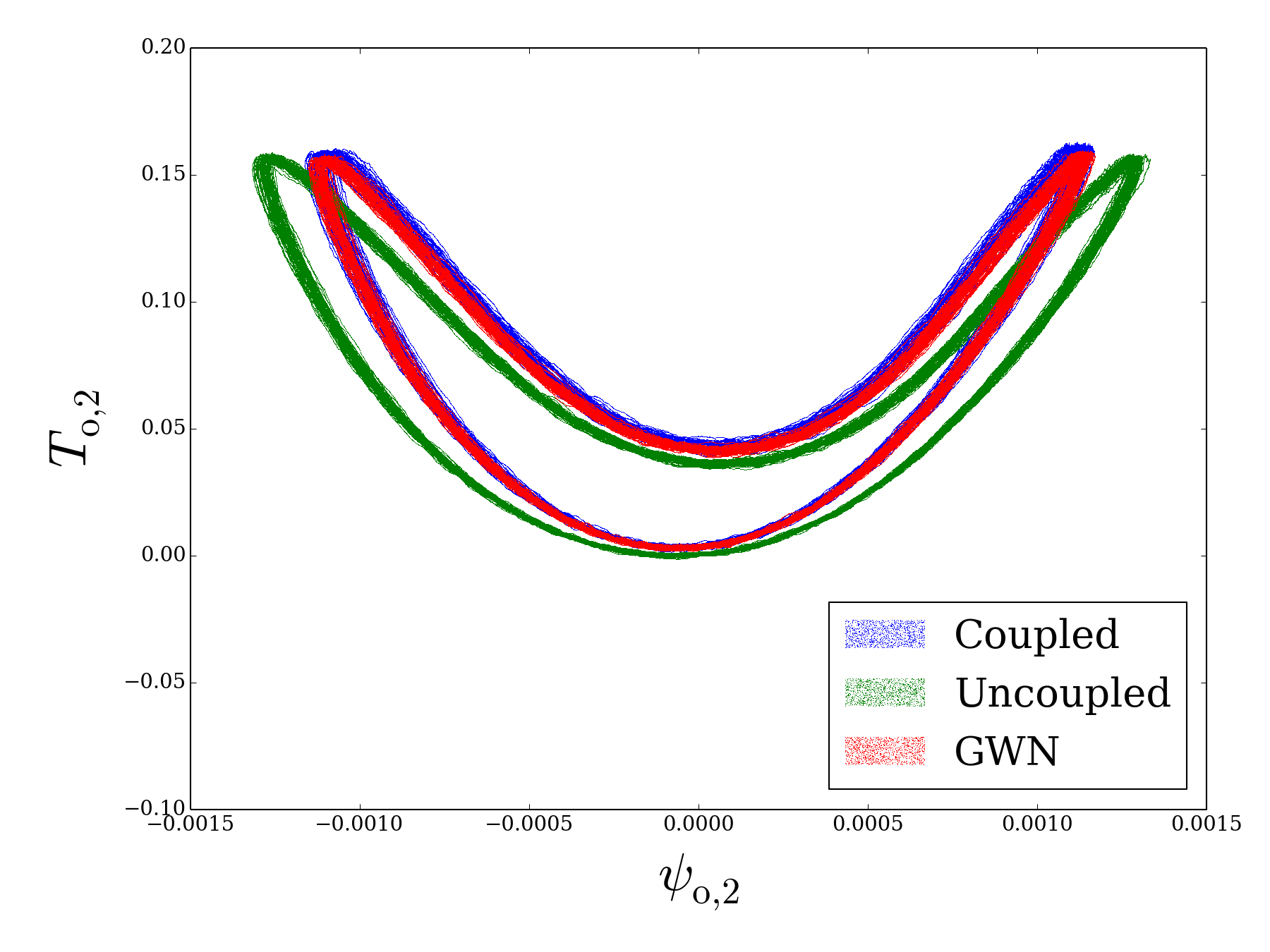}
    \caption{Strong coupling ($\varepsilon=1$).}
  \end{subfigure}

  \begin{subfigure}{\textwidth}
    \centering
    \includegraphics[width=.45\linewidth]{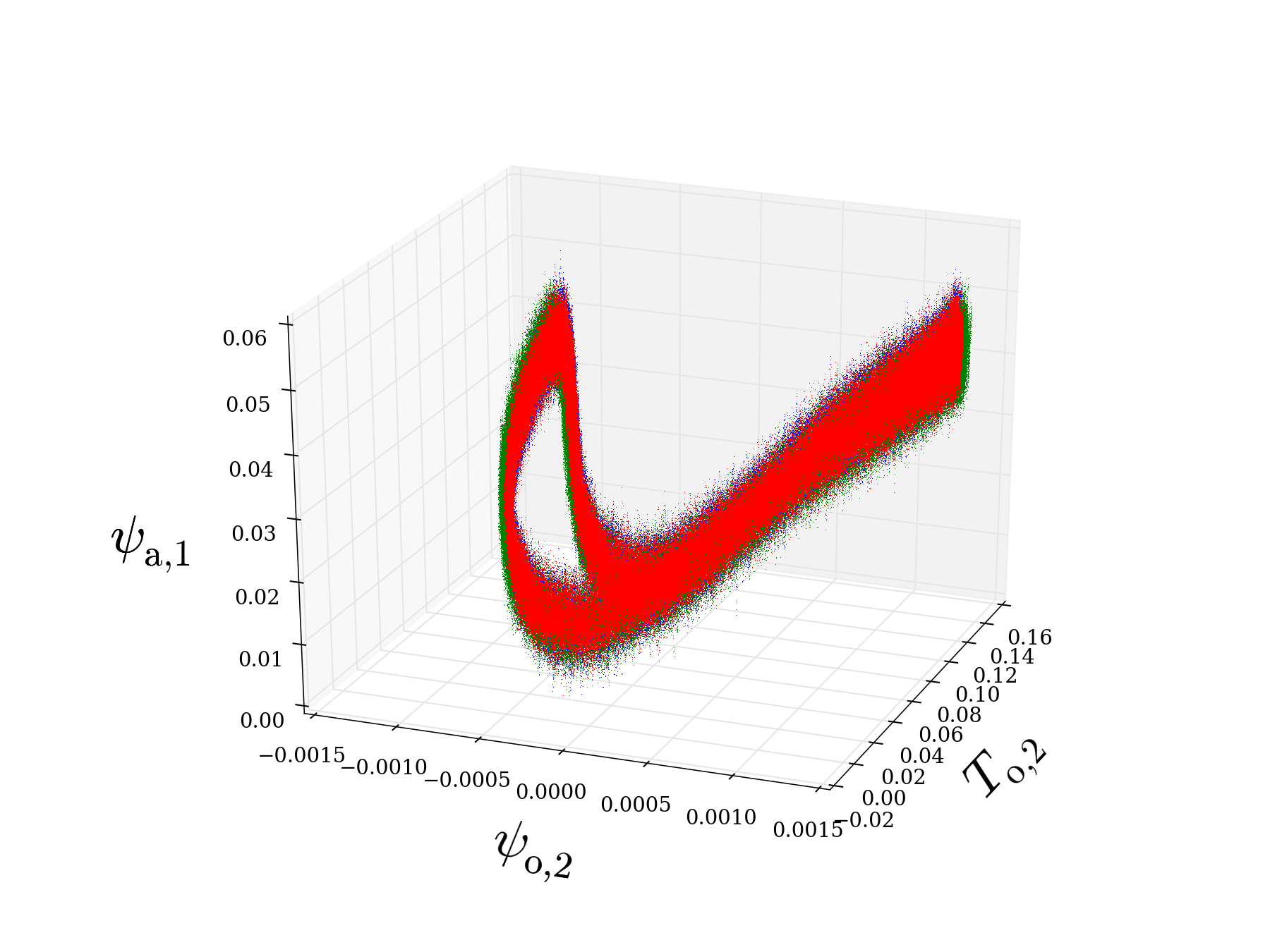}
    \includegraphics[width=.45\linewidth]{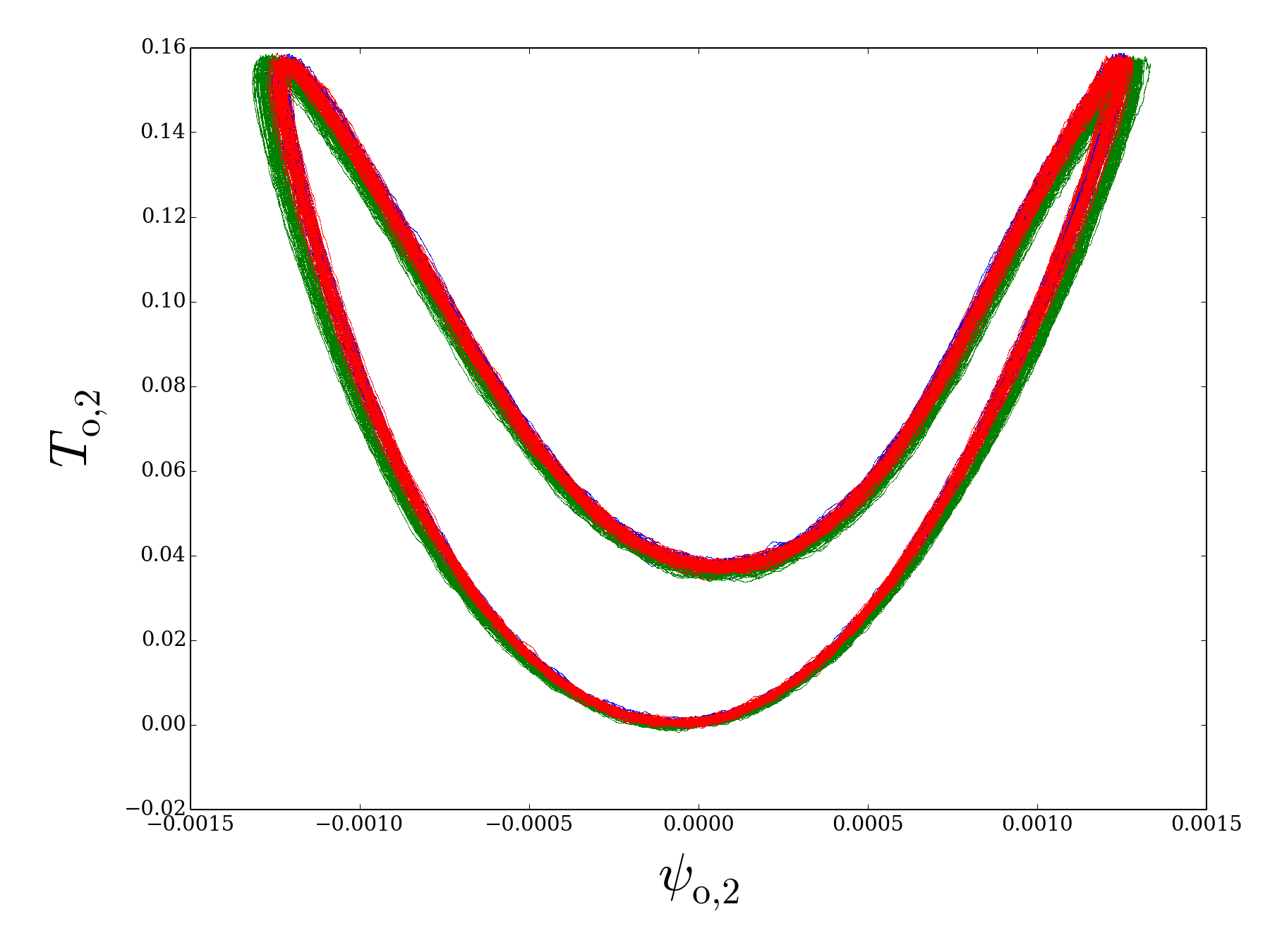}
    \caption{Weak coupling ($\varepsilon=0.5$).}
  \end{subfigure}

  \caption{Plots of attractors sections in the coordinates $\psi_{{\rm o},2}$, $\theta_{{\rm o},2}$ and $\psi_{{\rm a},1}$ in adimensional units for case 1. The coupled and uncoupled system are represented, as well as the Gaussian white noise (GWN) parameterization. See Table~\ref{tab:params} for the parameters values used. \label{fig:attractors_case1}}
\end{figure*}

\begin{figure*}
  \begin{subfigure}{\textwidth}
    \centering
    \includegraphics[width=.45\linewidth]{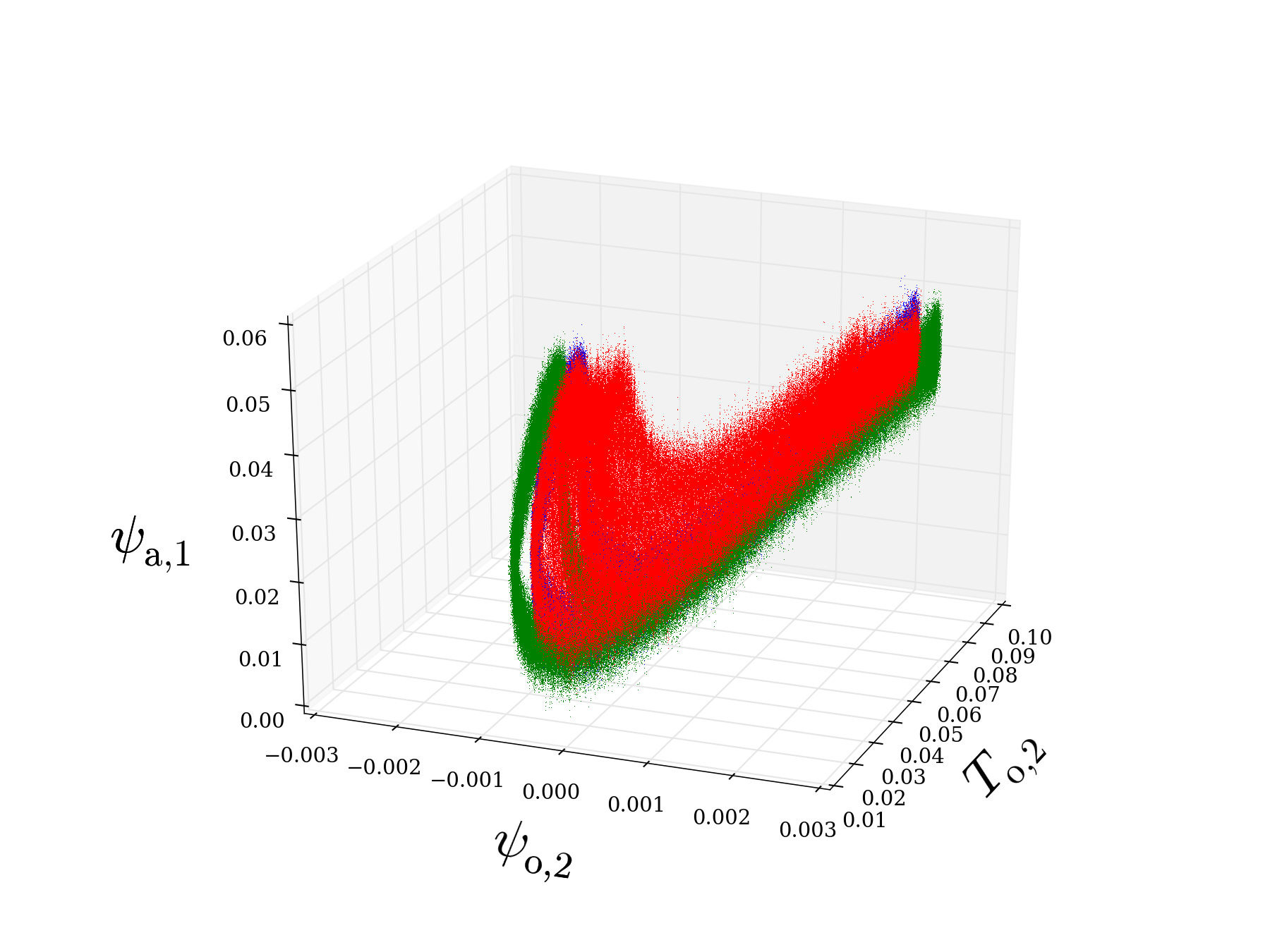}
    \includegraphics[width=.45\linewidth]{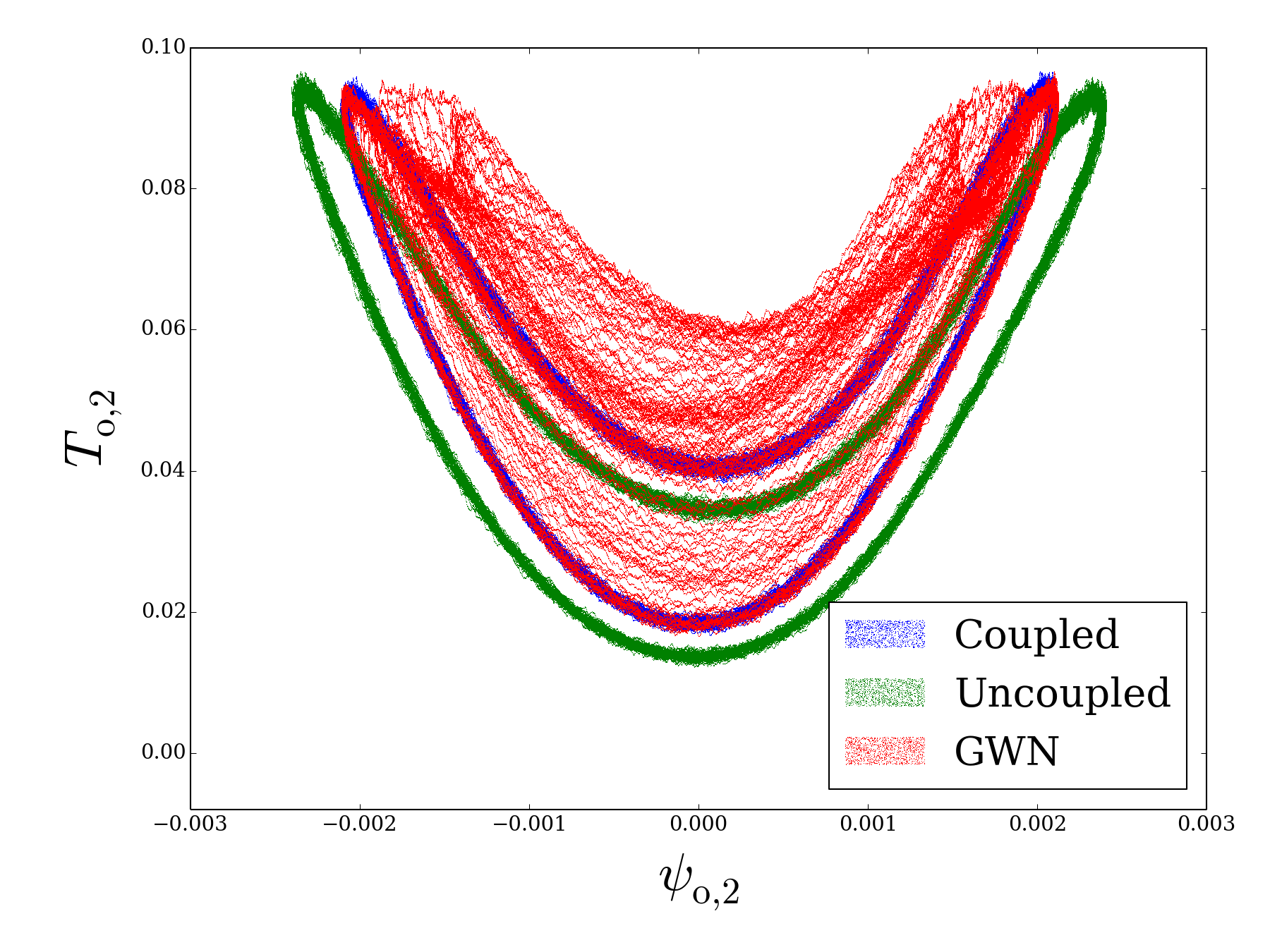}
    \caption{Strong coupling ($\varepsilon=1$).}
  \end{subfigure}
    \begin{subfigure}{\textwidth}
    \centering
    \includegraphics[width=.45\linewidth]{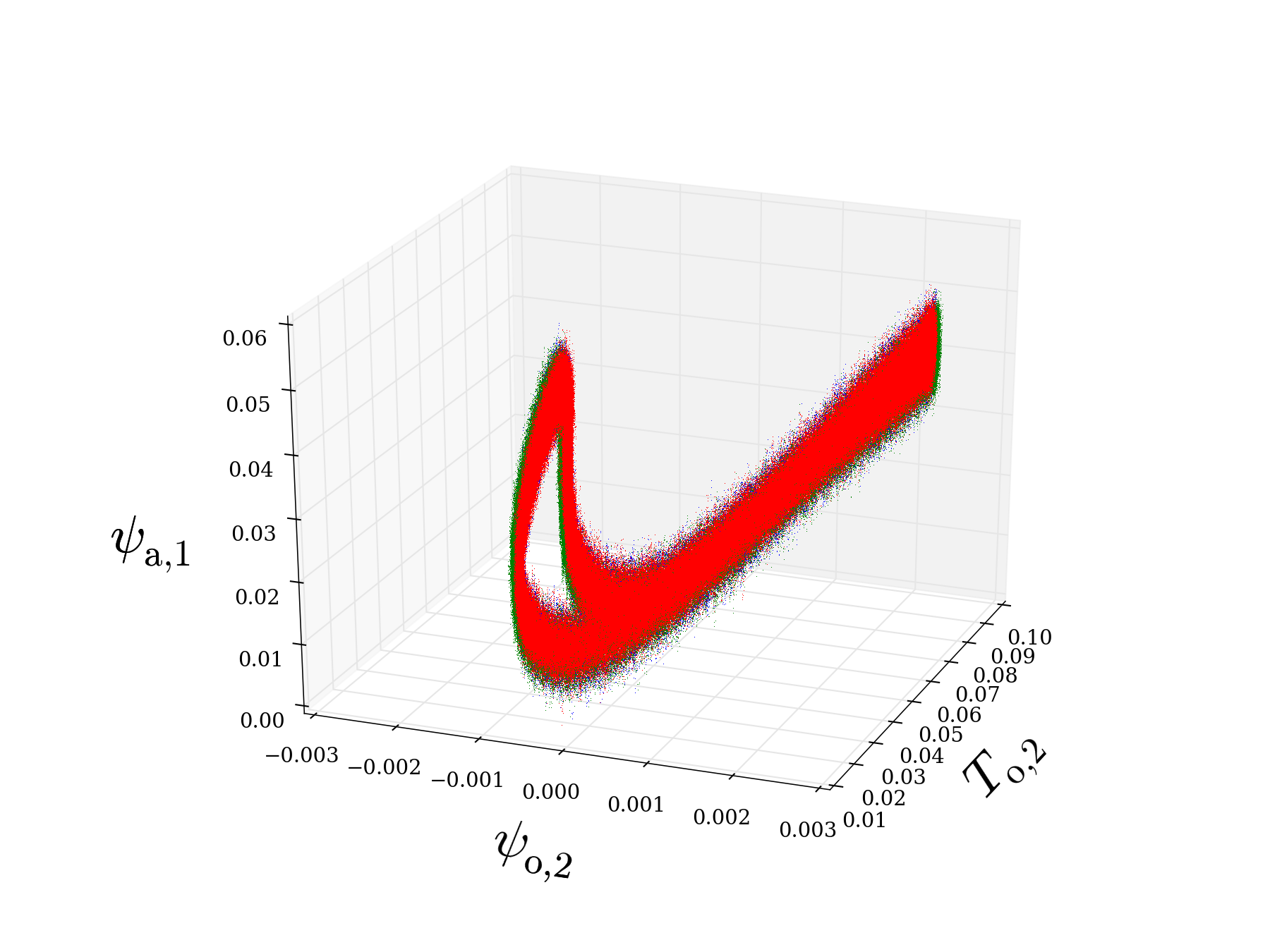}
    \includegraphics[width=.45\linewidth]{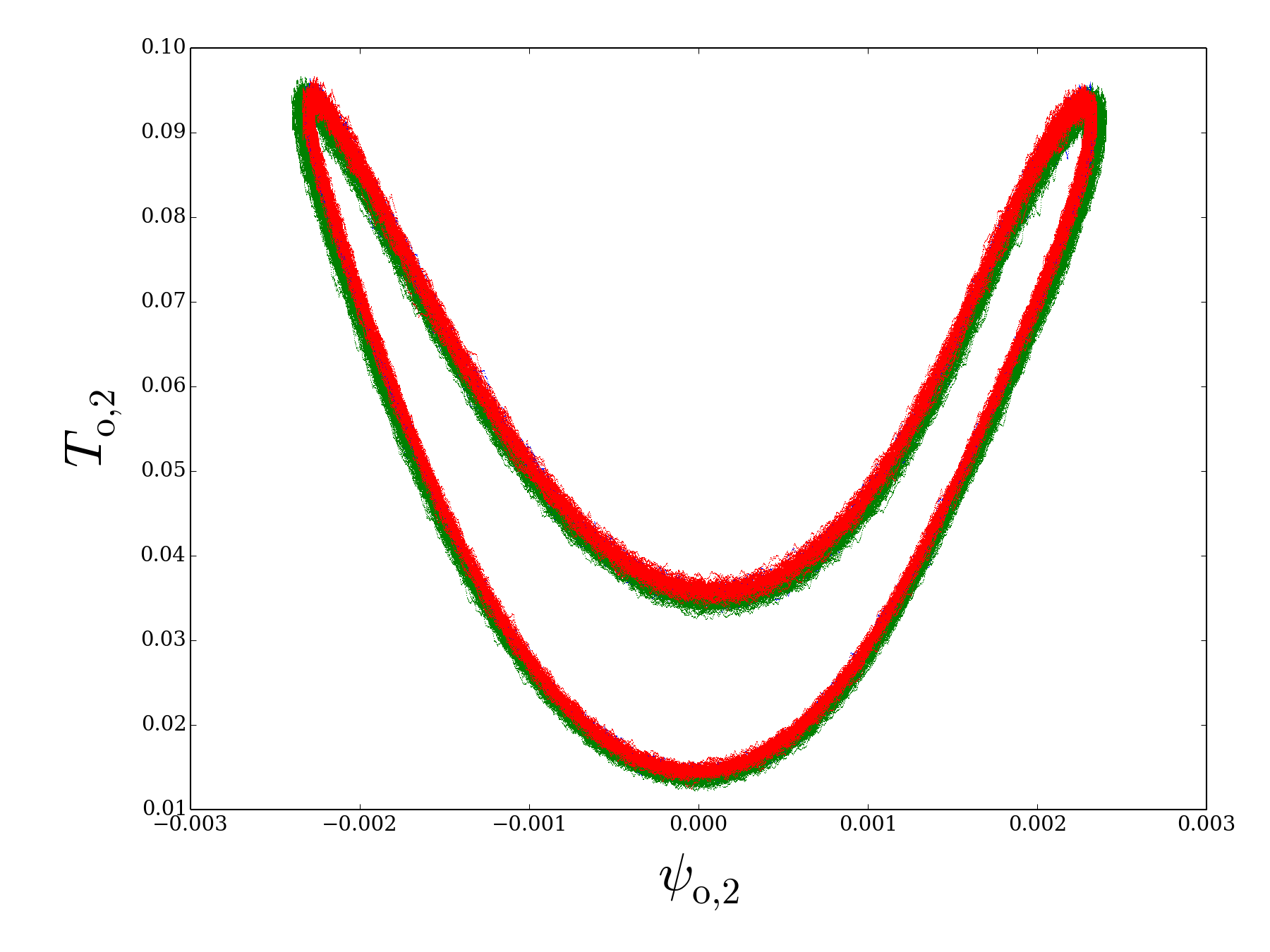}
    \caption{Weak coupling ($\varepsilon=0.5$).}

  \end{subfigure}
  \caption{Same as Figure~\ref{fig:attractors_case1} but for case 2. \label{fig:attractors_case2}}
\end{figure*}

\begin{figure*}
  \begin{subfigure}{\textwidth}
    \centering

    \includegraphics[width=.45\linewidth]{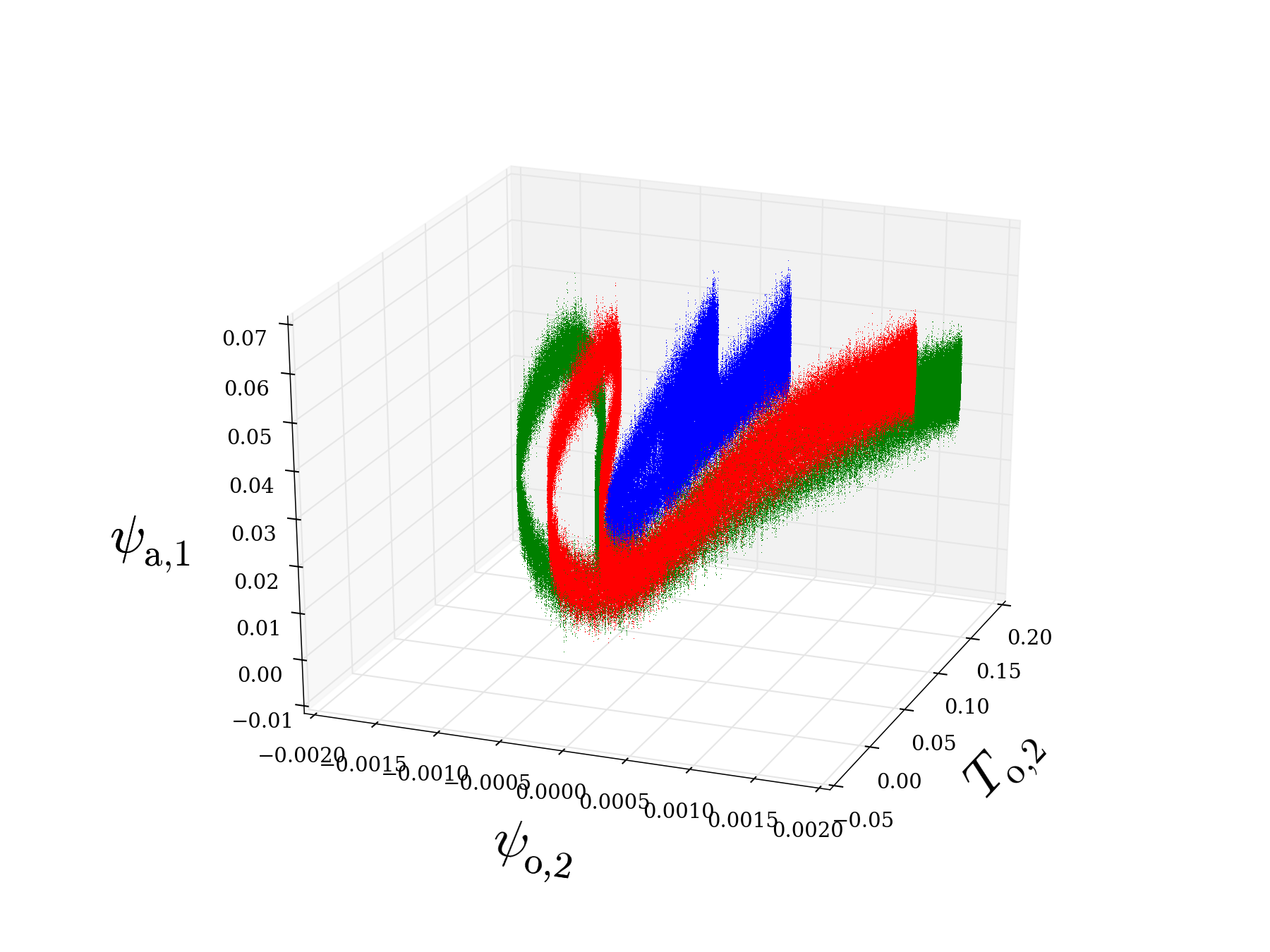}
    \includegraphics[width=.45\linewidth]{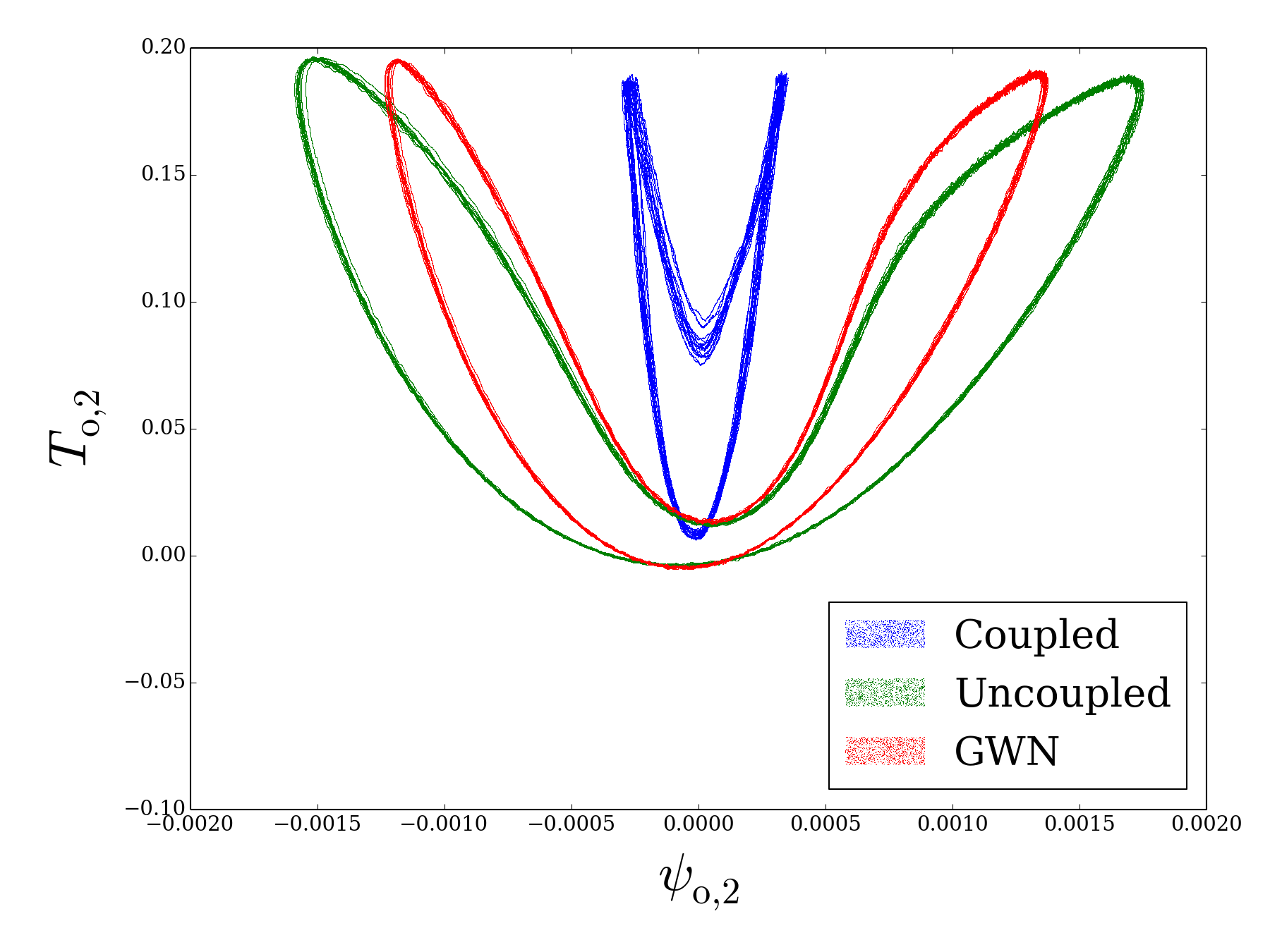}
    \caption{Strong coupling ($\varepsilon=1$).}
  \end{subfigure}
    \begin{subfigure}{\textwidth}
    \centering
    \includegraphics[width=.45\linewidth]{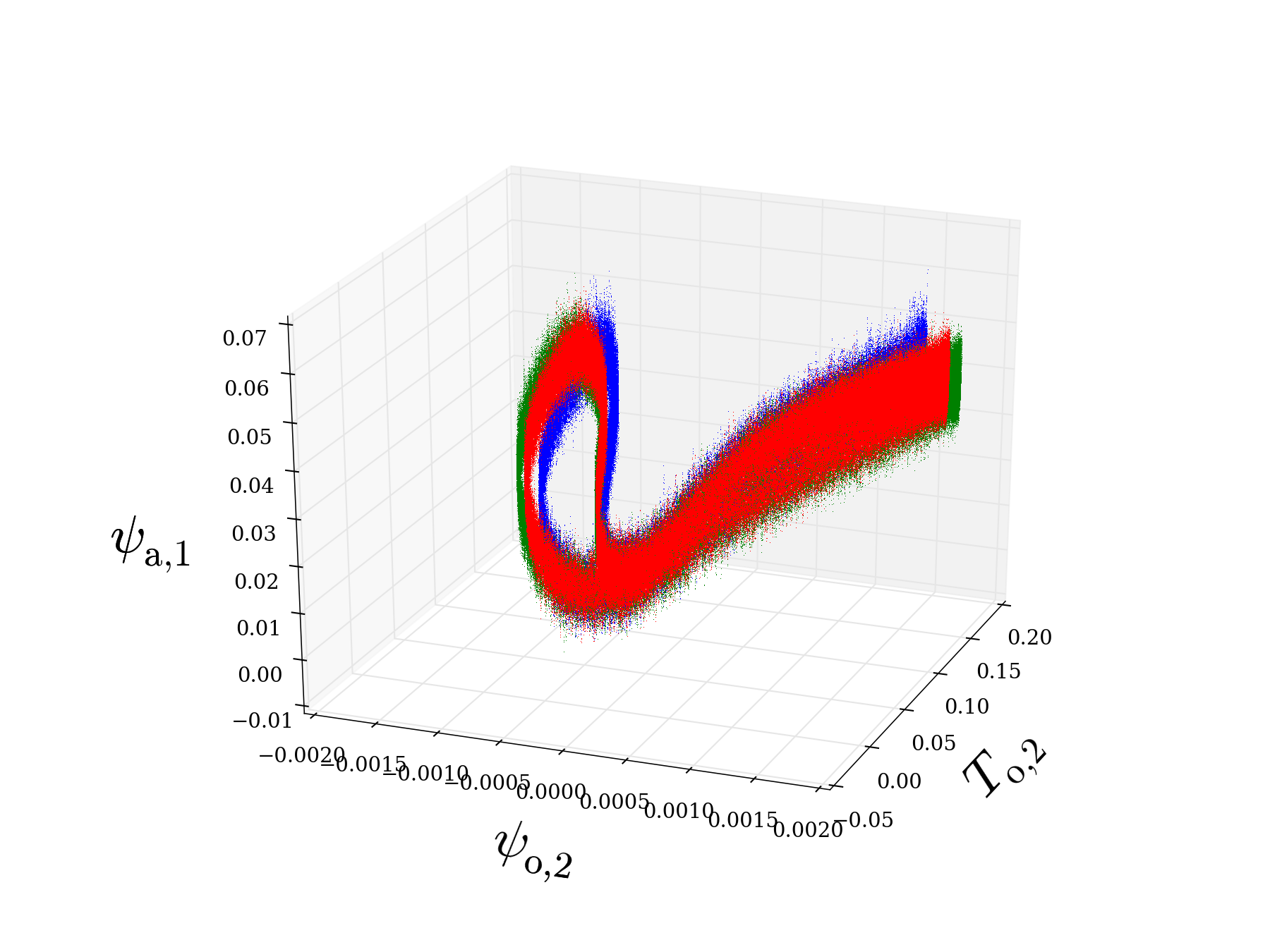}
    \includegraphics[width=.45\linewidth]{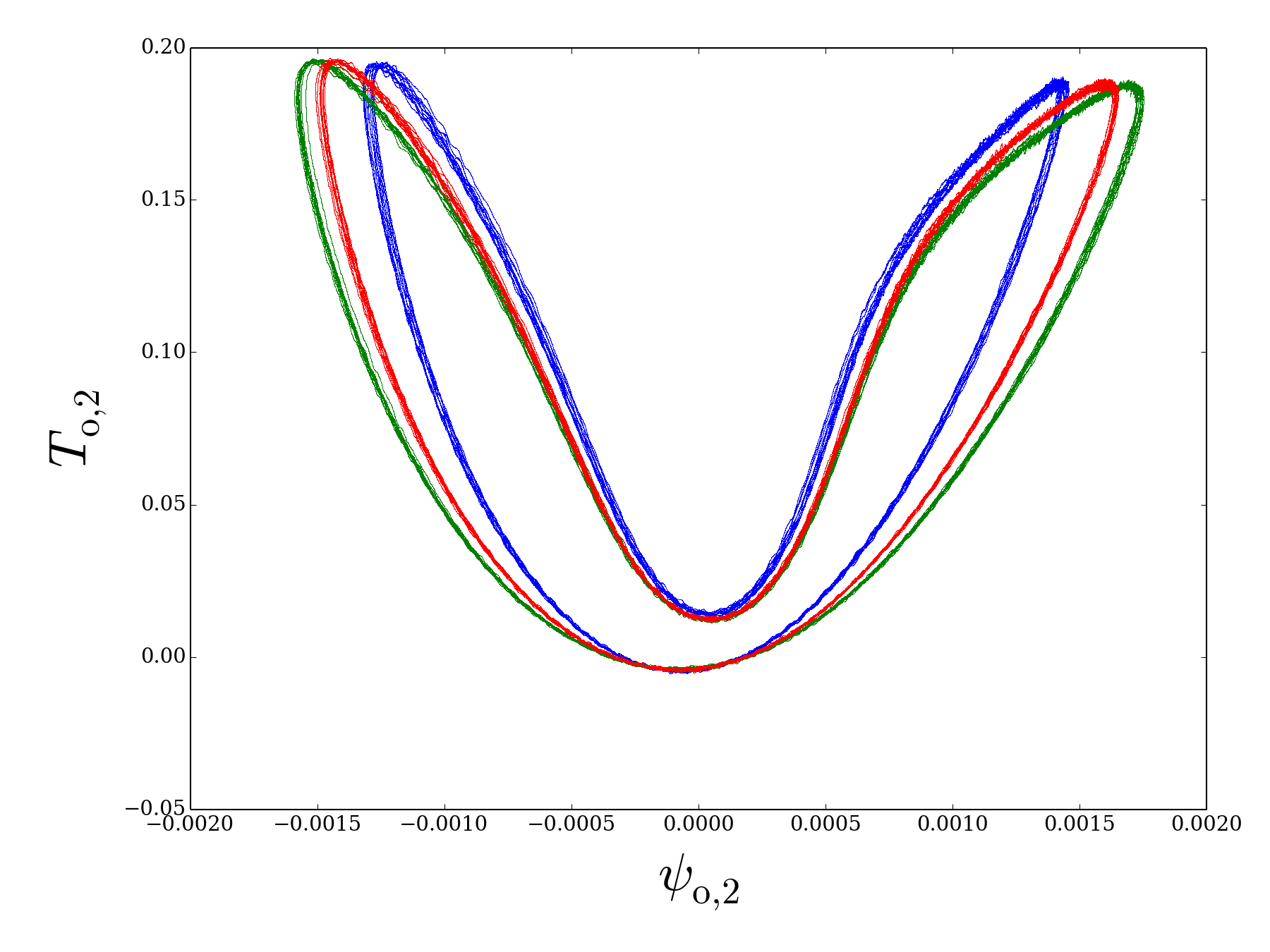}
    \caption{Weak coupling ($\varepsilon=0.5$).}

  \end{subfigure}
  \caption{Same as Figure~\ref{fig:attractors_case1} but for case 3. \label{fig:attractors_case3}}
\end{figure*}

The effect of the GWN parameterization is depicted on Figs.~\ref{fig:attractors_case1},~\ref{fig:attractors_case2} and~\ref{fig:attractors_case3} for respectively cases 1, 2 and 3. These are 3-dimensional representations of a section of the attractor, with the coefficients of the largest relevant scales $\psi_{{\rm o},2}$, $\theta_{{\rm o},2}$ and $\psi_{{\rm a},1}$, selected as axes. It mainly shows how this parameterization corrects the LFV signal. For the sake of clarity, only the Gaussian white noise parameterization is shown in these figures. However, for such a qualitative representation, the two parameterizations give similar results.

On Figure~\ref{fig:attractors_case1}, we see that for both $\varepsilon=1$ and $\varepsilon=0.5$, the LFV is well corrected by the method. However, it is not the case on Figure~\ref{fig:attractors_case2}~(a) where the strong coupling sub-case is showing long timespans with the correct low-frequency variability but also shorter timespans where the parameterized system seems to wander away from the LFV. This particular behavior could be explained by the presence of a repeller nearby that the parameterized system visits on rare occasions. This feature is interesting to explore in the future. However, in the weak coupling case (Figure~\ref{fig:attractors_case2}~(b)), the LFV is well corrected. The third case is depicted on Figure~\ref{fig:attractors_case3}. As shown on this figure, large differences between the coupled and the uncoupled dynamics exist and therefore the parameterizations is only able to partially correct the LFV dynamics.\\

\begin{figure*}
  \begin{subfigure}{0.49\textwidth}
    \centering
    \includegraphics[width=\linewidth]{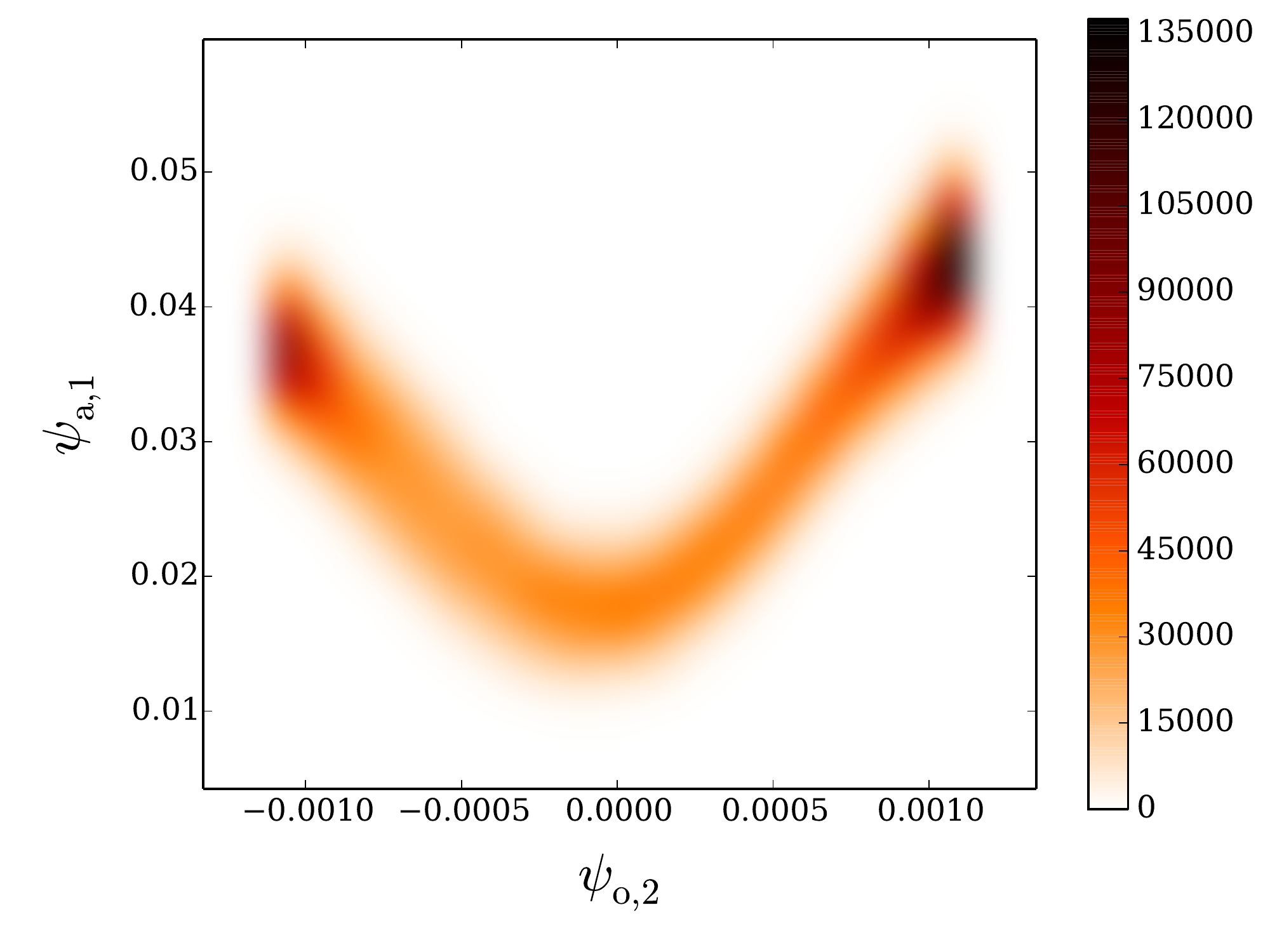}
    \caption{PDF of the coupled system.}
  \end{subfigure}
  \begin{subfigure}{0.49\textwidth}
    \centering
    \includegraphics[width=\linewidth]{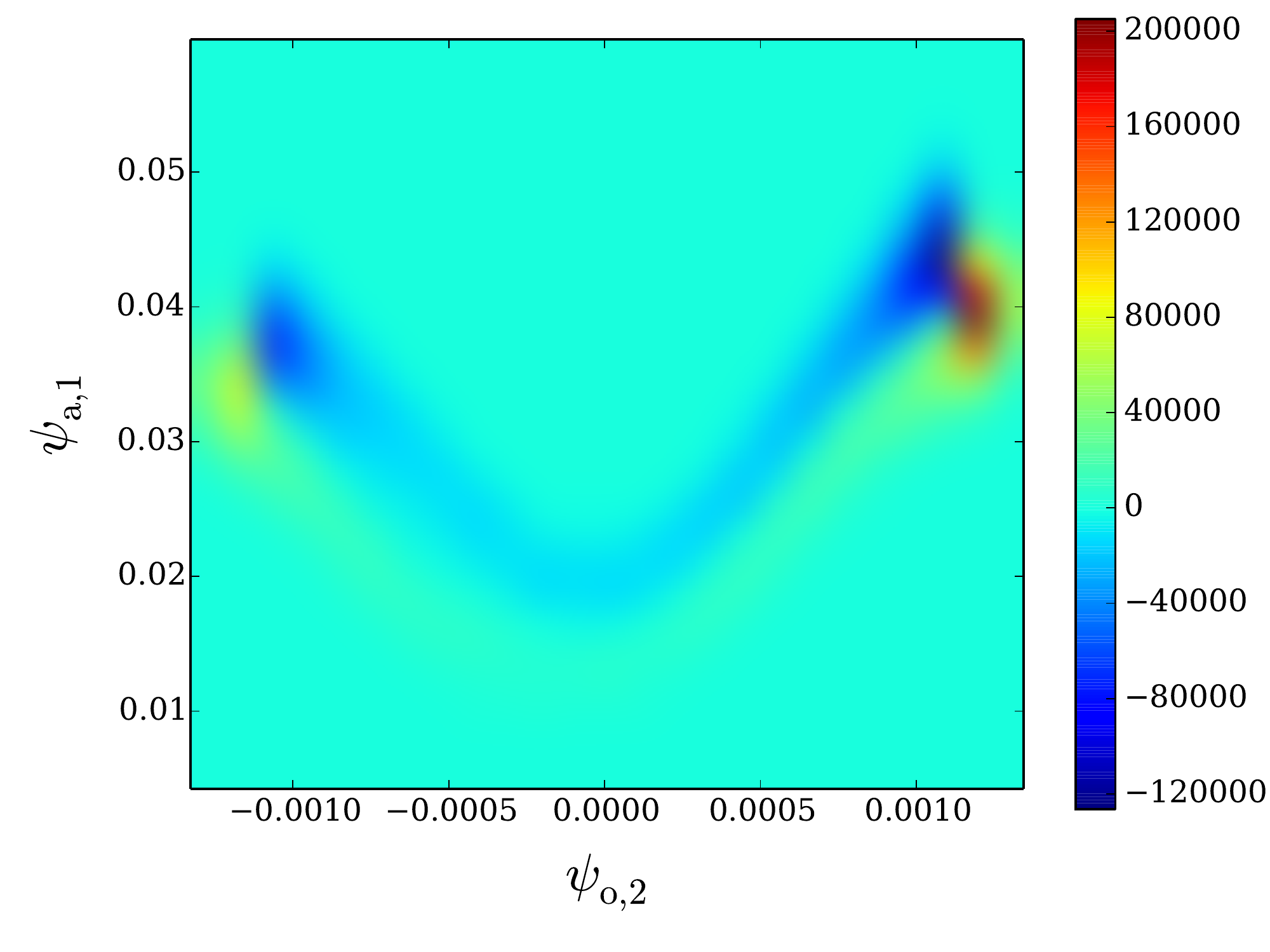}
    \caption{Anomaly of the uncoupled system PDF w.r.t the coupled system.}
  \end{subfigure}
  \begin{subfigure}{0.49\textwidth}
    \centering
    \includegraphics[width=\linewidth]{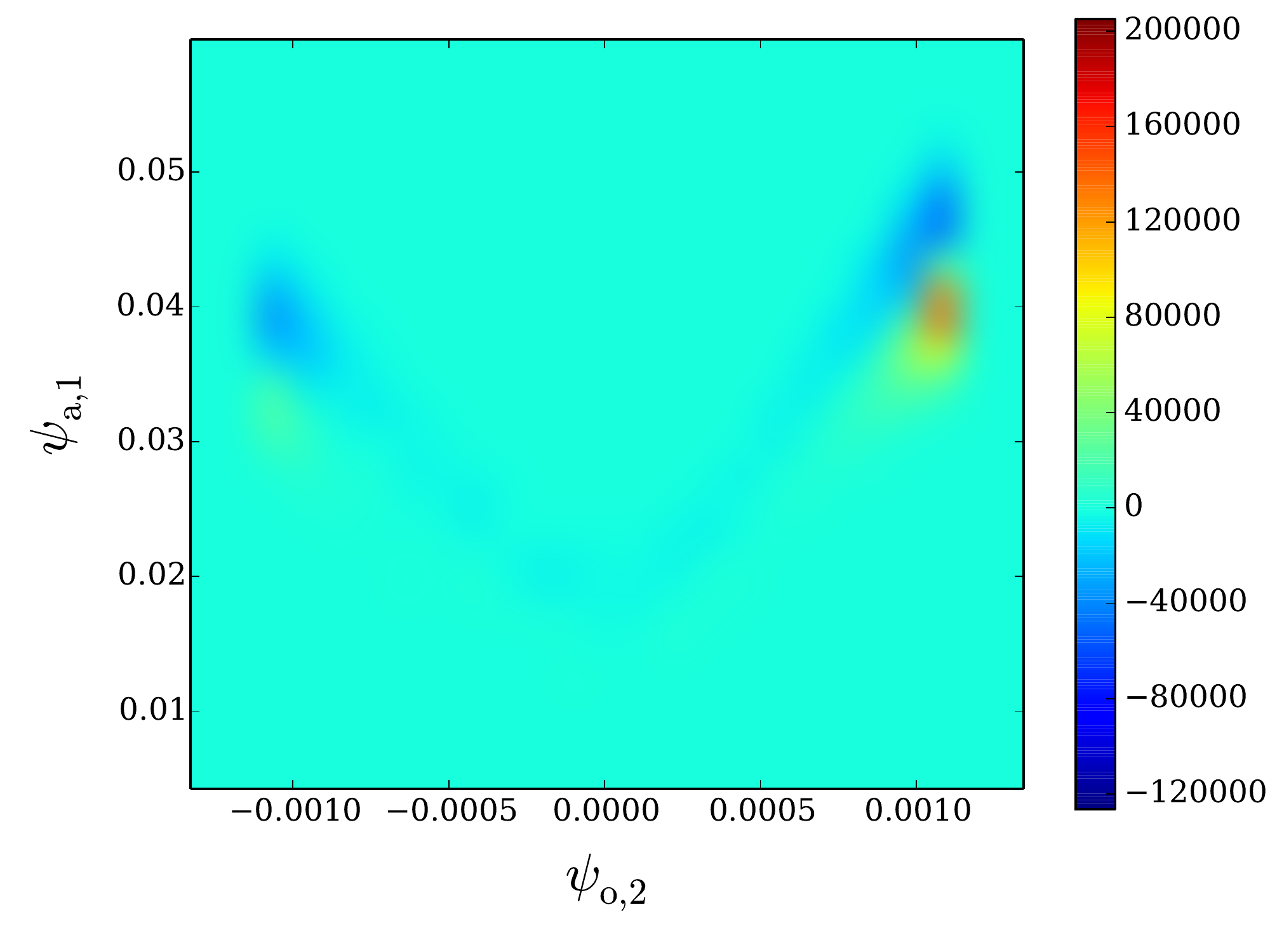}
    \caption{Anomaly of the Gaussian white noise parameterization PDF w.r.t the coupled system.}
  \end{subfigure}
  \begin{subfigure}{0.49\textwidth}
    \centering
    \includegraphics[width=\linewidth]{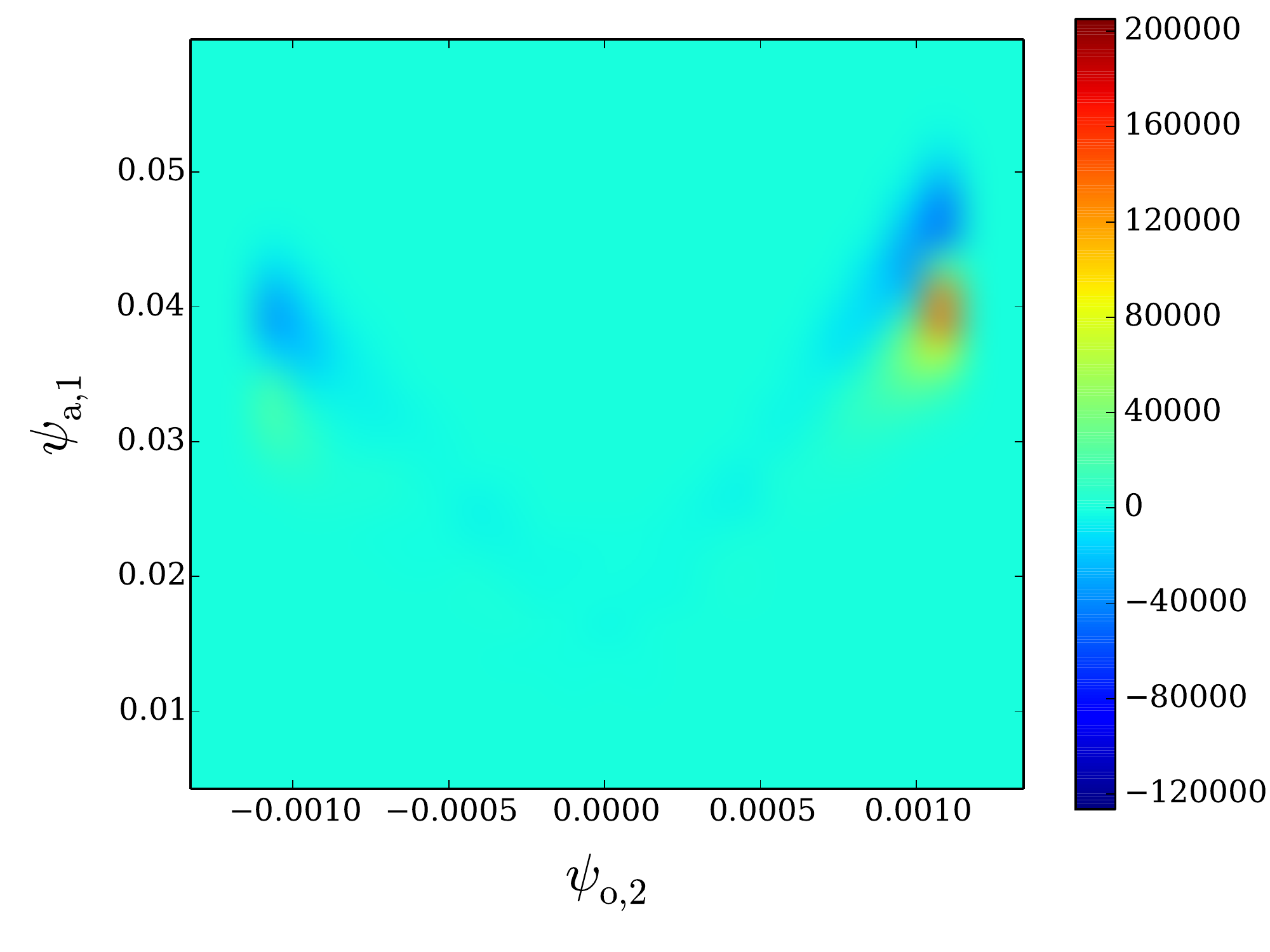}
    \caption{Anomaly of the Ornstein-Uhlenbeck parameterization PDF w.r.t the coupled system.}
  \end{subfigure}
  \caption{Two-dimensional probability density functions (PDFs) of $\psi_{{\rm a},1}$ and $\psi_{{\rm o},2}$ for case 1 ($\varepsilon=1$). \label{fig:2d_dist_case1s}}
\end{figure*}

\begin{figure*}
  \begin{subfigure}{0.49\textwidth}
    \centering
    \includegraphics[width=\linewidth]{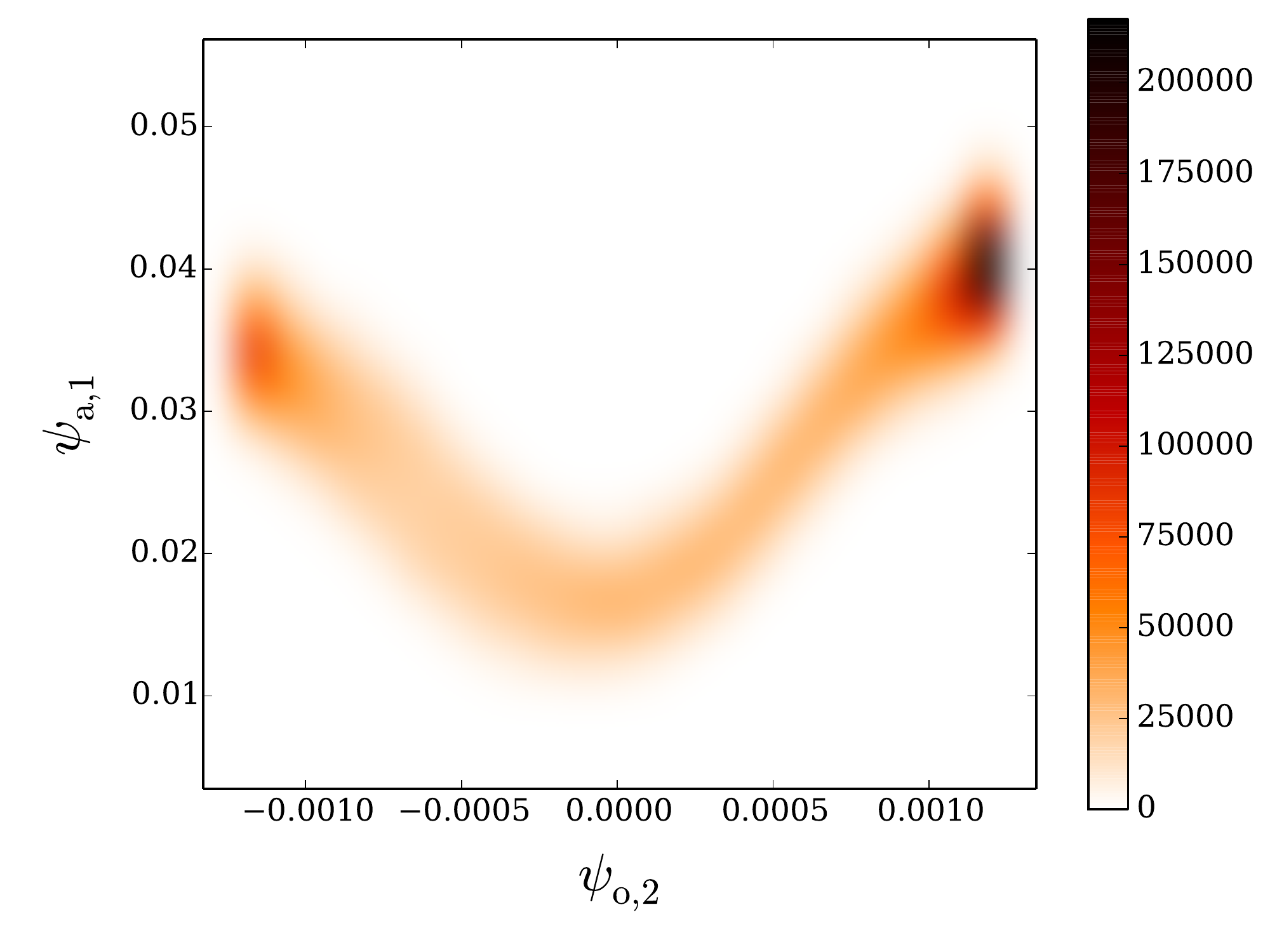}
    \caption{PDF of the coupled system.}
  \end{subfigure}
  \begin{subfigure}{0.49\textwidth}
    \centering
    \includegraphics[width=\linewidth]{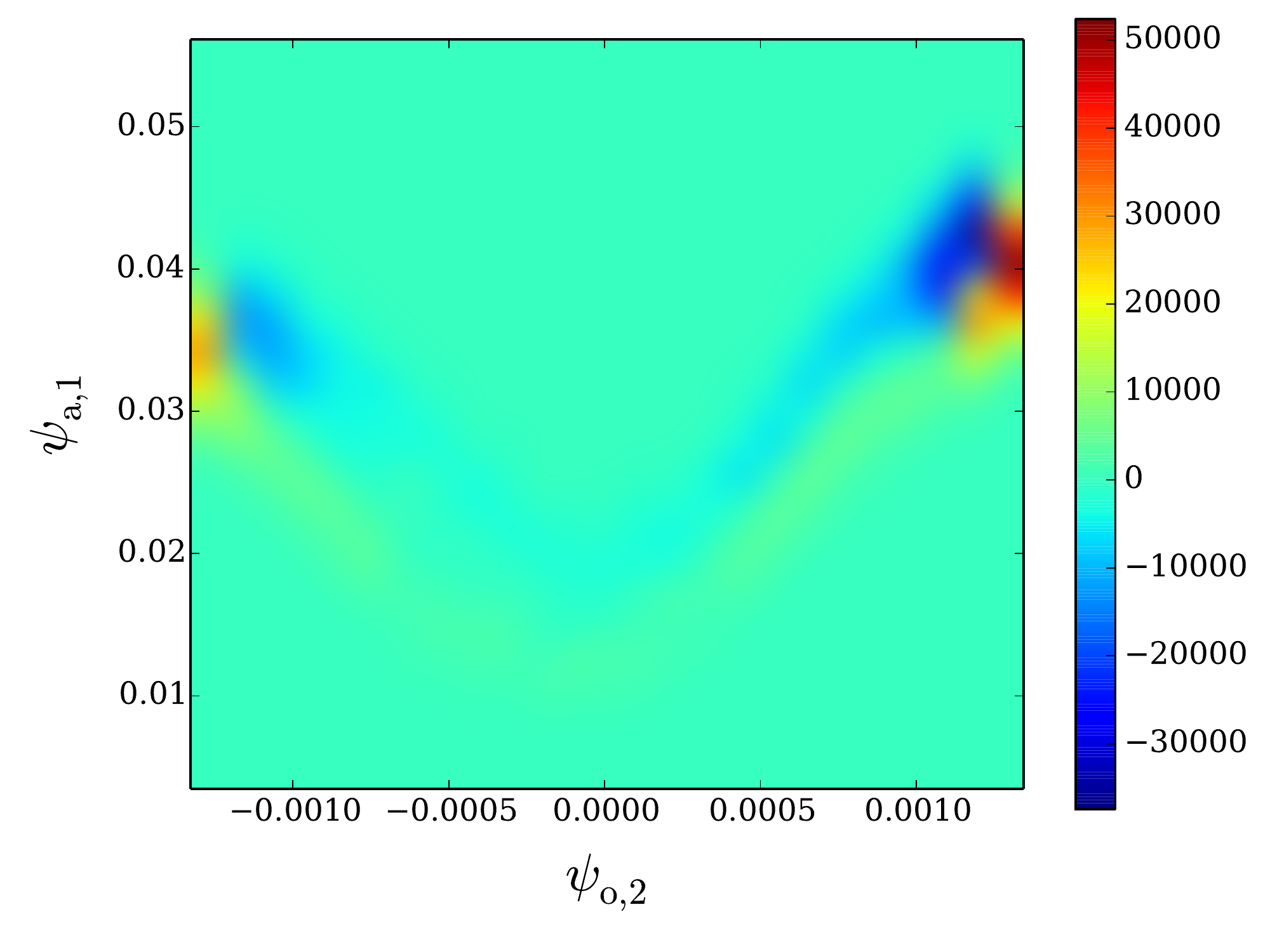}
    \caption{Anomaly of the uncoupled system PDF w.r.t the coupled system.}
  \end{subfigure}
  \begin{subfigure}{0.49\textwidth}
    \centering
    \includegraphics[width=\linewidth]{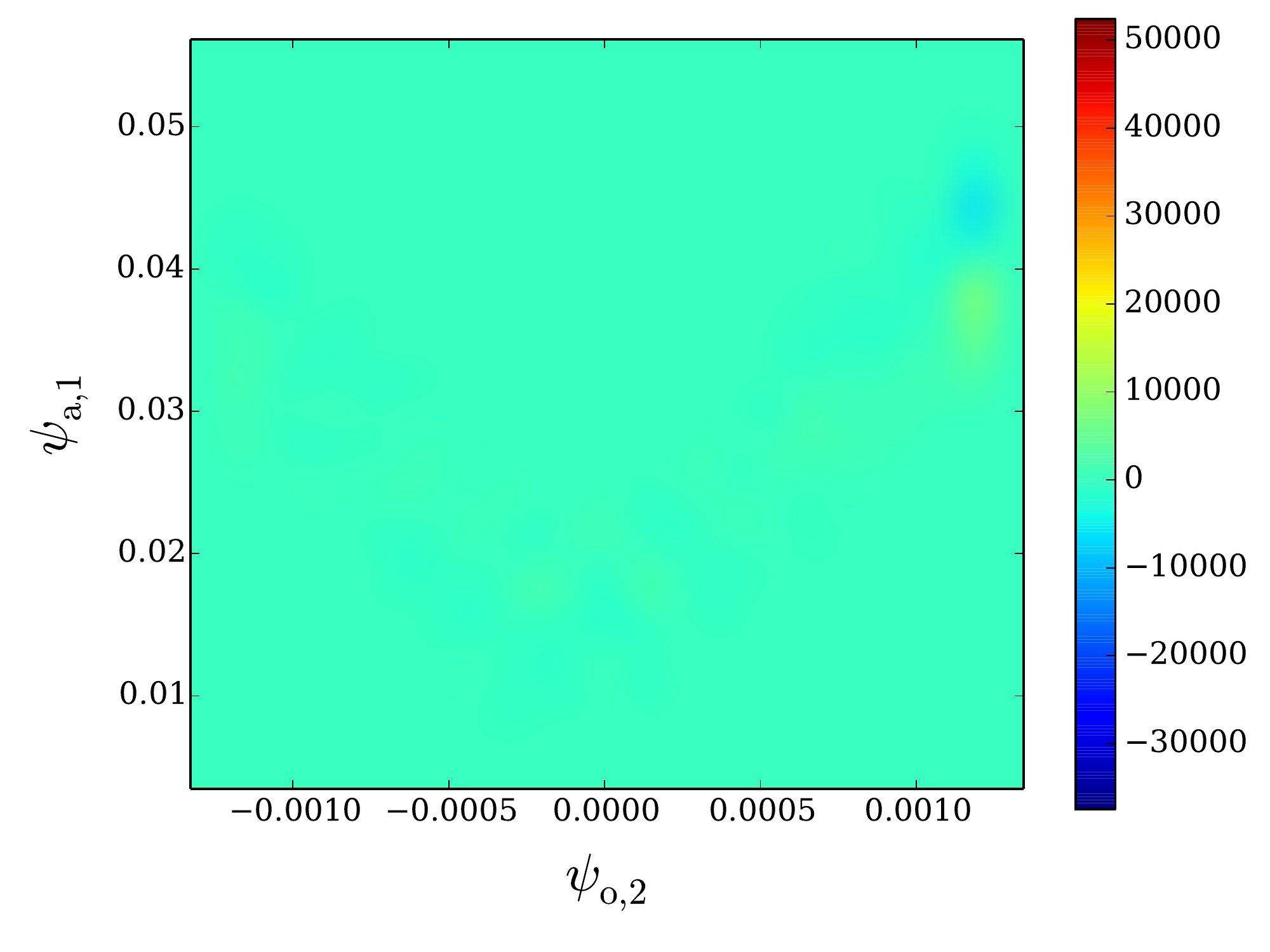}
    \caption{Anomaly of the Gaussian white noise parameterization PDF w.r.t the coupled system.}
  \end{subfigure}
  \begin{subfigure}{0.49\textwidth}
    \centering
    \includegraphics[width=\linewidth]{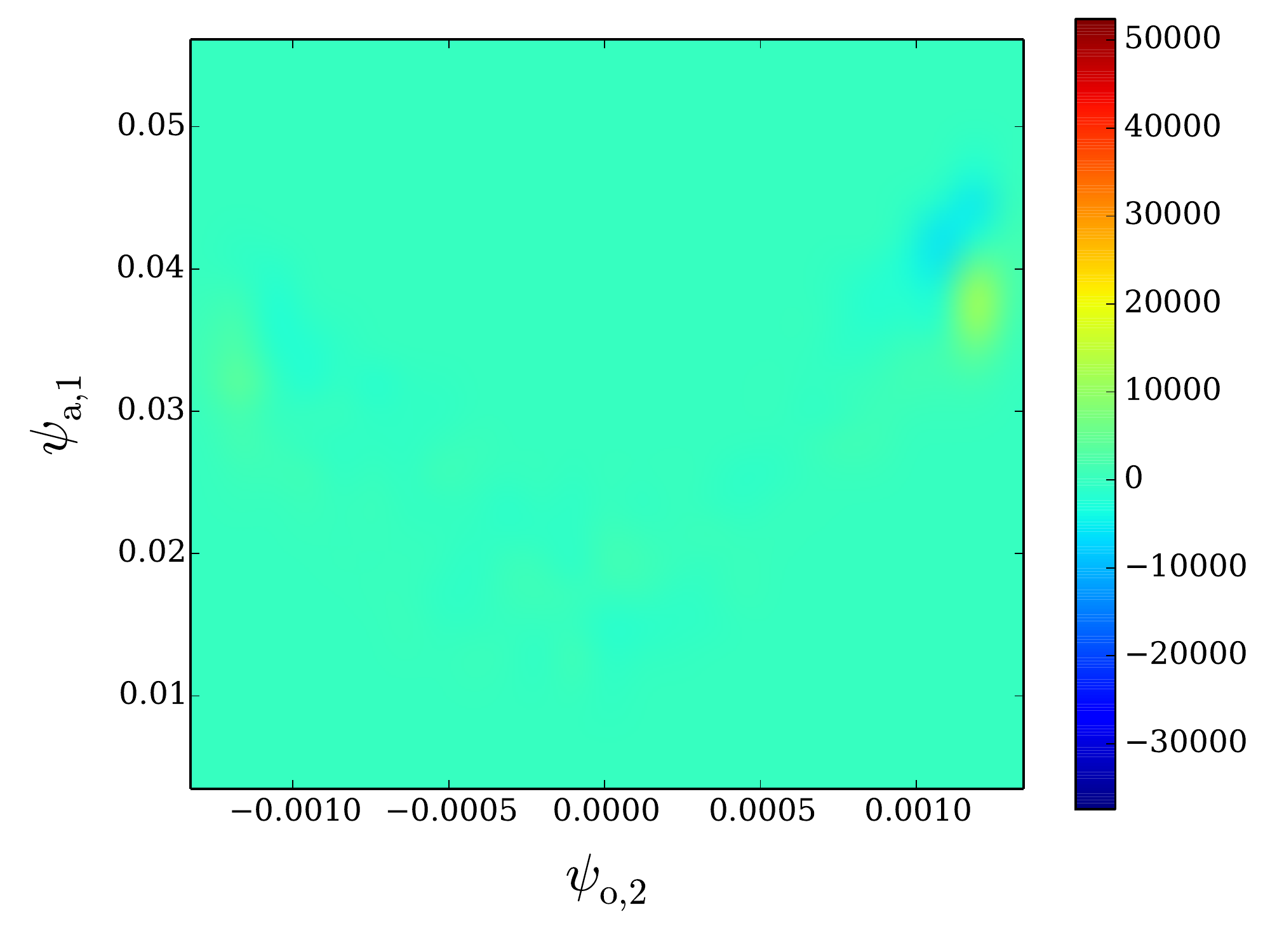}
    \caption{Anomaly of the Ornstein-Uhlenbeck parameterization PDF w.r.t the coupled system.}
  \end{subfigure}
  \caption{Two-dimensional probability density functions (PDFs) of $\psi_{{\rm a},1}$ and $\psi_{{\rm o},2}$ for case 1 ($\varepsilon=0.5$). \label{fig:2d_dist_case1w}}
\end{figure*}

\begin{figure*}
  \begin{subfigure}{0.49\textwidth}
    \centering
    \includegraphics[width=\linewidth]{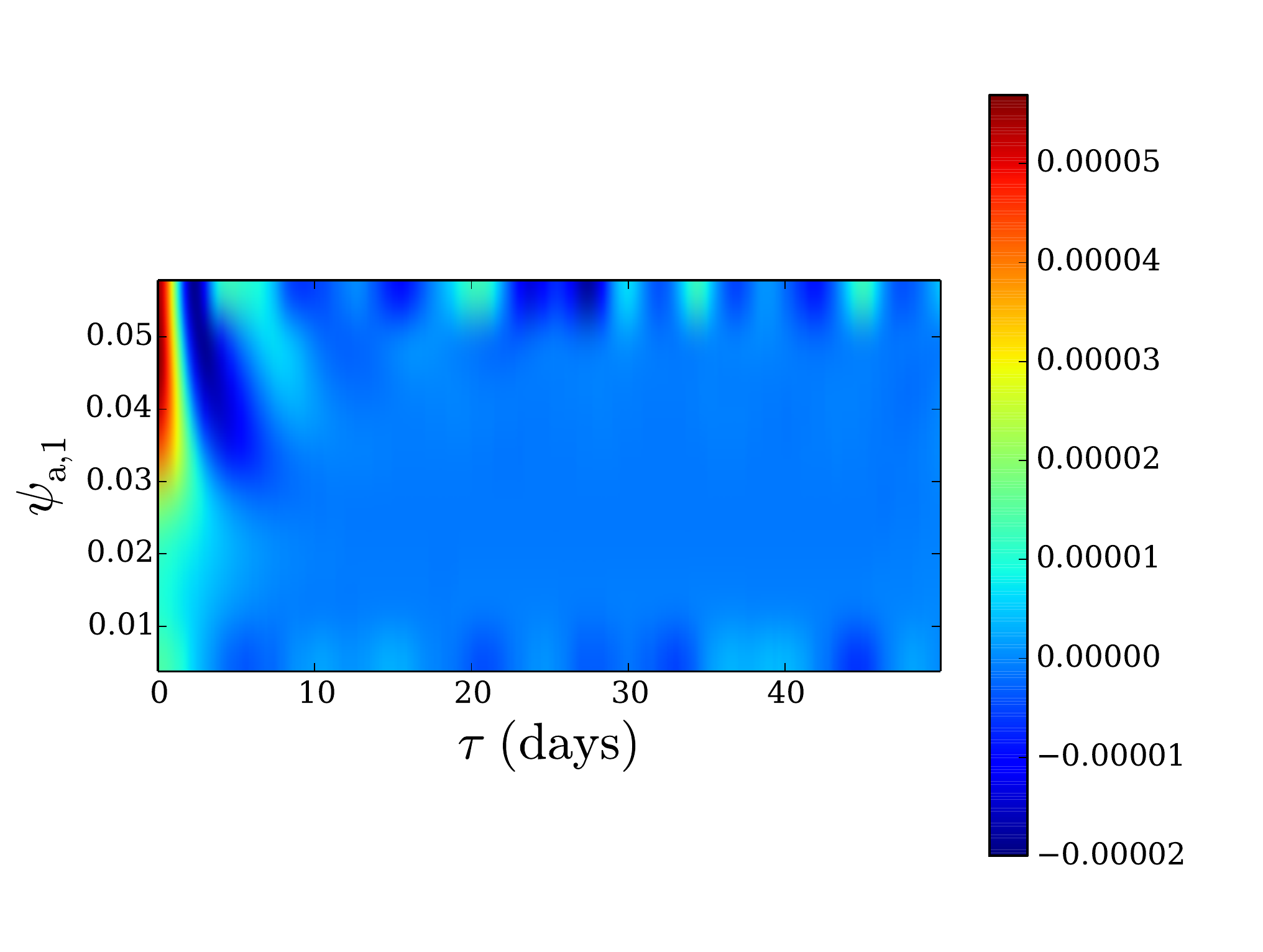}
    \caption{Autocorrelation of $\psi_{{\rm a},8}$ in the coupled system.}
  \end{subfigure}
    \begin{subfigure}{0.49\textwidth}
    \centering
    \includegraphics[width=\linewidth]{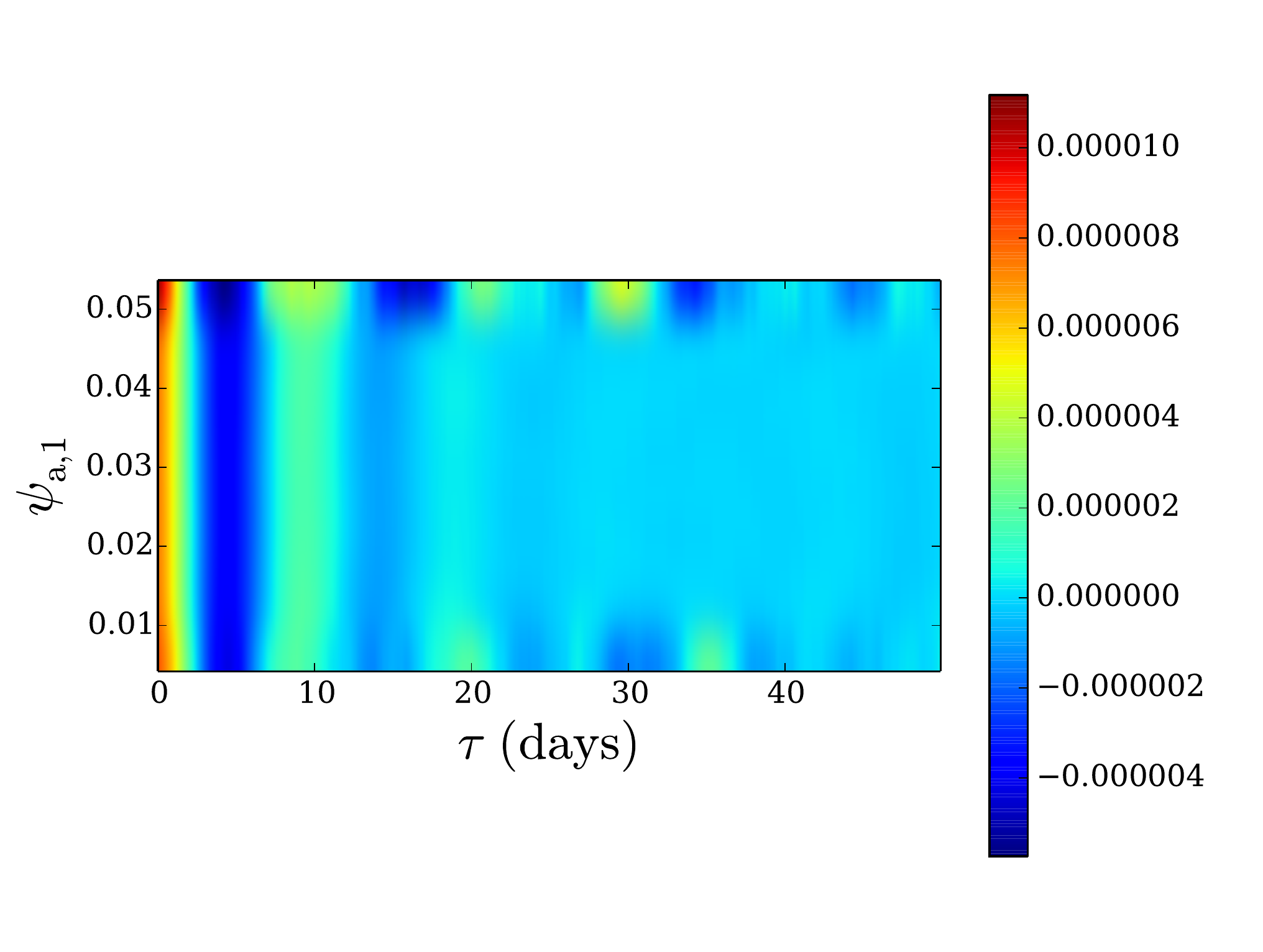}
    \caption{Autocorrelation of $\psi_{{\rm a},8}$ in the uncoupled system.}
  \end{subfigure}
  \caption{Autocorrelation as a function of $\psi_{{\rm a},1}$ and $\tau$ for case 1 ($\varepsilon=1$). \label{fig:corrdisc}}
\end{figure*}

We have also computed the two-dimensional probability density function (PDF) in the large-scale variables $\psi_{{\rm a},1}$ and $\psi_{{\rm o},1}$ for case 1, for both strong and weak couplings. These PDFs are respectively depicted on Figs.~\ref{fig:2d_dist_case1s} and~\ref{fig:2d_dist_case1w}. In both cases, it confirms that the parameterizations correct the LFV, with a slightly better result for the Ornstein-Uhlenbeck modeling of the noise. We also note that the biggest differences between the PDFs are located in some particular regions of the attractors. This is mainly due to the long term correlations of the unresolved variables in these regions, as illustrated on Figure~\ref{fig:corrdisc}. The decorrelation of $\psi_{{\rm a},8}$, an unresolved variable, is much longer in these regions (around $\psi_{{\rm a},1}\approx 0.04$) for the coupled system than for the uncoupled one. Those long term correlations cannot be reproduced by the parameterizations since they rely uniquely on the unperturbed dynamics\footnote{It is possible to consider the next orders in the response theory, but it is not the scope of the present work.}.

\begin{figure*}
  \centering
  \includegraphics[width=.45\linewidth]{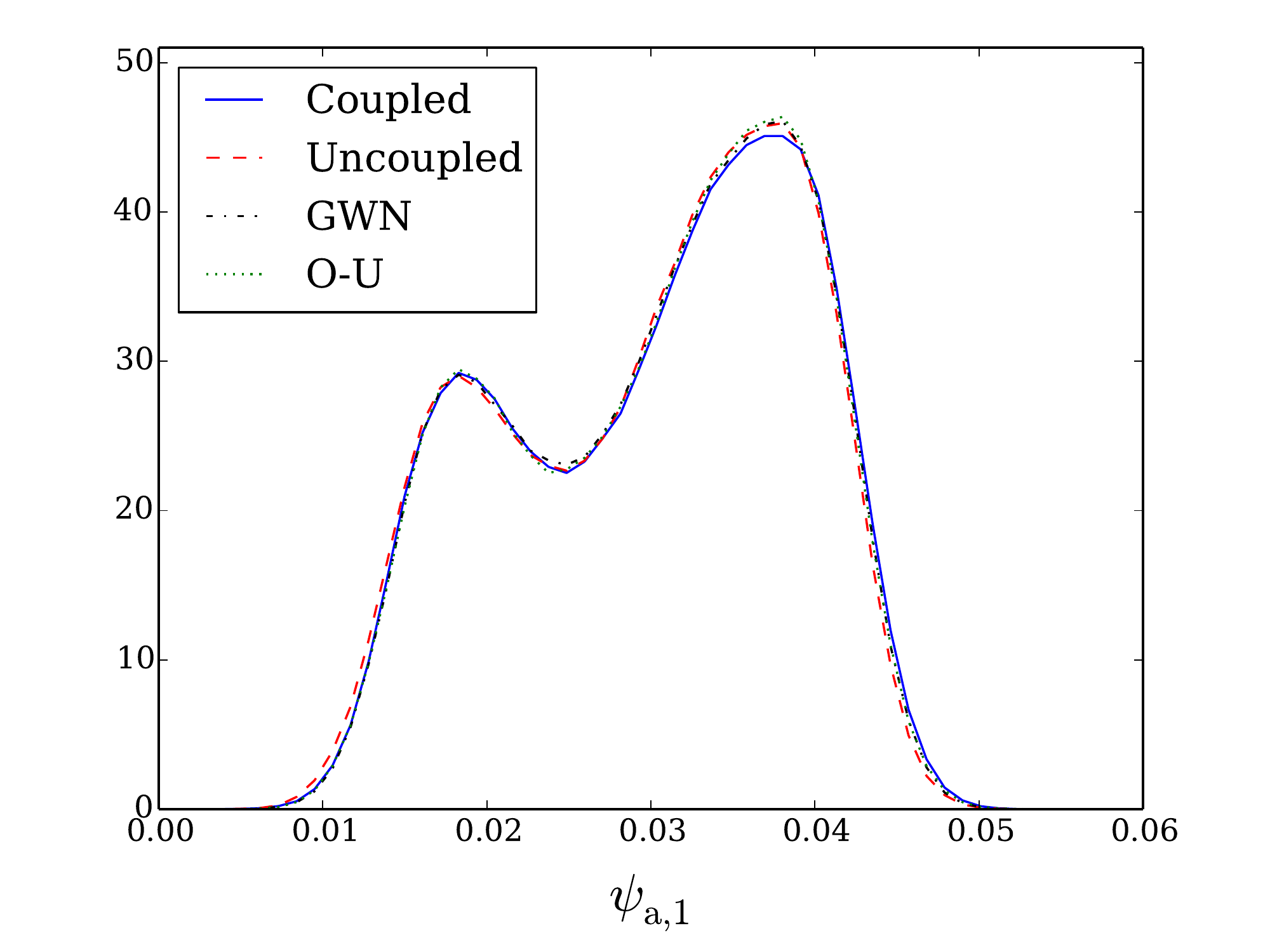}
  \includegraphics[width=.45\linewidth]{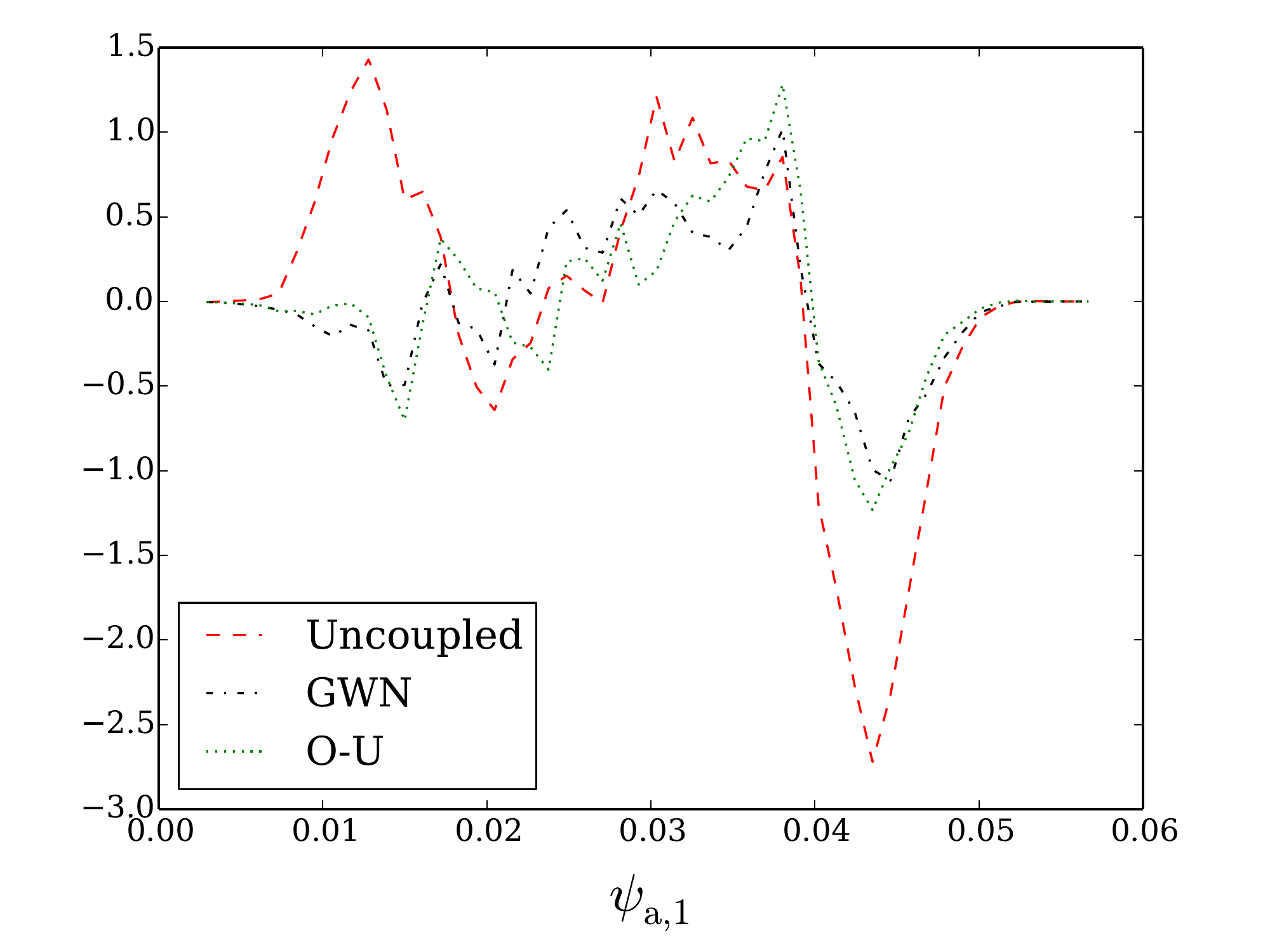}
  \includegraphics[width=.45\linewidth]{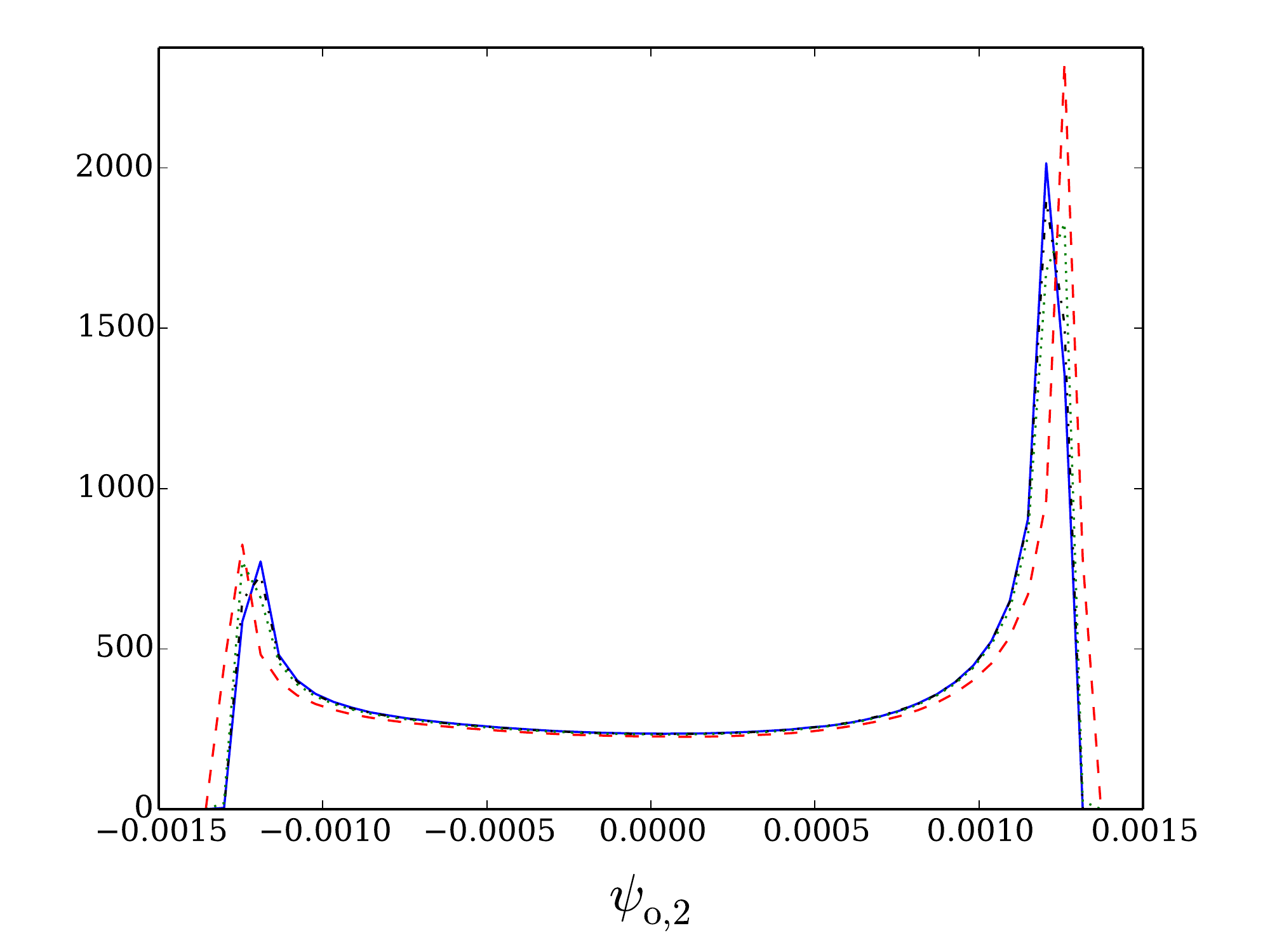}
  \includegraphics[width=.45\linewidth]{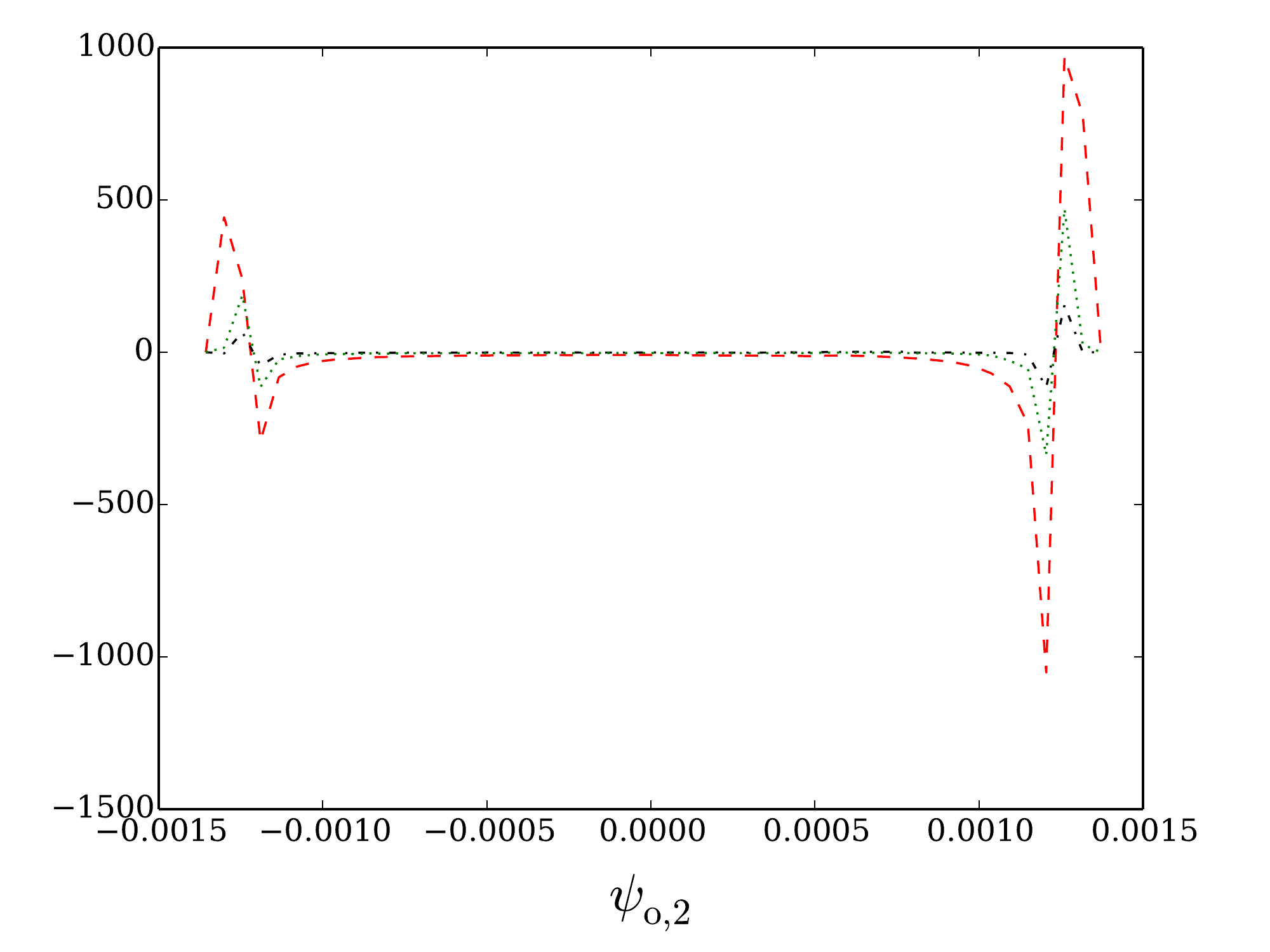}
  \includegraphics[width=.45\linewidth]{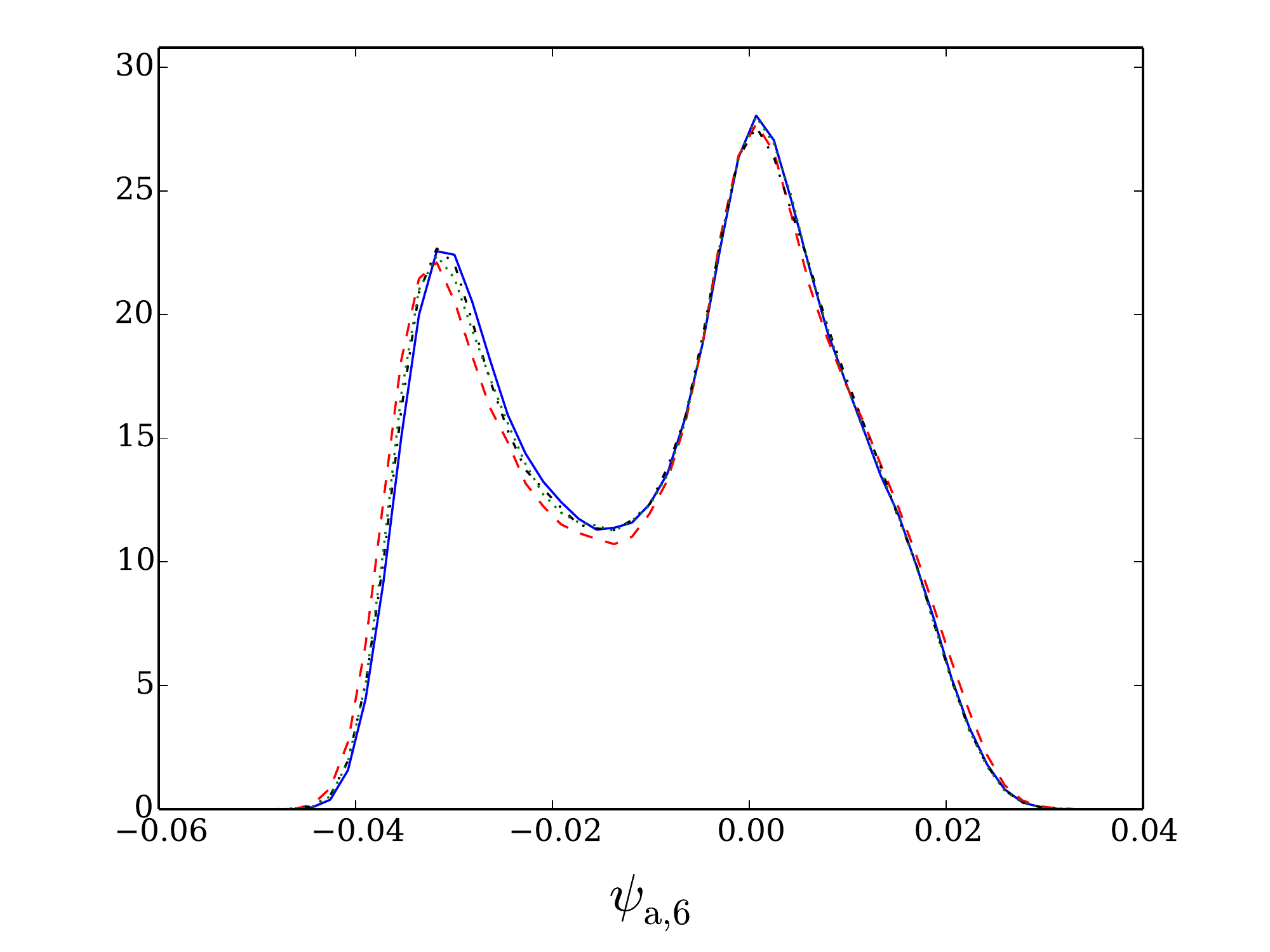}
  \includegraphics[width=.45\linewidth]{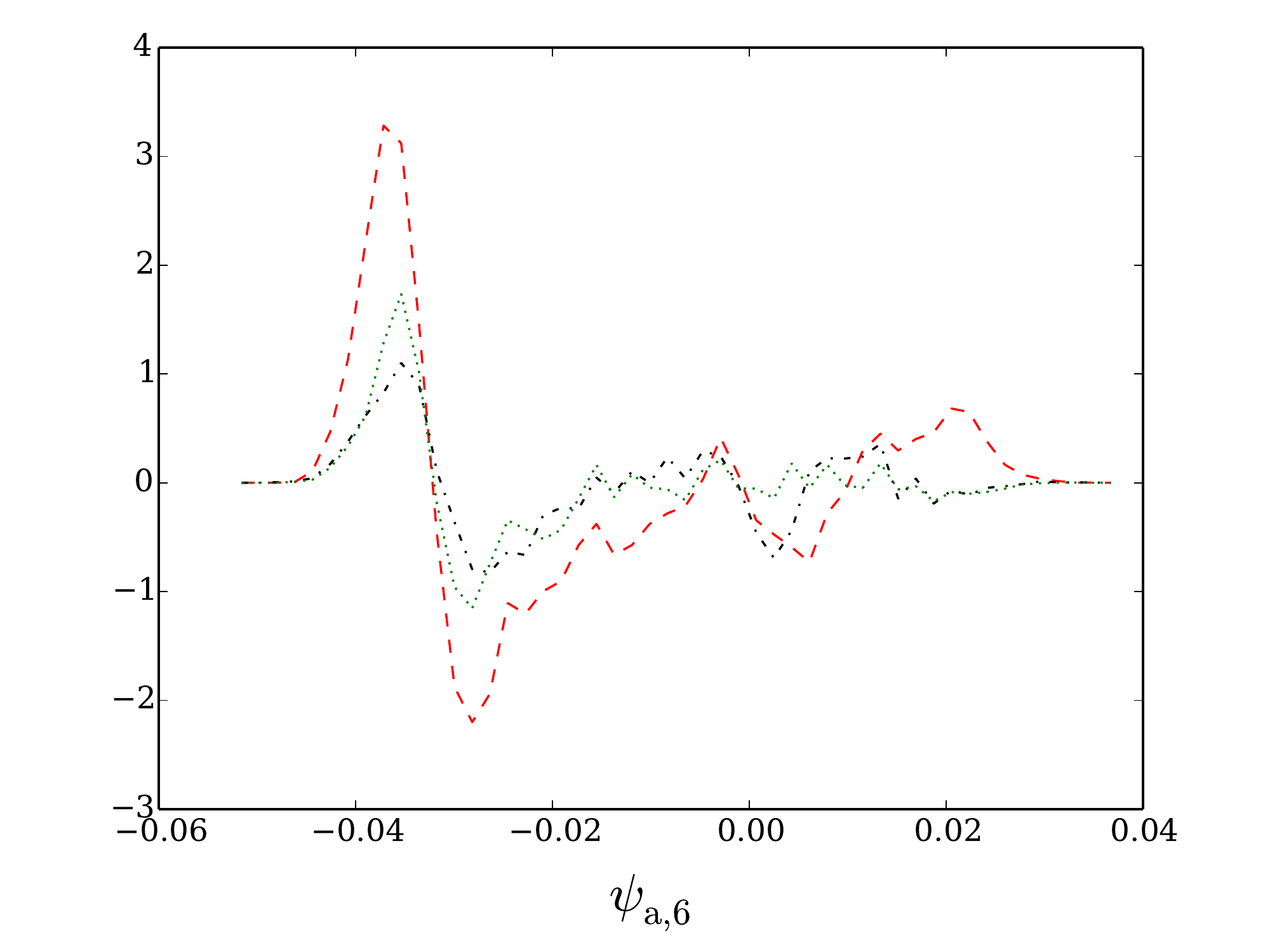}
  \caption{Probability density functions (PDFs) (left) and PDF anomalies (right) of some relevant variables for case 1 with $\varepsilon=0.5$. GWN and O-U refer to the Gaussian white noise and Ornstein-Uhlenbeck modelling of the noise terms, respectively.  \label{fig:dist_case2pert}}
\end{figure*}

 The one-dimensional probability density function of some important large-scale variables for case 1 with a weak coupling ($\varepsilon=0.5$) are shown on Figure~\ref{fig:dist_case2pert} and stress the potential of the method to improve the dynamics, provided that the weak coupling assumption is fulfilled.

\begin{figure*}
  \begin{subfigure}{0.49\textwidth}
    \centering
    \includegraphics[width=\linewidth]{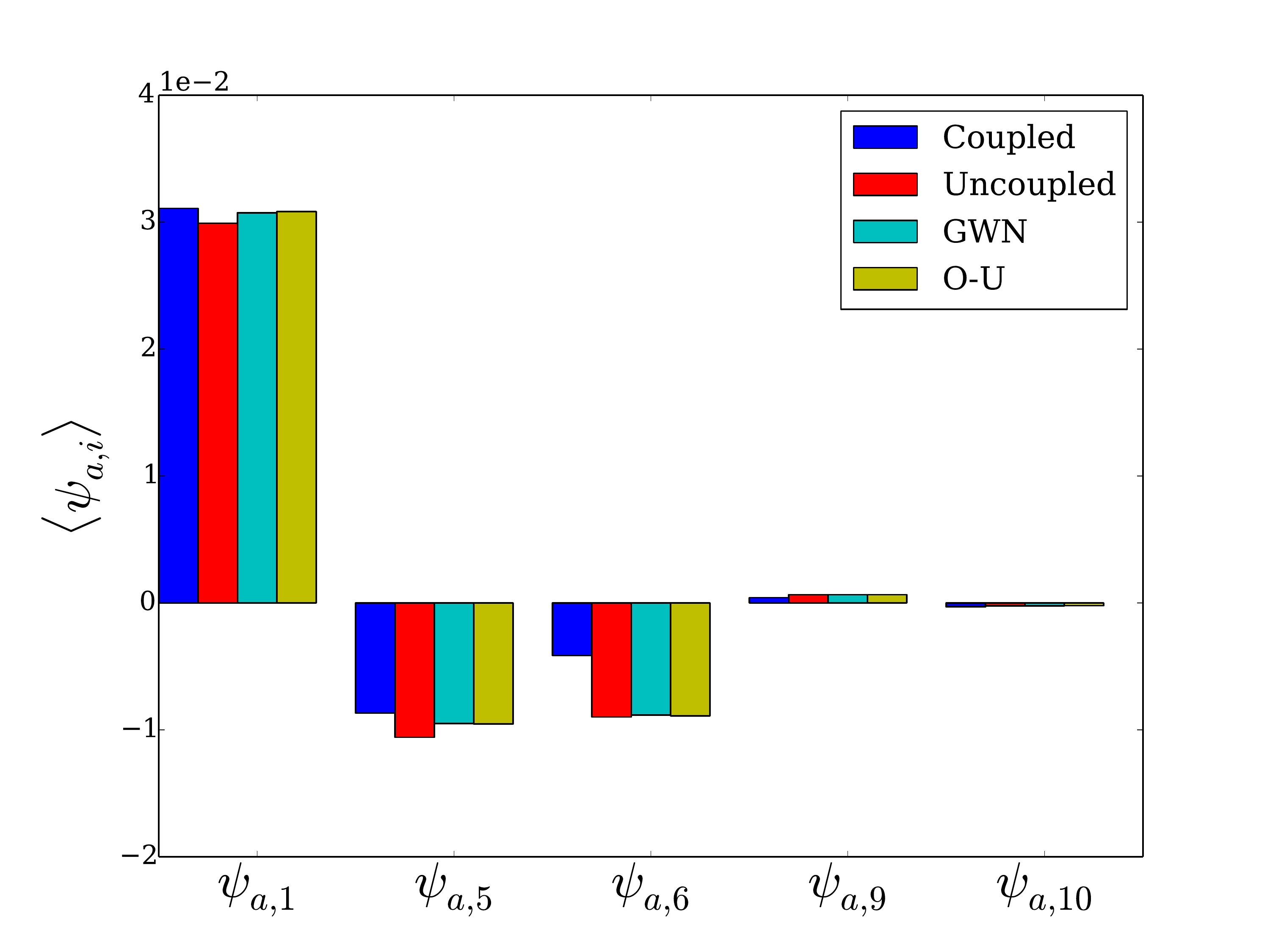}
    \caption{Atmospheric streamfunction variables $\psi_{a,i}$.}
  \end{subfigure}
  \begin{subfigure}{0.49\textwidth}
    \centering
    \includegraphics[width=\linewidth]{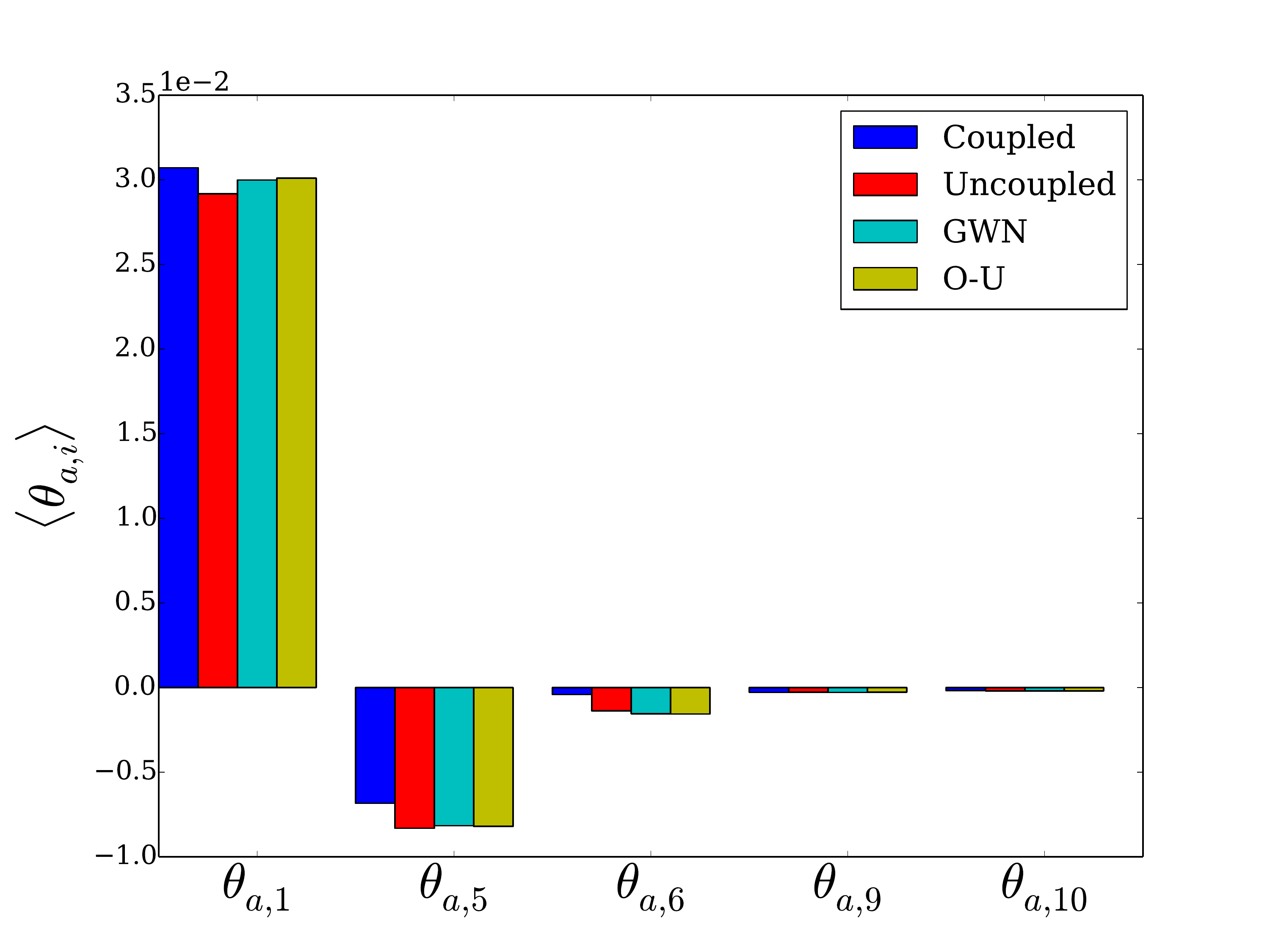}
    \caption{Atmospheric temperature variables $\theta_{a,i}$.}
  \end{subfigure}
  \begin{subfigure}{0.49\textwidth}
    \centering
    \includegraphics[width=\linewidth]{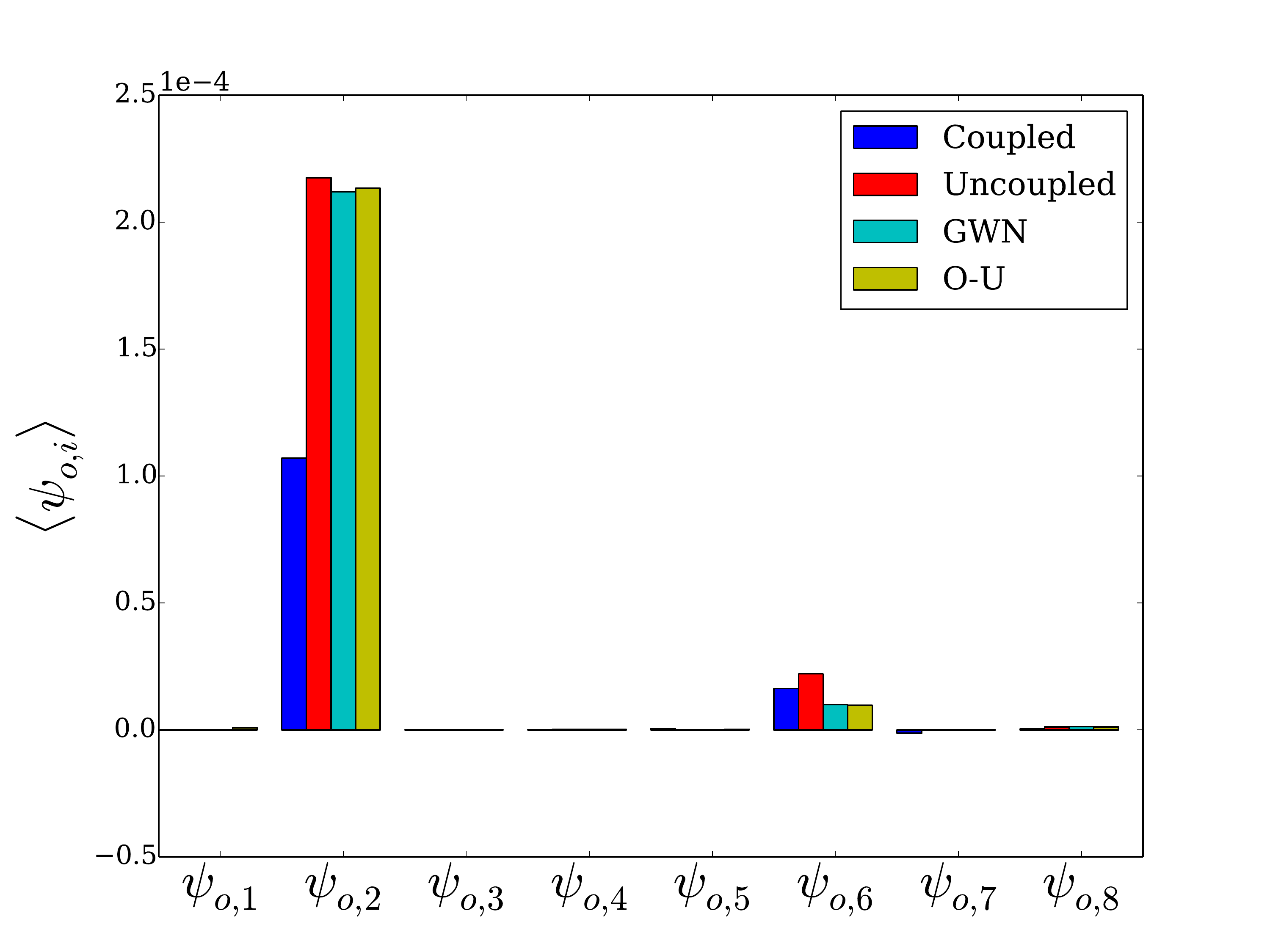}
    \caption{Oceanic streamfunction variables $\psi_{o,i}$.}
  \end{subfigure}
  \begin{subfigure}{0.49\textwidth}
    \centering
    \includegraphics[width=\linewidth]{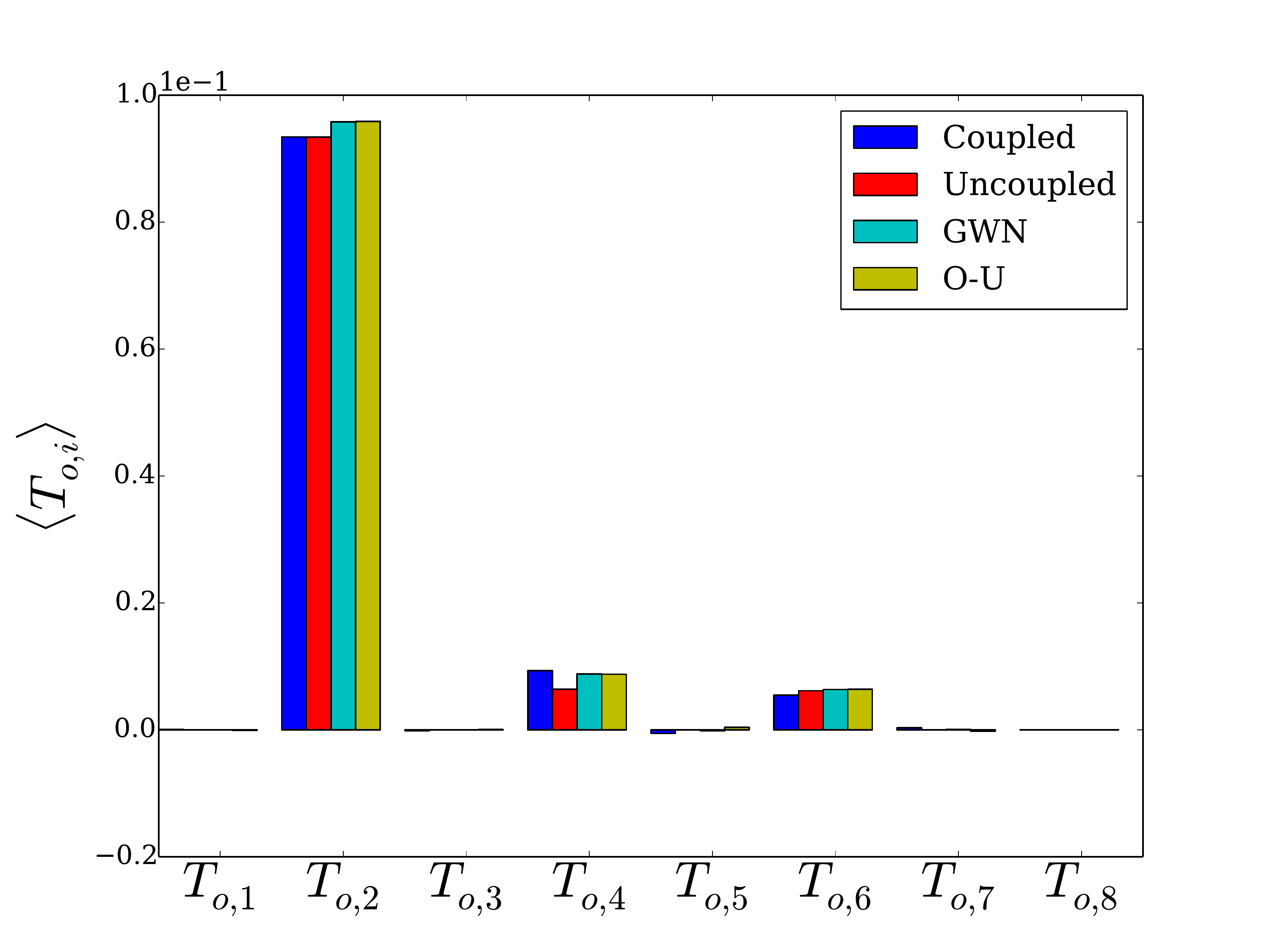}
    \caption{Oceanic temperature variables $T_{o,i}$.}
  \end{subfigure}
  \caption{Mean of each resolved variable in adimensional units for case 1 ($\varepsilon=1$). \label{fig:mean_dist_case1}}
\end{figure*}

\begin{figure*}
  \begin{subfigure}{0.49\textwidth}
    \centering
    \includegraphics[width=\linewidth]{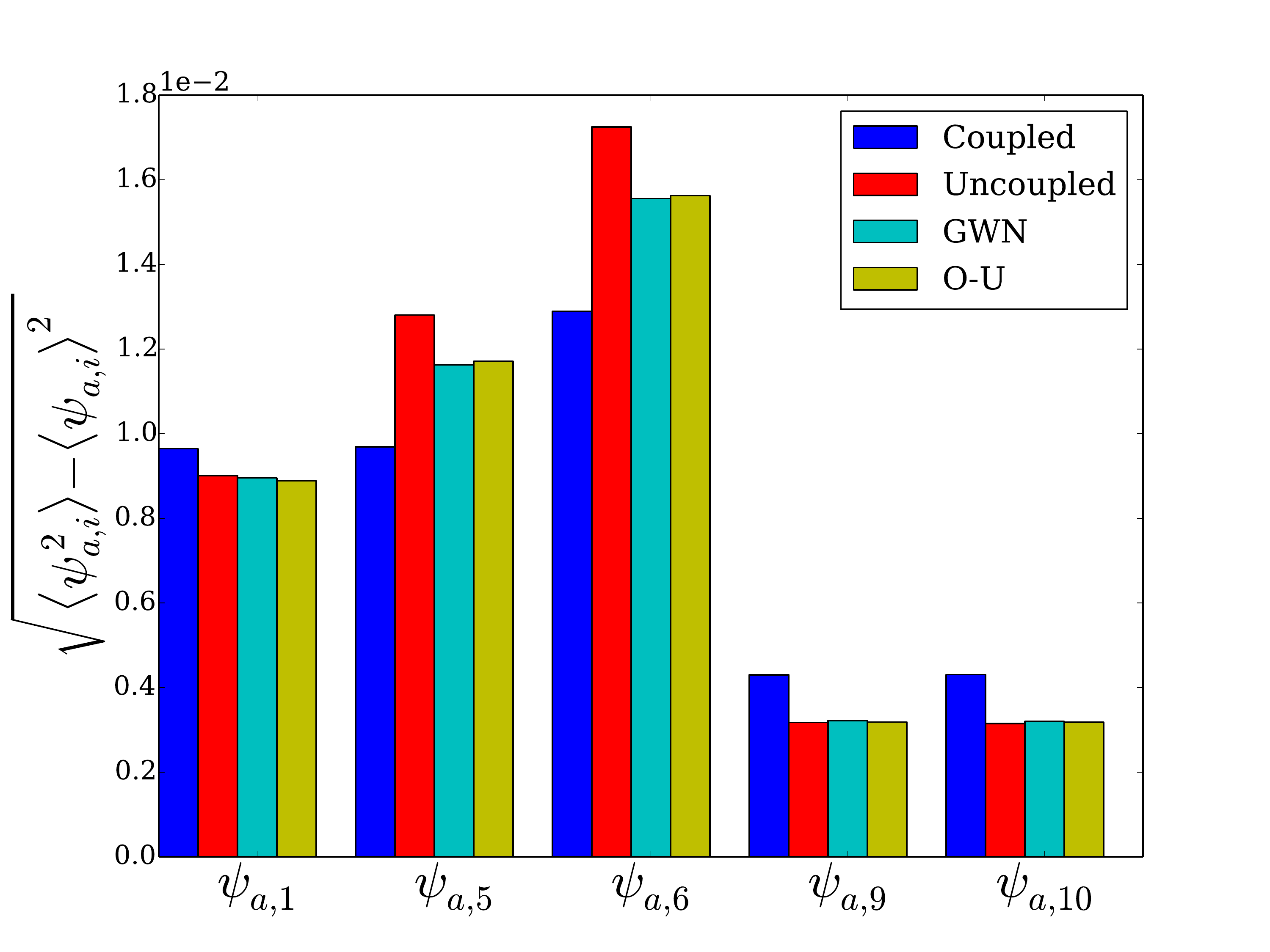}
    \caption{Atmospheric streamfunction variables $\psi_{a,i}$.}
  \end{subfigure}
  \begin{subfigure}{0.49\textwidth}
    \centering
    \includegraphics[width=\linewidth]{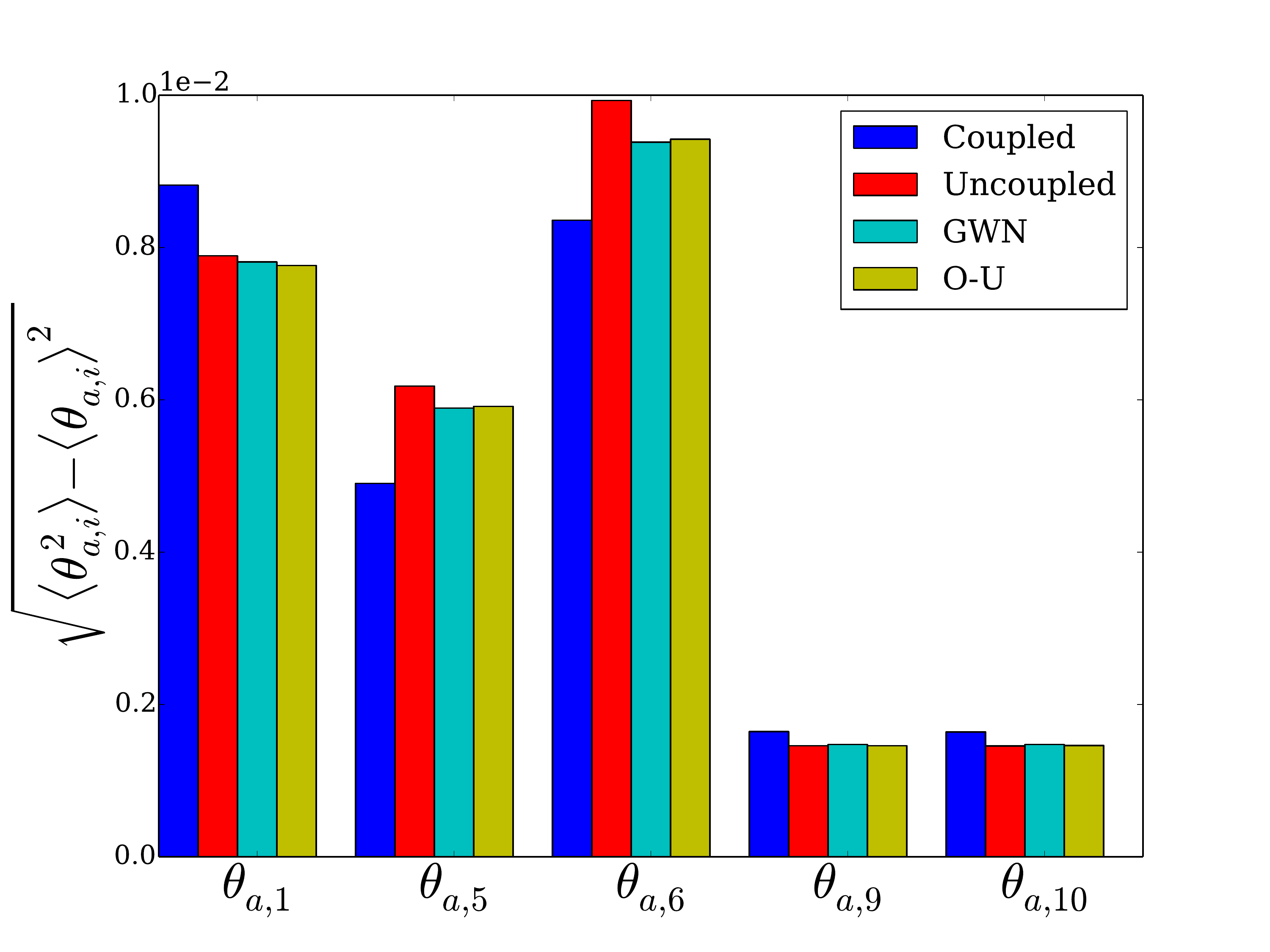}
    \caption{Atmospheric temperature variables $\theta_{a,i}$.}
  \end{subfigure}
  \begin{subfigure}{0.49\textwidth}
    \centering
    \includegraphics[width=\linewidth]{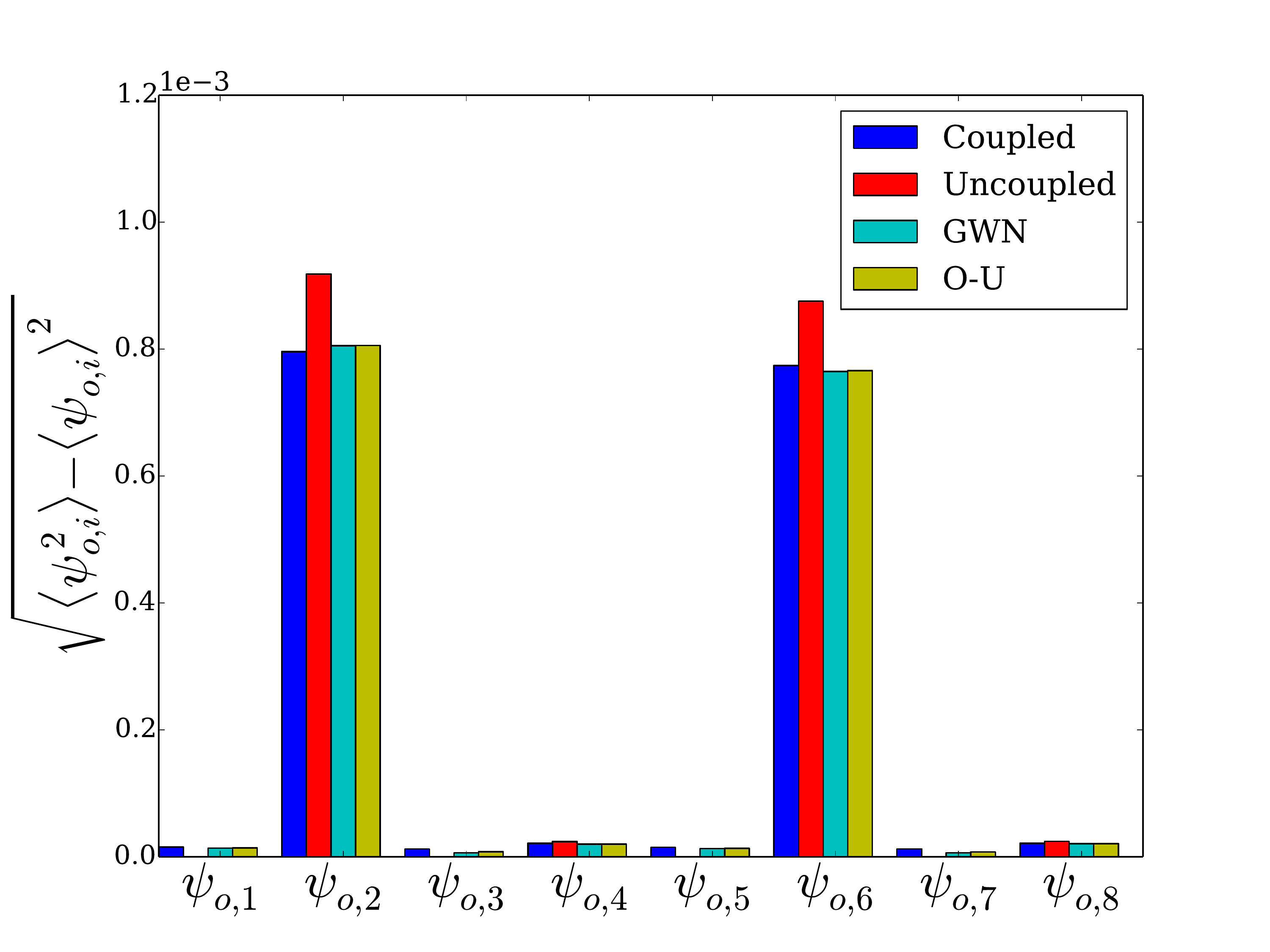}
    \caption{Oceanic streamfunction variables $\psi_{o,i}$.}
  \end{subfigure}
  \begin{subfigure}{0.49\textwidth}
    \centering
    \includegraphics[width=\linewidth]{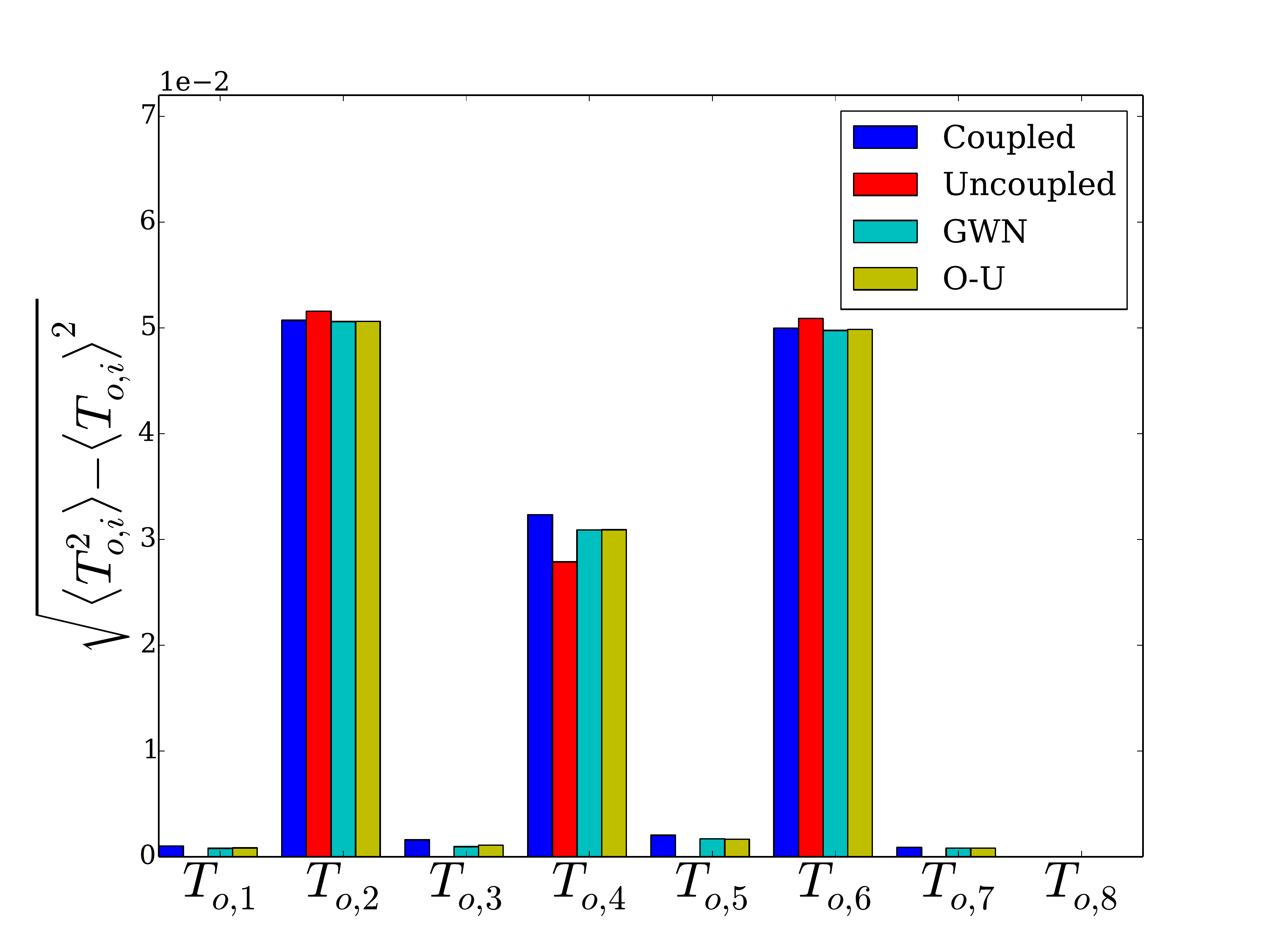}
    \caption{Oceanic temperature variables $T_{o,i}$.}
  \end{subfigure}
  \caption{Standard deviation of each resolved variable in adimensional units for case 1 ($\varepsilon=1$). \label{fig:var_dist_case1}}
\end{figure*}

\begin{figure*}
  \begin{subfigure}{0.49\textwidth}
    \centering
    \includegraphics[width=\linewidth]{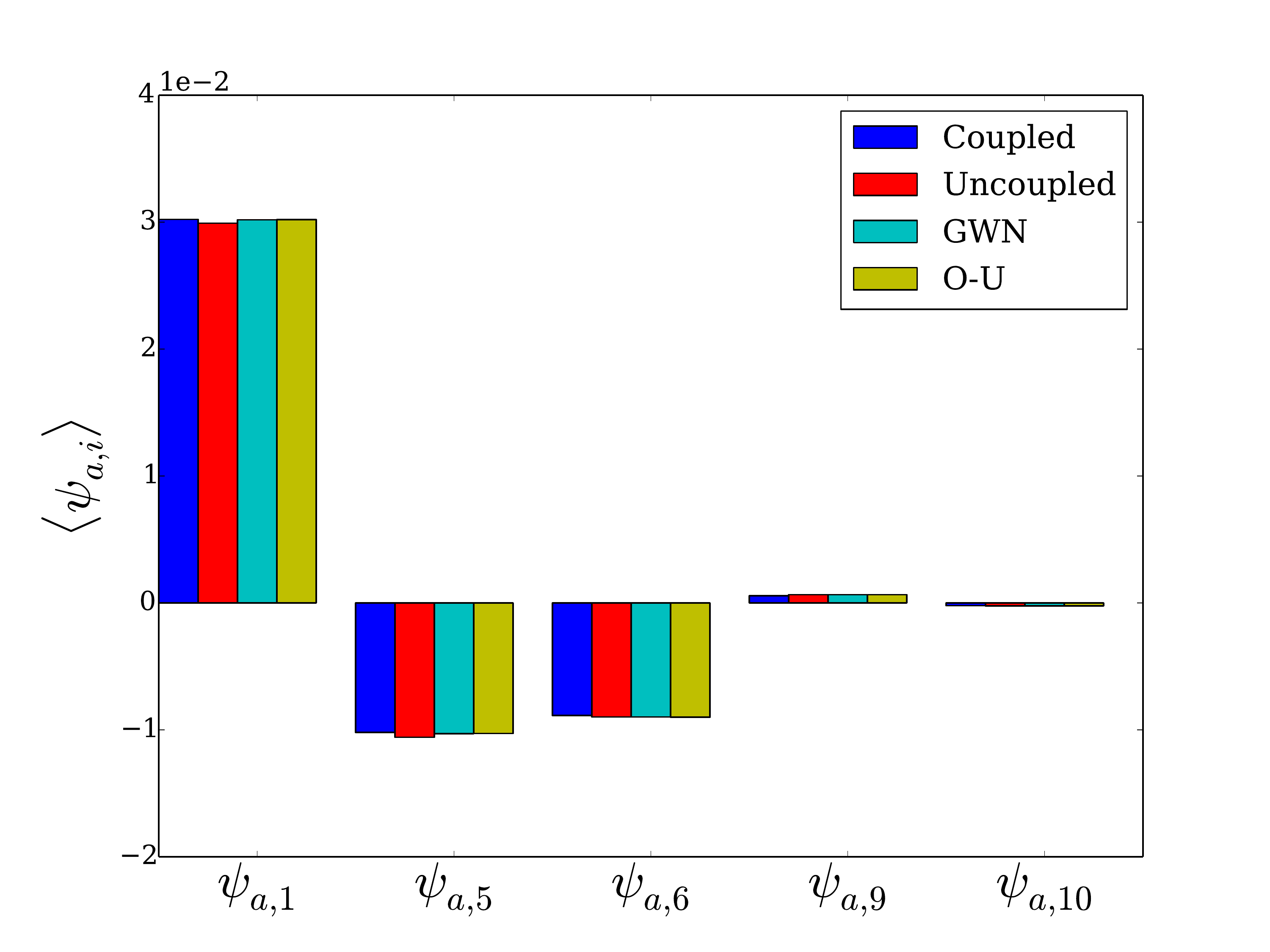}
    \caption{Atmospheric streamfunction variables $\psi_{a,i}$.}
  \end{subfigure}
  \begin{subfigure}{0.49\textwidth}
    \centering
    \includegraphics[width=\linewidth]{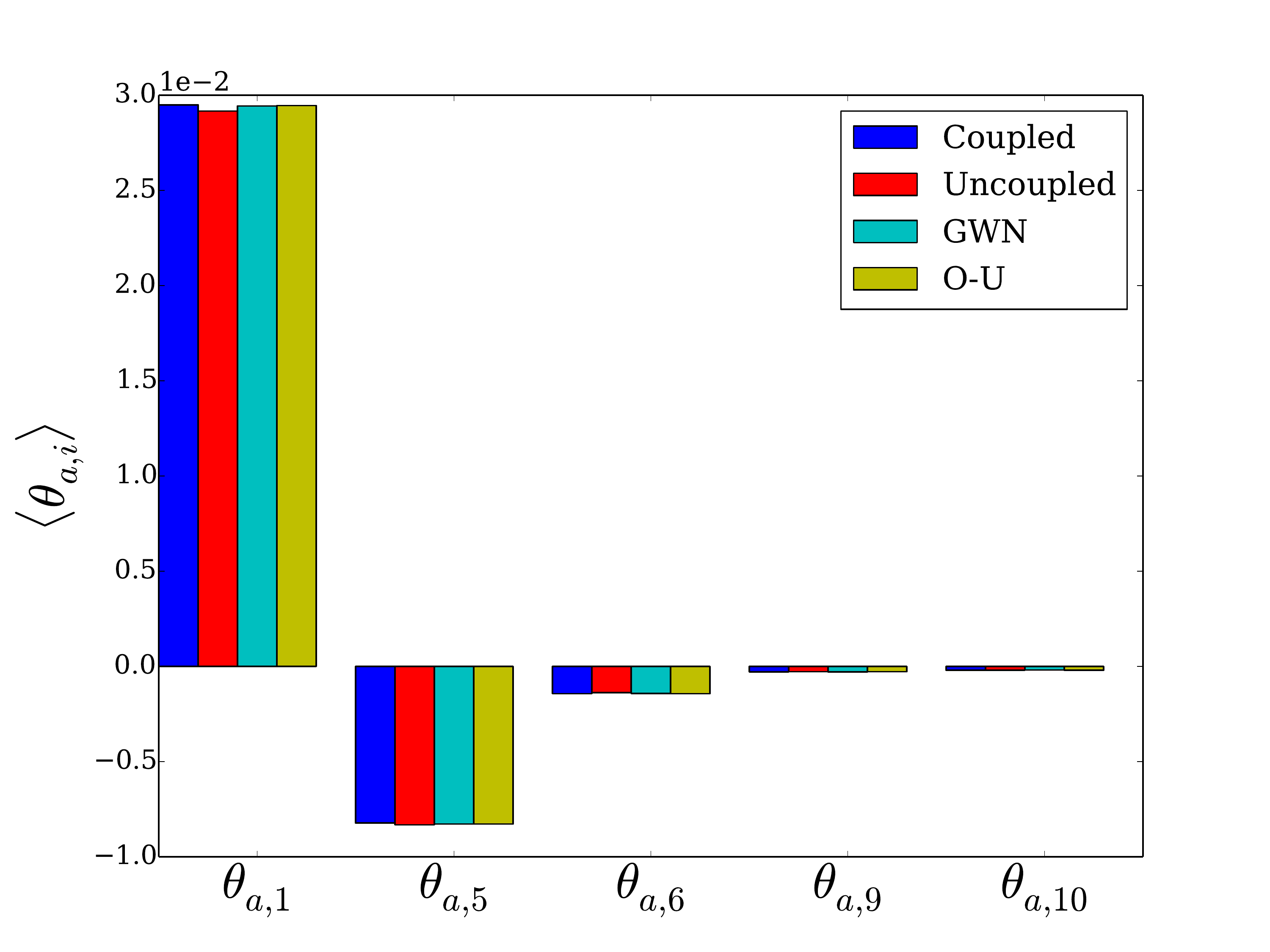}
    \caption{Atmospheric temperature variables $\theta_{a,i}$.}
  \end{subfigure}
  \begin{subfigure}{0.49\textwidth}
    \centering
    \includegraphics[width=\linewidth]{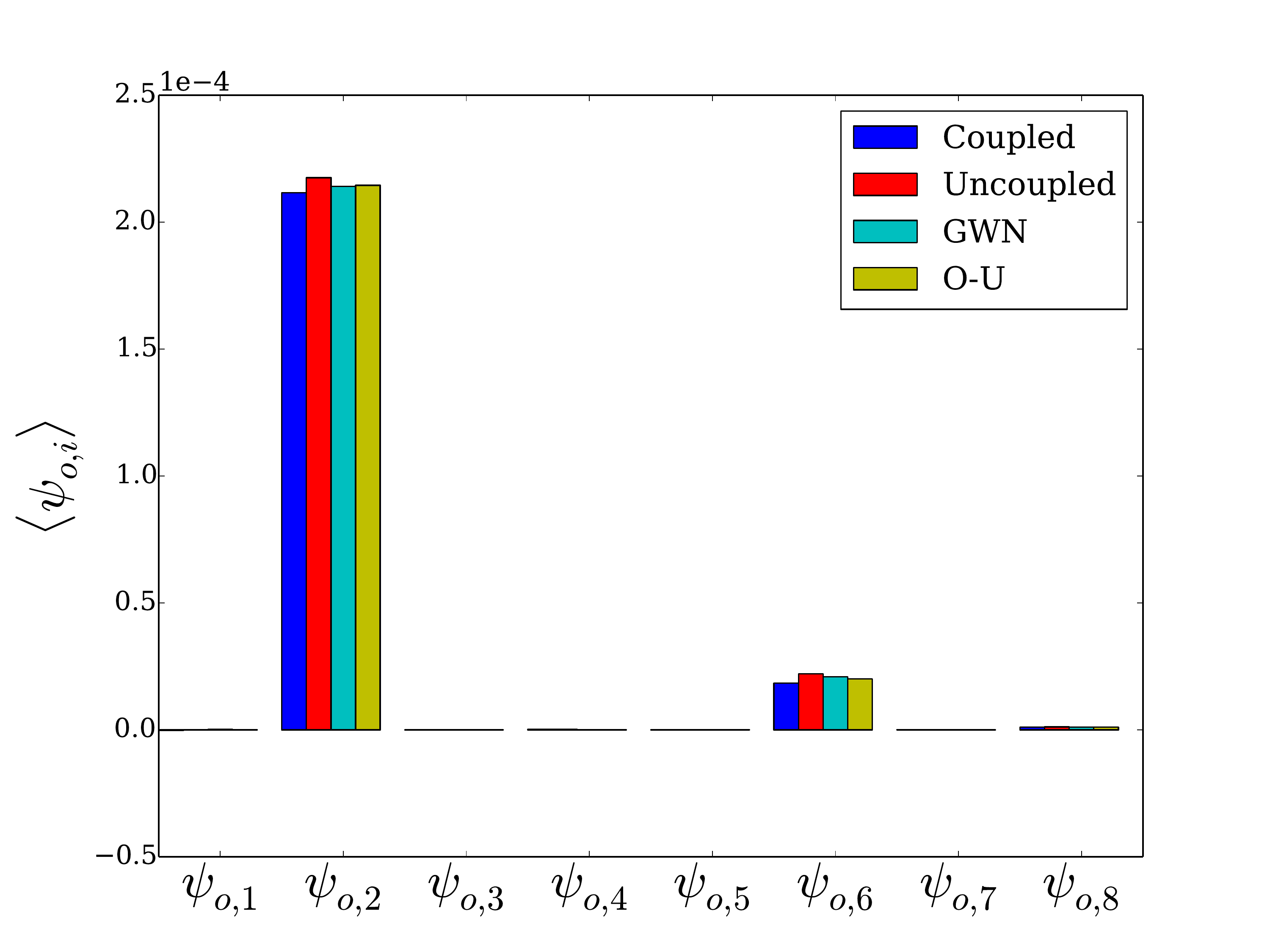}
    \caption{Oceanic streamfunction variables $\psi_{o,i}$.}
  \end{subfigure}
  \begin{subfigure}{0.49\textwidth}
    \centering
    \includegraphics[width=\linewidth]{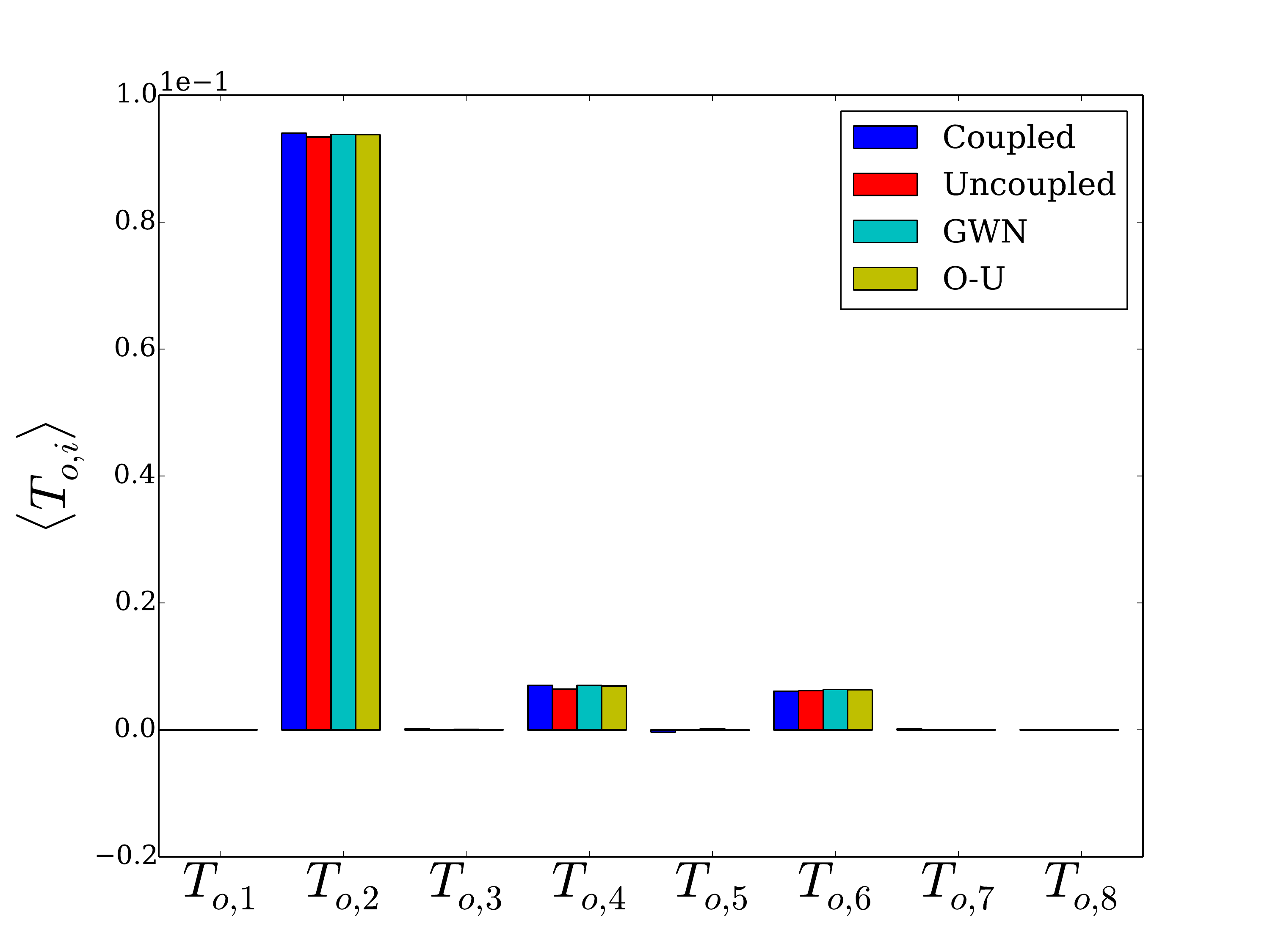}
    \caption{Oceanic temperature variables $T_{o,i}$.}
  \end{subfigure}
  \caption{Mean of each resolved variable in adimensional units for case 1 ($\varepsilon=0.5$). \label{fig:mean_dist_pert_case1}}
\end{figure*}

\begin{figure*}
  \begin{subfigure}{0.49\textwidth}
    \centering
    \includegraphics[width=\linewidth]{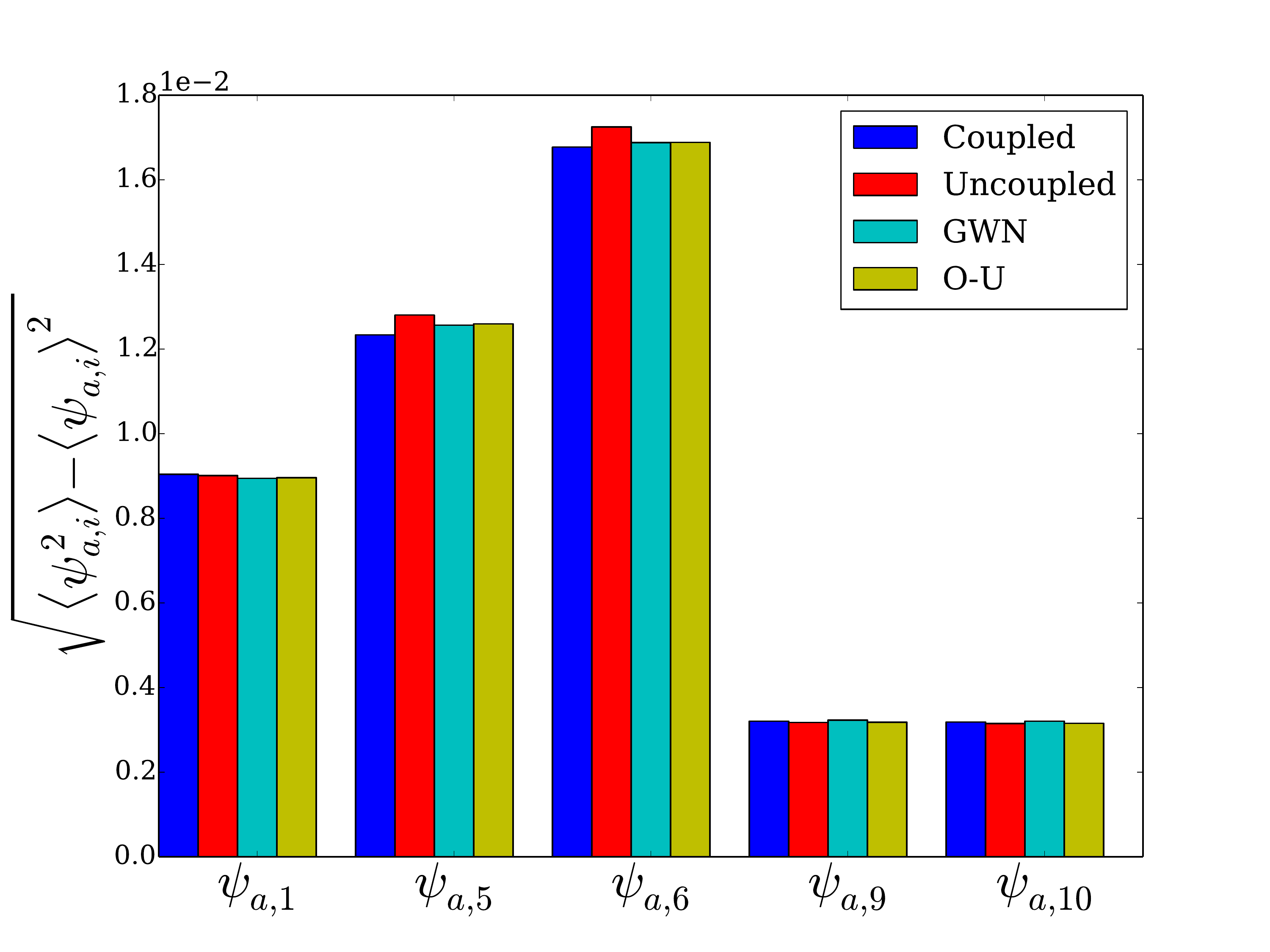}
    \caption{Atmospheric streamfunction variables $\psi_{a,i}$.}
  \end{subfigure}
  \begin{subfigure}{0.49\textwidth}
    \centering
    \includegraphics[width=\linewidth]{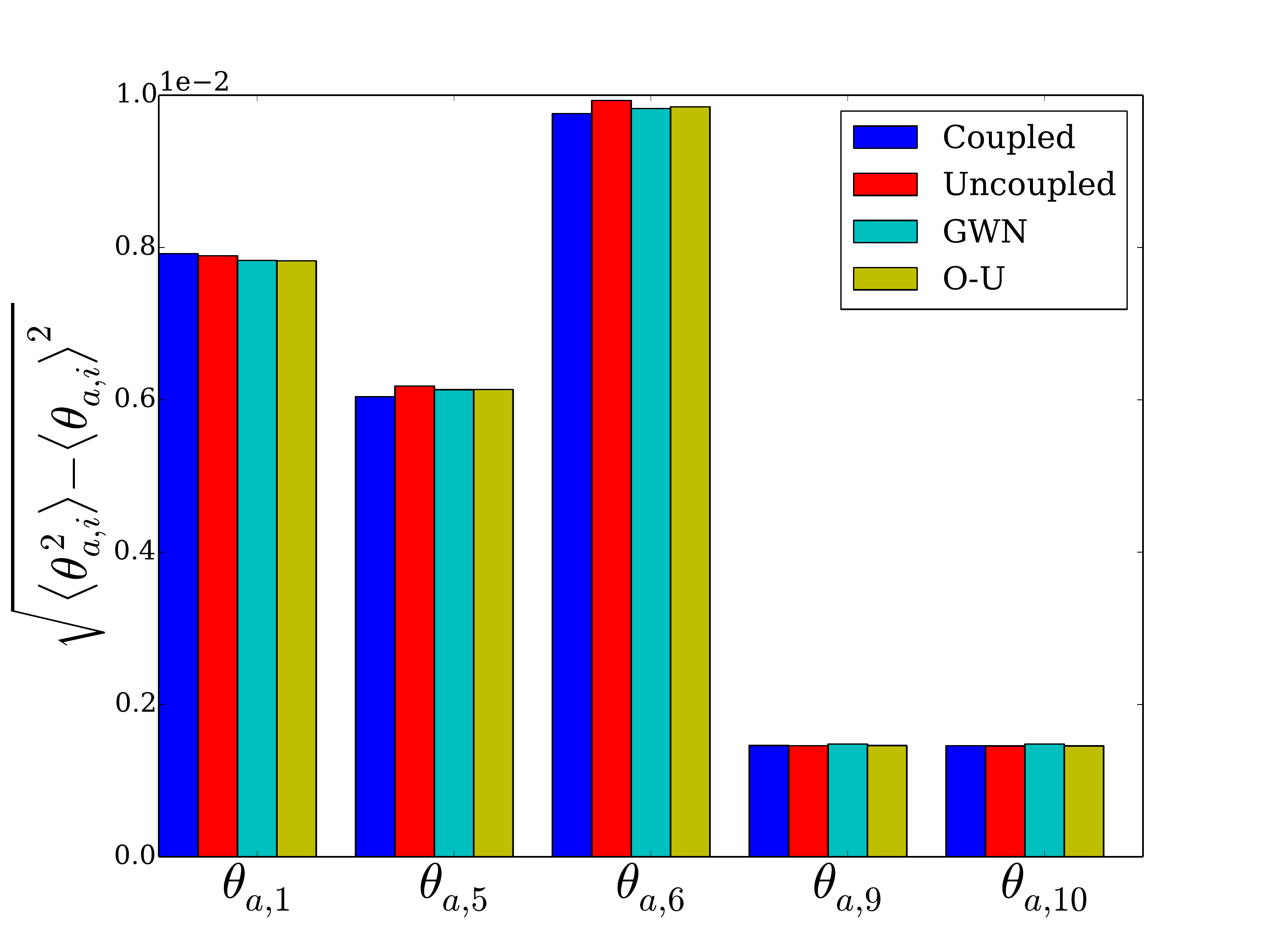}
    \caption{Atmospheric temperature variables $\theta_{a,i}$.}
  \end{subfigure}
  \begin{subfigure}{0.49\textwidth}
    \centering
    \includegraphics[width=\linewidth]{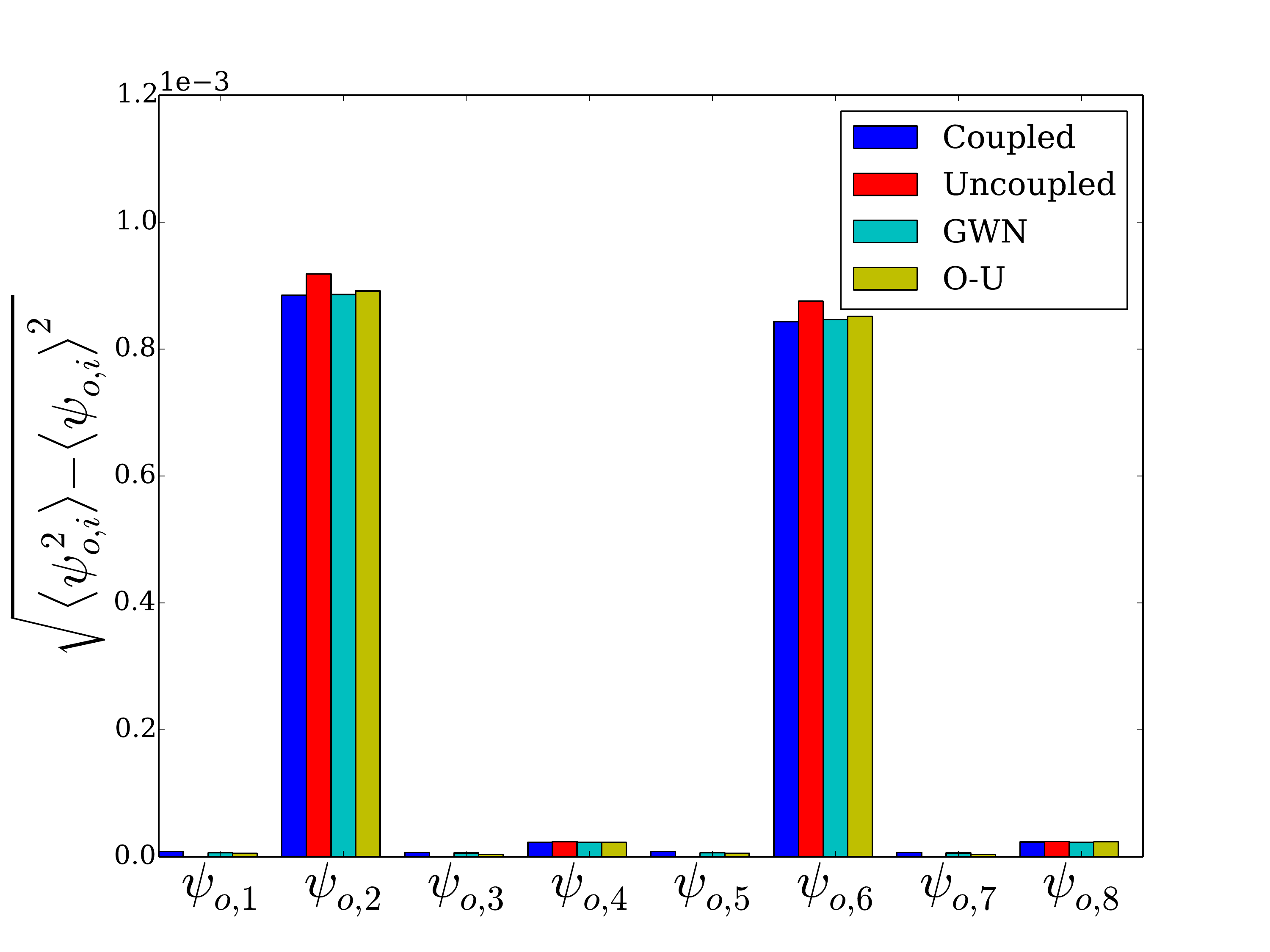}
    \caption{Oceanic streamfunction variables $\psi_{o,i}$.}
  \end{subfigure}
  \begin{subfigure}{0.49\textwidth}
    \centering
    \includegraphics[width=\linewidth]{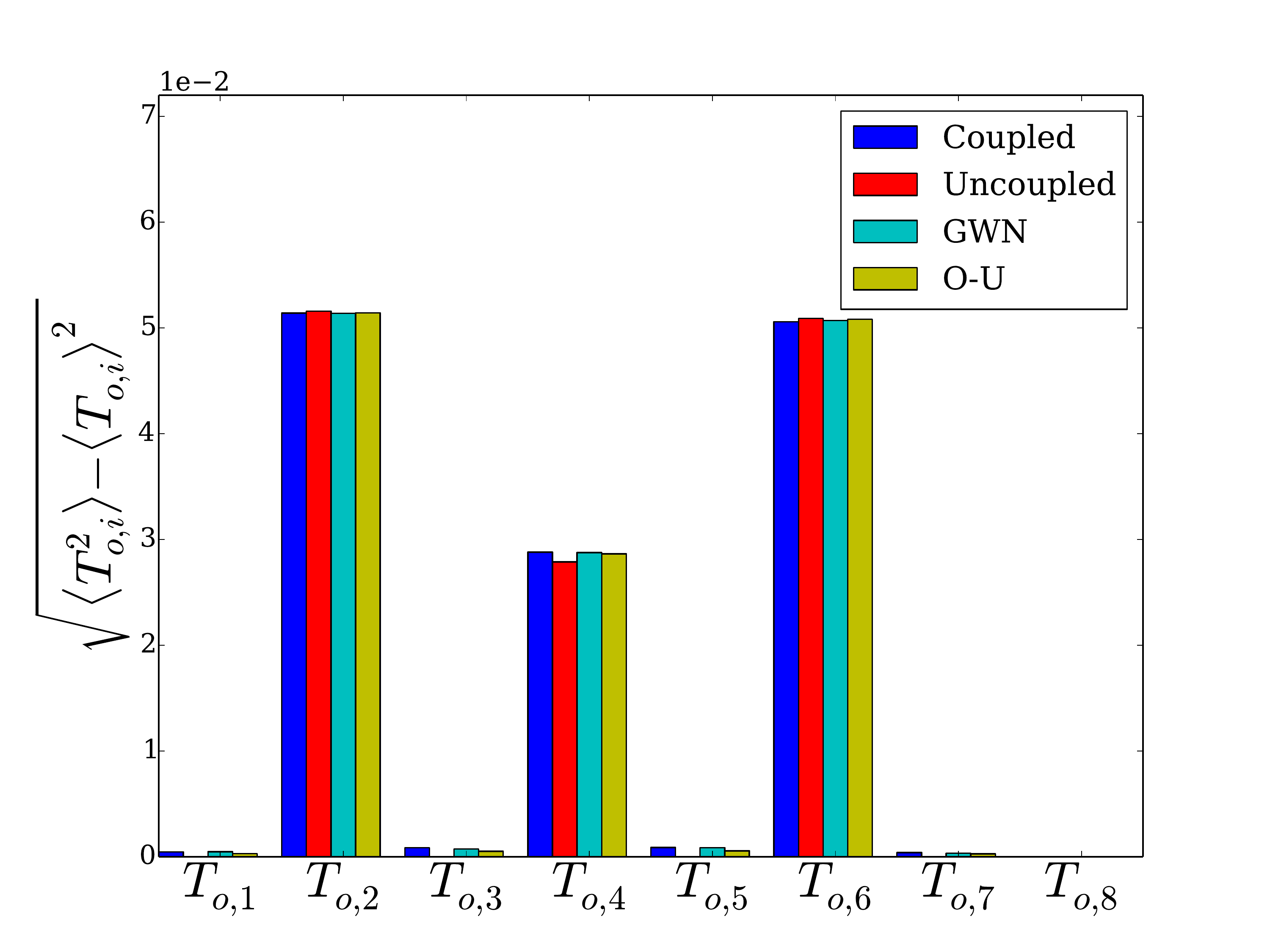}
    \caption{Oceanic temperature variables $T_{o,i}$.}
  \end{subfigure}
  \caption{Standard deviation of each resolved variable in adimensional units for case 1 ($\varepsilon=0.5$). \label{fig:var_dist_pert_case1}}
\end{figure*}

Another way to clarify the impact of the parameterization scheme is to compute the mean and the standard deviation of the resolved variables. The results for $\varepsilon=1$ and $\varepsilon=0.5$ are respectively depicted on Figs.~\ref{fig:mean_dist_case1},\ref{fig:var_dist_case1} and Figs.~\ref{fig:mean_dist_pert_case1},\ref{fig:var_dist_pert_case1}, for case 1. For $\varepsilon=1$, the correction of the means is very limited in the atmosphere but the standard deviations is well corrected. This trend is also confirmed in the ocean, where both parameterizations (GWN and O-U) cannot correct a huge bias (see Fig.~\ref{fig:mean_dist_case1}(c)) but on the other hand they induce the correction needed for the variability (Fig.~\ref{fig:var_dist_case1}(c)). The same conclusions hold for the weak coupling case $\varepsilon=0.5$, with the difference that there is roughly no model bias to correct.

The fact that the parameterizations correct better the variability than the climatological mean is reminiscent of a result obtained by~\citein{N2005} which states that, regarding these two quantities, no universal corrections of model errors can be obtained. More specifically, it was shown that a Markovian parameterization corrects the mean and the variability depending on the structure of the Jacobian and Hessian matrices. The author concludes that it is in general very difficult to correct simultaneously the error for both. In the present work, the non-Markovian parameterization is not able to correct both simultaneously. However, since the large LFV signal is the dominant feature of the model, we must emphasize that a good correction of the variability is here the desired outcome that we would expect from the parameterization scheme.

Finally, we have also tested the parameterizations for the three cases with a small ocean-atmosphere coupling, by setting the parameter $d$ to $10^{-9}$ s$^{-1}$. The difference between the coupled and uncoupled systems PDFs in the atmosphere was important and none of the parameterizations provided good results. This negative result is related to the absence of a LFV in these cases, and of the associated natural separation between ``slow'' and ``fast'' dynamics (see also next section).

\section{Discussion and conclusions}
\label{sec:conc}

The stochastic parameterization of the subgrid-scale processes in climate systems is a crucial ingredient to improve their natural variability when only long low-resolution runs are affordable. In this paper, we have considered a new method introduced by~\cite{WL2012,WL2013}, based on the Ruelle's response theory. In particular, we have applied this method to a midlatitude low-order coupled ocean-atmosphere system. This model possesses a decadal oscillation reminiscent of the North-Atlantic oscillation (NAO),  and allows for a natural separation of the phase space into a \emph{coupled} ocean-atmosphere low-frequency dynamics that we would like to reproduce accurately and a complementary dynamics consisting of the other oceanic and atmospheric modes that could be considered as perturbations. This natural separation is due to the presence of an invariant subspace of the Jacobian appearing in the model~\cite{CP1996,CP1997,CP1999}, which partition the phase space efficiently and allows to derive analytical expressions for the parameterization.\\

We tested this parameterization for different parameter sets and for different coupling strength between the resolved and unresolved components, quantified by a coupling parameter $\varepsilon$. We found a good agreement between the statistical properties of the full system and the resolved component of the reduced dynamics when $\varepsilon$ is sufficiently small (here in particlar we used $\varepsilon$ equal to $0.5$). When the full coupling strenght is restored ($\varepsilon = 1$), some encouraging results are still obtained for the largest values of the wind stress forcing explored as measured by the parameter $d=C/\rho H$. In this case we suspect that the stochastic parameterization is still correcting the variability within the ocean due to the smallness of the parameter $d$ (which in that case plays the role of a weak coupling parameter compared to the strenght of the other couplings). This, in turn, allows for correcting the dynamics within the atmosphere, which is considerably influenced by the ocean through the LFV developing along the coupled unstable periodic orbit of the system. When the wind stress friction parameter $d$ is small, the LFV disappears and the atmosphere dynamics is only mildly influenced by the ocean dynamics and heat transport, implying that any correction of the transport variability within the ocean will not improve the dynamics within the atmosphere. 

These results confirm that the hypothesis of weak coupling between atmospheric modes is not met in the present case (except when we enforce it with a lower value of $\varepsilon$). Nevertheless the fact that a purely atmospheric subgrid effects can be corrected for the largest wind stress forcing, even as byproduct of a better oceanic representation, is definitively a very encouraging result.\\

To summarize, the main results of the present work are: 

\begin{description}
  \item[(i)] The very good correction obtained by the response theory parameterization in the case of a weak coupling ($\varepsilon=0.5$).
  \item[(ii)] The good correction of the LFV in the strong coupling case and large value of $d$ ($\varepsilon=1$, $d>1.0 \times 10^{-8}$) as a result of the correction of the transport which benefit to the whole resolved component via feedback mechanisms.
  \item[(iii)] The absence of correction obtained for small value of $d$ ($d=1.0 \times 10^{-9}$), due to the mechanical decoupling of the dynamics between the ocean and the atmosphere.
\end{description}

In this work, a clear separation of ``structured'' (represented by an invariant set of variables of the systems) and ``unstructured'' dynamics in phase space is exploited, leading to a natural separation of scales. Finding such separations of scales in more sophisticated models can help building useful parameterizations preserving the large-scale slow dynamics of the model, and as a byproduct allowing long-term forecasts.\\

Finally, the method presented in this article is in principle not only limited to systems presenting a clear separation of scales. For instance, the general decomposition~(\ref{eq:oastruc}) - that allows for an arbitrary subgrid configuration in the atmosphere - could also be considered. In this setup, the noise is not necessarily additive as in the particular case~(\ref{eq:oamod}) considered here, but a combination of multiplicative and additive noises. In this context, the non-Markovian memory terms will also show an increased complexity. However, dealing with systems lacking a clear timescale separation requires the consideration of this general setup. This is a work in progress.

The constraint of weak coupling inherited from the method should also be investigated on other processes, on a case by case basis. In the general context of the quasi-geostrophic theory and the description of large-scale synoptic flows, a possibility is to consider the terms involving the divergent part of the velocity fields as weakly coupled terms~\cite{HW1980}. Another field of application is the impact of soil moisture in high-resolution numerical atmospheric models, considered as weakly coupled to the surface boundary layer~\cite{G2006,LS2005}. These fields of applications are certainly worth considering in the future.

\section*{Acknowledgment}

The authors would like to thank Valerio Lucarini and Jeroen Wouters for useful discussions and suggestions. They also thank the anonymous referees for their constructive and stimulating suggestions and comments. Some computations have been performed with Mathematica~\cite{M2014} and the figures have been prepared with the Matplotlib software~\cite{H2007}.
This work is supported by the Belgian Federal Science Policy Office under contracts BR/121/A2/STOCHCLIM.
\newpage
\FloatBarrier


\FloatBarrier
\appendix
\section{Computation of the functions for the parameterization based on response theory}
\label{sec:Appoa}
In this section, we compute the quantities $\bsrm{M}_1$, $\bsrm{g}(s)$ and $\bsrm{h}(\bsrm{X},s)$ for the system (\ref{eq:oamod}).
The equations of this system can be rewritten in a indices notation form:
\begin{align}
  \dot X_{{\rm a},i} & = F_{{\rm a},i} (\bsrm{X}) + \varepsilon \, \sum_{\alpha,\beta \in N^Y} \, C^{\rm a}_{i\alpha\beta} \, Y_\alpha \, Y_\beta \qquad , \quad  i \in N^{\rm a}\\
  \dot X_{{\rm o},i} & = F_{{\rm o},i} (\bsrm{X}) + \varepsilon \, \sum_{\alpha\in N^Y} R^{\rm o}_{i\alpha} \, Y_\alpha \qquad , \quad i \in N^{\rm o}\\
  \dot Y_\alpha & = \sum_{\beta\in N^Y} \, A_{\alpha\beta} \, Y_\beta + \sum_{\beta\in N^Y} \, B_{\alpha\beta} \, \xi_{Y,\beta}  \nonumber \\
  &  \quad + \varepsilon \, \left( \sum_{j\in N^{\rm a}} \sum_{\beta\in N^Y} \, V_{\alpha j\beta}^Y \, X_{{\rm a},j} \, Y_\beta + \sum_{j\in N^{\rm o}} \, R^Y_{\alpha j} \, X_{{\rm o},j}\right) \nonumber \\
  & \qquad\qquad\qquad\qquad , \quad \alpha \in N^{Y}
\end{align}
where $N^{\rm a}$, $N^{\rm o}$ and $N^Y$ are respectively the set of resolved atmospheric, resolved oceanic and unresolved variables indices. $\bsrm{B} = q_Y \, \bsrm{I}$ with $\bsrm{I}$ the identity matrix. Note that in the following, we have taken the convention to use Greek letters as indices for the unresolved variables, while the resolved variables indices are denoted by Roman letters.
The unperturbed $\bsrm{Y}$ dynamics is a multi-dimensional Ornstein-Uhlenbeck process and it solution is given by
\begin{equation}
  \label{eq:aoysol}
  \bsrm{Y}^t = \exp(\bsrm{A}t) \cdot \bsrm{Y}^0 +\int_0^t\exp[\bsrm{A}(t-\tau)] \cdot \bsrm{B} \cdot \dd \bsrm{W}_Y (\tau)
\end{equation}
where $\bsrm{W}_Y(t)$ is a multi-dimensional Wiener process. As we will deal with the stationary measure in the following, we are particularly interested in the stationary solution which reads:
\begin{equation}
  \label{eq:aoystat}
  \bsrm{Y}^t = \int_{-\infty}^t\exp[\bsrm{A}(t-\tau)] \cdot \bsrm{B} \cdot \dd \bsrm{W}_Y(\tau) .
\end{equation}
In the following, we will simplify the notation and write $\bsrm{E}(t) = \exp(\bsrm{A} t)$ and $\bsrm{W}(t) = \bsrm{W}_Y(t)$. We have thus for instance for the stationary solution:
\begin{equation}
  \label{eq:aoystat2}
  Y^t_\alpha = q_Y \, \sum_{\beta\in N^Y} \, \int_{-\infty}^t E_{\alpha\beta}{(t-\tau)} \, \dd W_\beta(\tau) \quad , \quad \beta \in N^Y .
\end{equation}
We can now turn on the computation of the aforementioned quantities.

\subsection{The term $\bsrm{M}_1$}
\label{sec:aom1}
The term $\bsrm{M}_1$ decomposes naturally as:
\begin{equation}
  \bsrm{M}_1 =\left[
  \begin{array}{c}
    \bsrm{M}_1^{\rm a} \\ \bsrm{M}_1^{\rm o}
  \end{array}
  \right]
\end{equation}
For the atmospheric part, we have:
\begin{equation}
  \label{eq:am1def}
  M_{1,i}^{\rm a} =  \varepsilon \,\left\langle \sum_{\alpha,\beta \in N^Y} \, C_{i\alpha\beta}^{\rm a} \, Y_\alpha\, Y_\beta \right\rangle_{\rho_{0,Y}} \quad , \quad i \in N^{\rm a}
\end{equation}
where $\rho_{0,Y}$ is the stationary measure of the aforementioned Ornstein-Uhlenbeck process. We thus have
\begin{multline}
  \label{eq:aom1res1}
  M_{1,i}^{\rm a} = \varepsilon \sum_{\alpha,\beta,\zeta,\kappa \in N^Y} \,  C_{i\alpha\beta}^{\rm a} \, q_Y^2 \, \left\langle \int_{-\infty}^t E_{\alpha\zeta}{(t-\tau)} \, \dd W_\zeta(\tau) \right. \\ \times \left. \int_{-\infty}^t E_{\beta\kappa}{(t-\tau)} \, \dd W_\kappa(\tau) \right\rangle
\end{multline}
which by the statistical independence of the $W_i$'s and by the correlation formula~\cite{G2009} becomes:
\begin{equation}
  \label{eq:aom1res2}
  M_{1,i}^{\rm a} = \varepsilon \sum_{\alpha,\beta\in N^Y} \, C_{i\alpha\beta}^{\rm a} \, \sigma^Y_{\alpha\beta}
\end{equation}
where 
\begin{align}
  \sigma^Y_{\alpha\beta} & = \Big\langle Y_\alpha \, Y_\beta \Big\rangle_{\rho_{0,Y}}    \quad , \quad \alpha,\beta \in N^Y  \\
  & = q_Y^2 \sum_{\zeta\in N^Y} \, \int_{-\infty}^t E_{\alpha\zeta}{(t-\tau)}\, E_{\beta\zeta}{(t-\tau)} \, \dd\tau \label{eq:OUcov}
\end{align}
is the covariance matrix of the Ornstein-Uhlenbeck process.\\

The ocean term is quite straightforward:
\begin{align}
  \label{eq:om1def}
  M_{1,i}^{\rm o} & = \varepsilon \left\langle\sum_{\alpha\in N^Y}\,  R_{i\alpha}^{\rm o} \, Y_\alpha\right\rangle_{\rho_{0,Y}} \\
  & =  \varepsilon \sum_{\alpha\in N^Y}\, R_{i\alpha}^{\rm o} \, \Big\langle Y_\alpha\Big\rangle_{\rho_{0,Y}} = 0   \quad , \quad i \in N^{\rm o}
\end{align}

\subsection{The function $\bsrm{g}(s)$}
\label{sec:aogs}

The function $\bsrm{g}(s)$ is defined as:
\begin{equation}
  \bsrm{g}(s) = \Big \langle \bsrm\Psi_X^\prime(\bsrm{Y}) \otimes \bsrm\Psi_X^\prime\Big(\bsrm\phi^s_Y(\bsrm{Y})\Big) \Big \rangle_{\rho_{0,Y}}
\end{equation}
where $\otimes$ is here the \emph{vector} outer product\footnote{For two vectors $\bsrm{u}$ and $\bsrm{v}$, $(\bsrm{u}\otimes \bsrm{v})_{ij} = u_i\, v_j$.}.
By decomposing $\bsrm\Psi_X^\prime$ given by Eq.~(\ref{eq:psipdef}) into its atmospheric and oceanic components
\[\bsrm\Psi_X^\prime(\bsrm{Y}) = \left[
  \begin{array}{c}
    \bsrm\Psi_{X_{\rm a}}^\prime(\bsrm{Y}) \\
    \bsrm\Psi_{X_{\rm o}}^\prime(\bsrm{Y})
  \end{array}
\right] = \left[
  \begin{array}{c}
    \bsrm\Psi_{X_{\rm a}}(\bsrm{Y}) -\bsrm{M}_1^{\rm a}\\
    \bsrm\Psi_{X_{\rm o}}(\bsrm{Y}) -\bsrm{M}_1^{\rm o}
  \end{array}
\right]\]
this function becomes
\begin{equation}
  \bsrm{g}(s) = \left[
    \begin{array}[c]{cc}
       \bsrm{g}_{\rm{a}}(s) & \bsrm{g}_{\rm{ao}}(s) \\
       \bsrm{g}_{\rm{oa}}(s)& \bsrm{g}_{\rm{o}}(s)
    \end{array}
    \right]
\end{equation}
with
\begin{align}
  \bsrm{g}_{\rm{a}}(s) & = \langle \bsrm\Psi^\prime_{X_{\rm a}}(\bsrm{Y}) \otimes \bsrm\Psi^\prime_{X_{\rm a}}\big(\bsrm\phi^s_Y(\bsrm{Y})\big) \rangle_{\rho_{0,Y}} \\ 
  \bsrm{g}_{\rm{ao}}(s) & = \langle \bsrm\Psi^\prime_{X_{\rm a}}(\bsrm{Y}) \otimes \bsrm\Psi^\prime_{X_{\rm o}}\big(\bsrm\phi^s_Y(\bsrm{Y})\big) \rangle_{\rho_{0,Y}} \\
  \bsrm{g}_{\rm{oa}}(s) & = \langle \bsrm\Psi^\prime_{X_{\rm o}}(\bsrm{Y}) \otimes \bsrm\Psi^\prime_{X_{\rm a}}\big(\bsrm\phi^s_Y(\bsrm{Y})\big) \rangle_{\rho_{0,Y}} \\
  \bsrm{g}_{\rm{o}}(s) & =  \langle \bsrm\Psi^\prime_{X_{\rm o}}(\bsrm{Y}) \otimes \bsrm\Psi^\prime_{X_{\rm o}}\big(\bsrm\phi^s_Y(\bsrm{Y})\big) \rangle_{\rho_{0,Y}}
\end{align}
The off-diagonal terms are equal to zero, because they are proportional to $\bsrm{Y}$ and $\bsrm{Y}^3$ and therefore their average vanishes\footnote{The average of an odd power of a stationary Ornstein-Uhlenbeck process vanishes.}. We are thus left with the matrix:
\begin{equation}
  \bsrm{g}(s) = \left[
    \begin{array}[c]{cc}
      \bsrm{g}_{\rm a}(s) & 0 \\
      0 & \bsrm{g}_{\rm o}(s)
    \end{array}
    \right]
\end{equation}
For the atmosphere, we have:
\begin{equation}
  \label{eq:gsadef}
  \bsrm{g}_{\rm a}(s) = \Big\langle \bsrm\Psi_{X_{\rm a}}^\prime(\bsrm{Y}) \otimes  \bsrm\Psi_{X_{\rm a}}^\prime \big(\bsrm\phi^s_Y(\bsrm{Y})\big) \Big\rangle_{\rho_{0,Y}} \quad .
\end{equation}
Using the definition~(\ref{eq:psipdef}), we get
\begin{align}
  \bsrm{g}_{\rm a}(s) & = \Big\langle \left(\bsrm\Psi_{X_{\rm a}}(\bsrm{Y}) - \bsrm{M}_1^{\rm a}\right) \otimes \left(\bsrm\Psi_{X_{\rm a}} \big(\bsrm\phi^s_Y(\bsrm{Y})\big)  - \bsrm{M}_1^{\rm a}\right) \Big\rangle_{\rho_{0,Y}} \nonumber  \\
  & = \Big\langle \bsrm\Psi_{X_{\rm a}}(\bsrm{Y}) \otimes  \bsrm\Psi_{X_{\rm a}} \big(\bsrm\phi^s_Y(\bsrm{Y})\big) \Big\rangle_{\rho_{0,Y}} - \bsrm{M}_1^{\rm a} \otimes \bsrm{M}_1^{\rm a}   \label{eq:gsadefdep}
\end{align}
We thus focus on the term:
\begin{multline}
  \Big\langle \Psi_{X_{\rm a},i}(Y) \, \Psi_{X_{\rm a},j} \big(\phi^s_Y(Y)\big) \Big\rangle_{\rho_{0,Y}} =  \\ \sum_{\alpha,\beta,\kappa,\zeta\in N^Y} \varepsilon^2 \, C_{i\alpha\beta}^{\rm a} \, C_{j\kappa\zeta}^{\rm a} \Big\langle Y_\alpha \, Y_\beta \, Y_\kappa^s \, Y_\zeta^s \Big\rangle_{\rho_{0,Y}} \quad , \quad i,j\in N^{\rm a}
\end{multline}
where $Y_\alpha^s = \phi^s_{Y,\alpha}(\bsrm{Y})$ (with $\bsrm\phi^s_Y$ the flow of $\dbsrm{Y}=F_Y(\bsrm{Y})$) and where $Y_\alpha = Y_\alpha^0$.
This term can be rewritten:
\begin{multline}
  \label{eq:gsasimp}
  \Big\langle \Psi_{X_{\rm a},i}(Y) \,  \Psi_{X_{\rm a},j} \big(\phi^s_Y(Y)\big) \Big\rangle_{\rho_{0,Y}} =  \varepsilon^2 \sum_{\alpha,\beta,\kappa,\zeta\in N^Y} C_{i\alpha\beta}^{\rm a} \, C_{j\kappa\zeta}^{\rm a} \\ \times \Big( \big\langle Y_\alpha \, Y_\beta \big\rangle_{\rho_{0,Y}} \, \big\langle Y^s_\kappa \, Y^s_\zeta \big\rangle_{\rho_{0,Y}}
  + \big\langle Y_\alpha \, Y_\kappa^s \rangle_{\rho_{0,Y}} \, \big\langle Y_\beta \, Y^s_\zeta \big\rangle_{\rho_{0,Y}} \\
  + \big\langle Y_\alpha \, Y_\zeta^s \big\rangle_{\rho_{0,Y}} \, \big\langle Y_\beta \, Y^s_\kappa \big\rangle_{\rho_{0,Y}} \Big) 
\end{multline}
\begin{multline}
  =  M_{1,i}^{\rm a} \, M_{1,j}^{\rm a} + \varepsilon^2 \sum_{\alpha,\beta,\kappa,\zeta\in N^Y} C_{i\alpha\beta}^{\rm a} \, C_{j\kappa\zeta}^{\rm a} \\
  \times \sum_{\mu,\nu\in N^Y} \Big(  \sigma^Y_{\alpha \mu} \, E^{\rm T}_{\mu \kappa }({s})  \, \sigma^Y_{\beta \nu }\, E^{\rm T}_{ \nu \zeta}({s}) \\
  + \sigma^Y_{\alpha \mu} \, E^{\rm T}_{\mu \zeta}({s})  \,  \sigma^Y_{\beta \nu} \, E^{\rm T}_{\nu \kappa}({s})   \Big)
\end{multline}
where we have used Eq.~(\ref{eq:aom1res2}) and 
\begin{align}
  \big\langle Y_\alpha \, Y_\beta^s \big\rangle_{\rho_{0,Y}} & = q_Y^2 \sum_{\mu\in N^Y} \int_{-\infty}^0 E_{\alpha\mu}({-\tau}) \, E_{\beta\mu}{(s-\tau)} \, \dd\tau \nonumber \\ 
  & = \sum_{\mu,\nu \in N^Y} E^{\rm T}_{\nu \beta}({s}) \, q_Y^2 \, \int_{-\infty}^0 E_{\alpha \mu}({-\tau}) \, E^{\rm T}_{\mu\nu}({-\tau}) \, \dd\tau \nonumber  \\ 
  & = \sum_{\nu\in N^Y} \sigma^Y_{\alpha \nu} E^{\rm T}_{\nu \beta}({s})\quad , \quad \alpha,\beta\in N^Y   \label{eq:Ycorr}
\end{align}
where $\bsrm{E}^{\rm T}$ is the transpose of $\bsrm{E}$ and we have $\bsrm{E}(t)\cdot \bsrm{E}(\tau) = \bsrm{E}({t+\tau})$ by definition.
Finally we thus get:
\begin{multline}
  g_{{\rm a},ij}(s) = \varepsilon^2 \sum_{\alpha,\beta,\kappa,\zeta\in N^Y} \sum_{\mu,\nu\in N^Y} C_{i\alpha\beta}^{\rm a} \, C_{j\kappa\zeta}^{\rm a} \\
  \times  \Big(  \sigma^Y_{\alpha \mu} \, E^{\rm T}_{\mu \kappa }({s})  \, \sigma^Y_{\beta \nu }\, E^{\rm T}_{ \nu \zeta}({s}) + \sigma^Y_{\alpha \mu} \, E^{\rm T}_{\mu \zeta}({s})  \,  \sigma^Y_{\beta \nu} \, E^{\rm T}_{\nu \kappa}({s})   \Big)
\end{multline}
which, with $\left(\bsrm\sigma^Y\right)^{\rm T} = \bsrm\sigma^{Y}$, can be written in the form of Eq.~(\ref{eq:gsa}).

For the ocean, the function is defined by the expression:
\begin{equation}
  \label{eq:gsodef}
  \bsrm{g}_{{\rm o}}(s) = \Big\langle \bsrm\Psi_{X_{\rm o}}^\prime(\bsrm{Y}) \otimes \bsrm\Psi_{X_{\rm o}}^\prime \big(\bsrm\phi^s_Y(\bsrm{Y})\big) \Big\rangle_{\rho_{0,Y}} 
\end{equation}
and since $\bsrm{M}_1^{\rm o}=0$, we have $\bsrm\Psi_{X_{\rm o}}^\prime(\bsrm{Y})=\bsrm\Psi_{X_{\rm o}}(\bsrm{Y})$ and it follows that
\begin{align}
  g_{{\rm o},ij}(s) & = \sum_{\alpha,\beta\in N^Y} \Big\langle \varepsilon^2 \, R_{i\alpha}^{\rm o} \, Y_\alpha \, R_{j\beta}^{\rm o} \, Y_\beta^s \Big\rangle_{\rho_{0,Y}} \quad, \quad i,j \in N^{\rm o}\\
  & = \sum_{\alpha,\beta\in N^Y} \varepsilon^2 \, R_{i\alpha}^{\rm o}  \, R_{j\beta}^{\rm o} \,  \Big\langle Y_\alpha \, Y_\beta^s \Big\rangle_{\rho_{0,Y}} \\
   & = \sum_{\alpha,\beta,\nu\in N^Y} \varepsilon^2 \, R_{i\alpha}^{\rm o}  \, R_{j\beta}^{\rm o} \, \sigma^Y_{\alpha \nu} \, E^{\rm T}_{\nu \beta}({s})
\end{align}
which gives the result (\ref{eq:gso}).

\subsection{The function $\bsrm{h}(\bsrm{X},s)$}
\label{sec:aohXs}
The function $\bsrm{h}(\bsrm{X},s)$ decomposes straightforwardly into
\begin{equation}
  \bsrm{h}(\bsrm{X},s)= \left[
  \begin{array}{c}
    \bsrm{h}_{\rm a}(\bsrm{X},s) \\
    \bsrm{h}_{\rm o}(\bsrm{X},s)
  \end{array}
  \right]
\end{equation}
For the atmosphere, the function $\bsrm{h}_{\rm a}(\bsrm{X},s)$ is given by the expression:
\begin{multline}
  h_{{\rm a},i}(\bsrm{X},s) = \sum_{\alpha,\beta,\zeta,\nu\in N^Y} \sum_{m\in N^{\rm a},l\in N^{\rm o}} \Big\langle \varepsilon^2 \, \Big(V_{\alpha m\beta}^Y \, X_{{\rm a},m} \, Y_\beta \\ + R_{\alpha l}^Y \, X_{{\rm o},l}\Big) \,\partial_{Y_\alpha} \Big[ C_{i\zeta\nu}^{\rm a} \, Y^s_\zeta \, Y^s_\nu \Big] \Big\rangle_{\rho_{0,Y}} \quad , \quad i \in N^{\rm a}\\
  = \sum_{\alpha,\beta,\zeta,\nu\in N^Y} \sum_{m\in N^{\rm a}}\Big\langle \varepsilon^2 \, V^Y_{\alpha m\beta} \, X_{{\rm a},m} \, Y_\beta \, C_{i\zeta\nu}^{\rm a} \, \partial_{Y_\alpha} \Big[ Y^s_\zeta \, Y^s_\nu \Big] \Big\rangle_{\rho_{0,Y}} \\
  + \sum_{\alpha,\zeta,\nu\in N^Y} \sum_{l\in N^{\rm o}} \Big\langle \varepsilon^2 \, R^Y_{\alpha l} \, X_{{\rm o},l} \, C_{i\zeta\nu}^{\rm a} \, \partial_{Y_\alpha} \Big[ Y^s_\zeta \, Y^s_\nu \Big] \Big\rangle_{\rho_{0,Y}} \label{eq:hXsadef}
\end{multline}
where $Y_\alpha^s = \phi^s_{Y,\alpha}(\bsrm{Y})$ (with $\bsrm\phi^s_Y$ the flow of $\dbsrm{Y}=F_Y(\bsrm{Y})$) and where $Y_\alpha = Y_\alpha^0$. To compute the derivative in this latter expression, we have to use the non-stationary solution (\ref{eq:aoysol}), and we get:
\begin{multline}
  \partial_{Y_\alpha} \Big[ Y_\zeta^s \, Y_\nu^s \Big] = \sum_{\mu\in N^Y} \Big( E_{\zeta\alpha}(s) \, E_{\nu\mu}(s) + E_{\nu\alpha}(s) \, E_{\zeta\mu}(s)\Big) \, Y_\mu \\
  + q_Y \sum_{\mu\in N^Y} \int_0^s \Big( E_{\zeta\alpha}(s) \, E_{\nu\mu}(s-\tau) \\ + E_{\nu\alpha}(s) \, E_{\zeta\mu}(s-\tau)\Big) \, \dd W_\mu(\tau) \label{eq:deryy}
\end{multline}
and since for $\zeta,\mu \in N^Y$
\[\left\langle Y_\zeta \, \int_0^s \ldots \, \dd W_\mu(\tau) \right\rangle = 0 \quad , \]
we have:
\begin{multline}
  \left\langle Y_\beta \, \partial_{Y_\alpha} \Big[ Y_\zeta^s \, Y_\nu^s \Big] \right\rangle_{\rho_{0,Y}} = \\ \sum_{\mu\in N^Y} \Big( E_{\zeta\alpha}(s) \, E_{\nu\mu}(s) + E_{\nu\alpha}(s) \, E_{\zeta\mu}(s)\Big) \, \Big\langle Y_\beta Y_\mu \Big\rangle_{\rho_{0,Y}} \label{eq:avyderyy}
\end{multline}
and
\begin{multline}
  \left\langle \partial_{Y_\alpha} \Big[ Y_\zeta^s \, Y_\nu^s \Big] \right\rangle_{\rho_{0,Y}} = \\ \sum_{\mu\in N^Y} \Big( E_{\zeta\alpha}(s) \, E_{\nu\mu}(s) + E_{\nu\alpha}(s) \, E_{\zeta\mu}(s)\Big) \, \Big\langle Y_\mu \Big\rangle_{\rho_{0,Y}} \quad .
 \label{eq:avderyy}
\end{multline}
Since $\Big\langle Y_\mu \Big\rangle_{\rho_{0,Y}} = 0$ and $\Big\langle Y_\beta Y_\mu \Big\rangle_{\rho_{0,Y}} = \sigma_{\beta\mu}^Y \, $, we get for Eq.~(\ref{eq:hXsadef}):
\begin{multline}
  h_{{\rm a},i}(\bsrm{X}_{\rm a},s)  = \varepsilon^2 \sum_{\alpha,\beta,\zeta,\nu\in N^Y} \sum_{m\in N^{\rm a}} V^Y_{\alpha m\beta} \, X_{{\rm a},m} \, \, C_{i\zeta\nu}^{\rm a} \\
  \times \Big( E_{\zeta\alpha}(s) \, E_{\nu\mu}(s) + E_{\nu\alpha}(s) \, E_{\zeta\mu}(s)\Big)\,\sigma_{\beta\mu}^Y \label{eq:hXsares}
\end{multline}
which is in fact Eq.~(\ref{eq:hXsa}).

For the ocean, the function is given by the expression:
\begin{multline}
  \label{eq:hXsodef}
  h_{{\rm o},i}(\bsrm{X}_{\rm o},s) = \sum_{\alpha,\beta,\zeta\in N^Y} \sum_{m\in N^{\rm a},l\in N^{\rm o}} \Big\langle \varepsilon^2 \, \Big(V_{\alpha m\beta}^Y \, X_{{\rm a},m} \, Y_\beta \\ + R_{\alpha l}^Y \, X_{{\rm o},l}\Big) \,\partial_{Y_\alpha} \Big[ R^{\rm o}_{i\zeta} \, Y^s_\zeta \Big] \Big\rangle_{\rho_{0,Y}} \quad , \quad i \in N^{\rm o}  \\
  = \sum_{\alpha,\beta,\zeta\in N^Y} \sum_{m\in N^{\rm a}} \Big\langle \varepsilon^2 \, V_{\alpha m\beta}^Y \, X_{{\rm a},m} \, Y_\beta \, R^{\rm o}_{i\zeta} \,\partial_{Y_\alpha}  Y^s_\zeta  \Big\rangle_{\rho_{0,Y}} \\
  + \sum_{\alpha,\zeta\in N^Y} \sum_{l\in N^{\rm o}} \Big\langle \varepsilon^2 \, R_{\alpha l}^Y \, X_{{\rm o},l}\, R^{\rm o}_{i\zeta} \,\partial_{Y_\alpha}  \, Y^s_\zeta \Big\rangle_{\rho_{0,Y}}
\end{multline}
and since $\partial_{Y_\alpha} Y_\zeta^s = E_{\zeta\alpha}(s)$, we have:
\begin{align}
  \left\langle Y_\beta \, \partial_{Y_\alpha} Y_\zeta^s \right\rangle_{\rho_{0,Y}} & = E_{\zeta\alpha}(s)  \left\langle Y_\beta \right\rangle_{\rho_{0,Y}} = 0   \label{eq:avdery} \\
\left\langle \partial_{Y_\alpha} Y_\zeta^s \right\rangle_{\rho_{0,Y}} & = E_{\zeta\alpha}(s)  \label{eq:avydery}
\end{align}
and it implies the result (\ref{eq:hXso}).
\section{A simple example}
\label{sec:simpex}

\begin{figure*}
  \begin{subfigure}{0.48\textwidth}
    \centering
    \includegraphics[width=\linewidth]{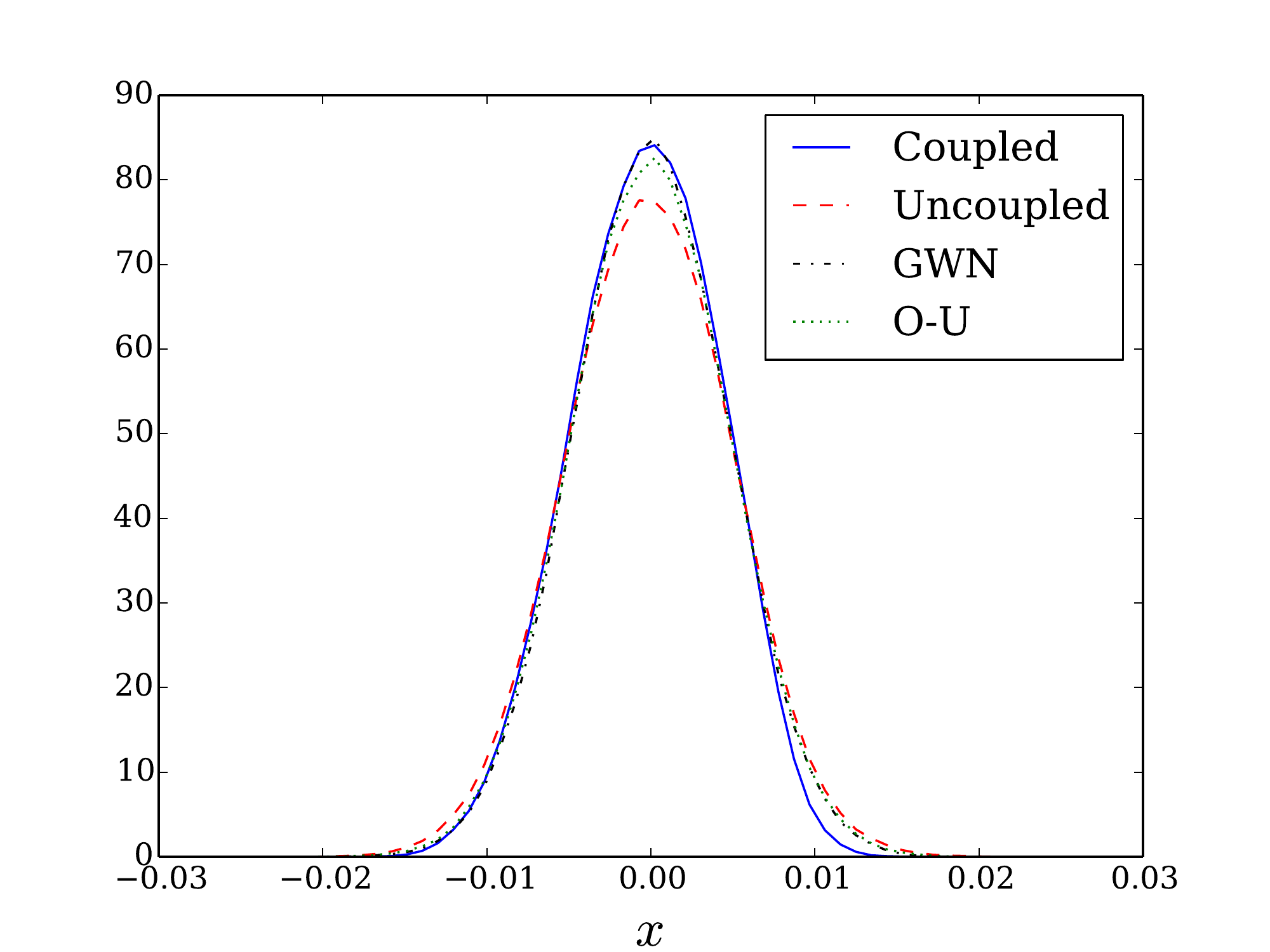}
    \caption{Probability density of $x$.}
  \end{subfigure}
  \begin{subfigure}{0.48\textwidth}
    \centering
    \includegraphics[width=\linewidth]{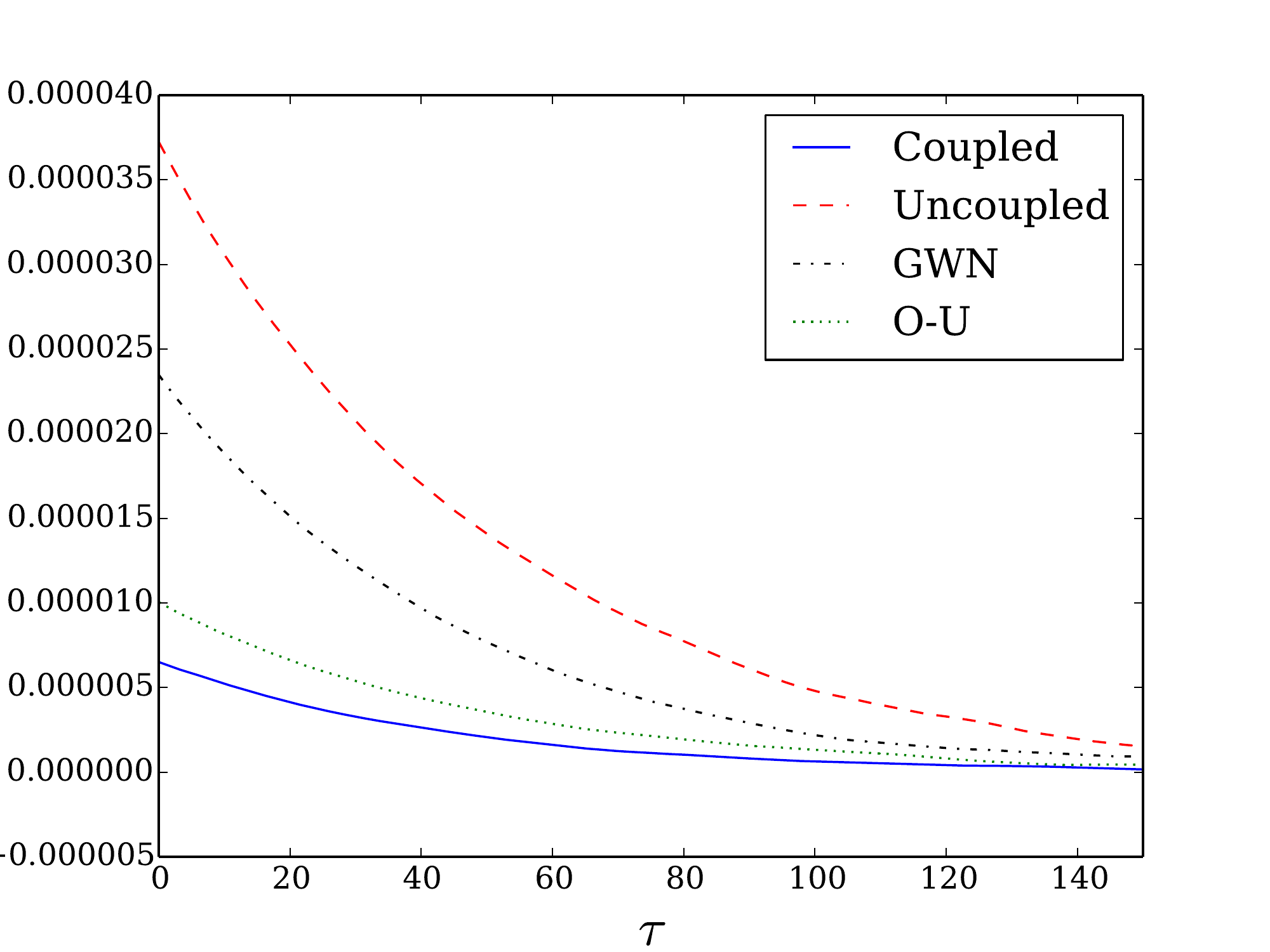}
    \caption{Autocorrelation function of $x$ as function of the lag-time $\tau$.}
  \end{subfigure}
  \begin{subfigure}{0.48\textwidth}
    \centering
    \includegraphics[width=\linewidth]{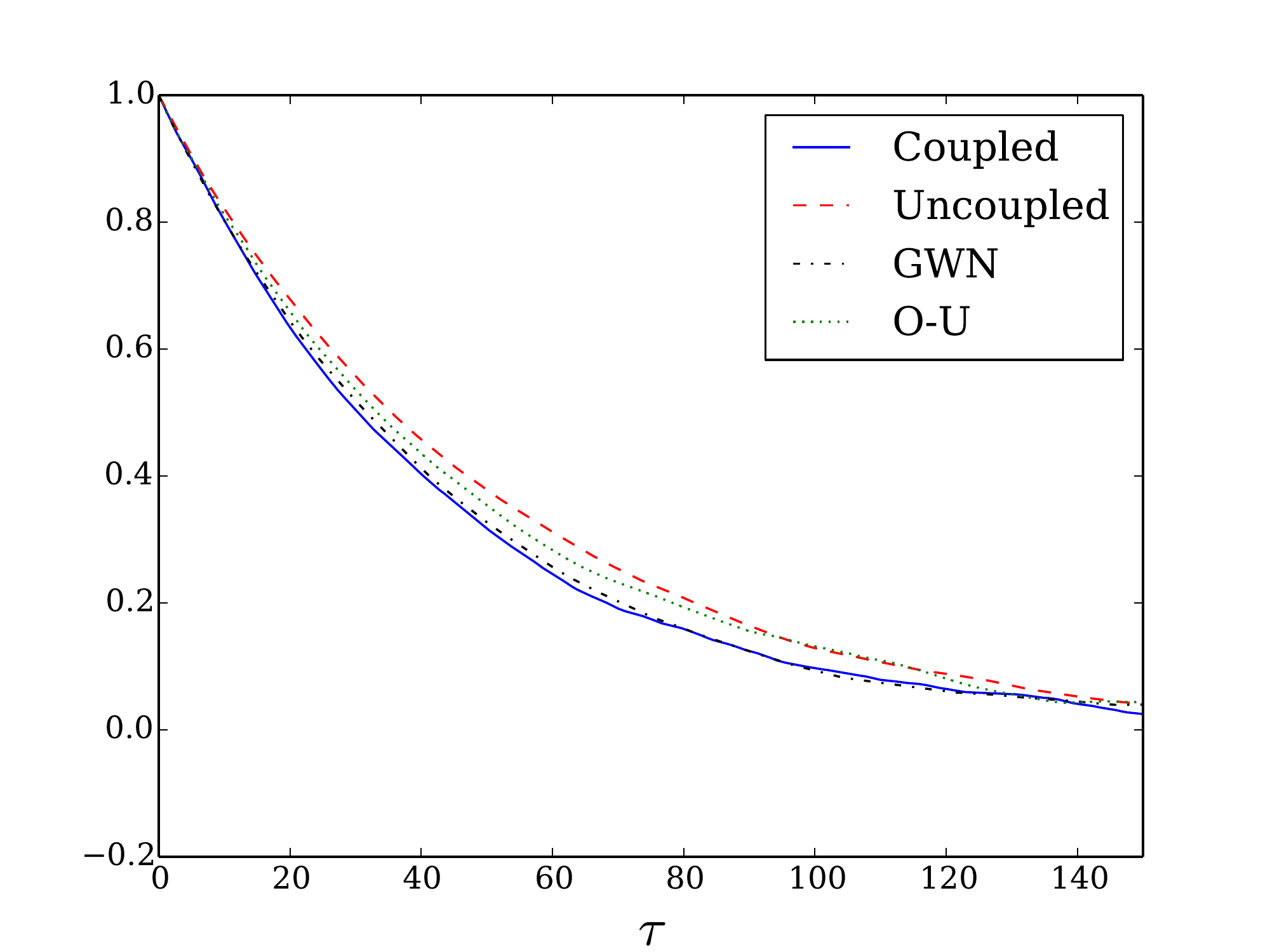}
    \caption{Normalized autocorrelation of $x$.}
  \end{subfigure}
  \caption{Probability density and autocorrelation of $x$. \label{fig:appsimp}}
\end{figure*}

In this section, we apply the results of the previous appendix to a simple 3-dimensional model:
\begin{equation}
  \label{eq:smod}
  \left\{
  \begin{array}{lcl}
    \dot x & = & b \, x + q \, \xi(t) + C \, y_1 y_2 \\
    \dot y_1 & = & a \, y_1 + \beta\, y_2 + q\, \xi_1(t) + V_1 \, x\, y_2  \\
    \dot y_2 & = & -\beta\, y_1 + a\, y_2 + q\, \xi_2(t) + V_2\, x\, y_1 .   
  \end{array}
  \right.
\end{equation}
with $a<b<0$ and where the $\xi$'s are Gaussian white noise processes. The variables $y_1$ and $y_2$ form the unresolved component to be reduced. The $x$ variable alone forms the resolved component. Applying the formula~(\ref{eq:M1a}), (\ref{eq:gsa}) and (\ref{eq:hXsa}) previously derived for the atmospheric component to the present case, with $\varepsilon=1$ and the covariance matrix
\begin{equation}
  \bsrm\sigma^Y = \left[
    \begin{array}{cc}
      -q^2/2a & 0 \\
      0 & -q^2/2a
    \end{array}
    \right]
\end{equation}
and the exponential matrix
\begin{equation}
  \bsrm{E}(t) = e^{at} \, \left[
    \begin{array}{cc}
      \cos(\beta t) & \sin(\beta t) \\
      -\sin(\beta t) & \cos(\beta t) 
    \end{array}
    \right]
\end{equation}
one gets:
\begin{align}
  M_1 & = 0 \label{eq:sM1} \\
  g(s) & = C^2 \, \frac{q^4}{4 a^2} \, e^{2 a s} \cos(2\beta s) \label{eq:sgs} \\
  h(x,s) & = -(V_1+V_2) \, x \, C \, \frac{q^2}{2 a} \, e^{2 a s} \cos(2\beta s) \label{eq:shXs} 
\end{align}
The terms $M_2$ and $M_3$ are computed using respectively $g(s)$ and $h(X,s)$.
As in Sec.~\ref{sec:red_add}, we test the method with two different parameterizations for $M_2$: one with a Gaussian white noise (GWN) and the other with a two-dimensional Ornstein-Uhlenbeck (O-U) process:
\begin{equation*}
  \left\{
  \begin{array}{lcl}
    \dot y_1 & = & a \, y_1 + \beta\, y_2 + q\, \xi_1(t) \\
    \dot y_2 & = & -\beta\, y_1 + a\, y_2 + q\, \xi_2(t)  
  \end{array}
  \right.
\end{equation*}
We integrated the model with a stochastic Heun scheme with a timestep $\Delta t=0.01$. The function $h(x,s)$ is computed at each timestep and memory term $M_3$ is re-evaluated at the same frequency.
The results are shown on Figs.~\ref{fig:appsimp} (a), (b) and (c) for the parameters
\begin{multline*}
  a = -0.05 \, ,  \quad
      b   =   -0.02 \, ,  \quad
      \beta   =   0.5  \, ,  \quad
      C   =   -20.5 \, ,  \quad \\
      V_1   =   40.2 \, ,  \quad
      V_2   =   56.2 \, ,  \quad
      q   =   0.001
\end{multline*}
 Both parameterizations improve well the probability density of the resolved variable $x$. However, the Gaussian white noise parameterization seems to improve the probability density and the decorrelation time better while the Ornstein-Uhlenbeck parameterization improves the variance better.

\bibliographystyle{plainnat}
\setlength{\bibsep}{4pt}
\bibliography{art_irm}

\end{document}